\documentclass[12pt,english]{extarticle}
\usepackage[T1]{fontenc}
\usepackage[latin9]{inputenc}
\usepackage[a4paper]{geometry}
\usepackage[active]{srcltx}
\usepackage{color}
\usepackage{bm}
\usepackage{amsmath}
\usepackage{amsthm}
\usepackage{amssymb}
\usepackage{graphicx}
\usepackage{setspace}
\usepackage[authoryear]{natbib}
\PassOptionsToPackage{normalem}{ulem}
\usepackage{ulem}
\makeatletter
\theoremstyle{plain}
\newtheorem{thm}{\protect\theoremname}
\newenvironment{lyxlist}[1]
	{\begin{list}{}
		{\settowidth{\labelwidth}{#1}
		 \setlength{\leftmargin}{\labelwidth}
		 \addtolength{\leftmargin}{\labelsep}
		 }}
	{\end{list}}
\theoremstyle{plain}
\newtheorem{assumption}{\protect\assumptionname}
\theoremstyle{plain}
\newtheorem*{cor*}{\protect\corollaryname}
\theoremstyle{plain}
\newtheorem{prop}{\protect\propositionname}

\@ifundefined{date}{}{\date{}}
\bibpunct{(}{)}{,}{a}{,}{,}

\usepackage{babel}

\usepackage[multiple]{footmisc}

\usepackage{chngcntr}
\usepackage{apptools}
\AtAppendix{\counterwithin{lemma}{section}}

\usepackage{scrextend}

\makeatletter
\def\thmhead@plain#1#2#3{%
  \thmname{#1}\thmnumber{\@ifnotempty{#1}{ }\@upn{#2}}%
  \thmnote{ {\the\thm@notefont#3}}}
\let\thmhead\thmhead@plain
\makeatother

\usepackage{pgfpages}

\makeatother

\usepackage{babel}
\providecommand{\assumptionname}{Assumption}
\providecommand{\corollaryname}{Corollary}
\providecommand{\propositionname}{Proposition}
\providecommand{\theoremname}{Theorem}

\begin{document}
\title{\vspace{20pt}
Subgeometrically ergodic autoregressions\thanks{The authors thank the Academy of Finland for financial support, and
Co-Editor Donald Andrews and two anonymous referees for useful comments
and suggestions. Contact addresses: Mika Meitz, Department of Economics,
University of Helsinki, P. O. Box 17, FI\textendash 00014 University
of Helsinki, Finland; e-mail: mika.meitz@helsinki.fi. Pentti Saikkonen,
Department of Mathematics and Statistics, University of Helsinki,
P. O. Box 68, FI\textendash 00014 University of Helsinki, Finland;
e-mail: pentti.saikkonen@helsinki.fi.}\vspace{20pt}
}
\author{Mika Meitz\\\small{University of Helsinki} \and Pentti Saikkonen\\\small{University of Helsinki}\vspace{20pt}
}
\date{First version April 2019, revised February 2020}
\maketitle
\begin{abstract}
\noindent In this paper we discuss how the notion of subgeometric
ergodicity in Markov chain theory can be exploited to study stationarity
and ergodicity of nonlinear time series models. Subgeometric ergodicity
means that the transition probability measures converge to the stationary
measure at a rate slower than geometric. Specifically, we consider
suitably defined higher-order nonlinear autoregressions that behave
similarly to a unit root process for large values of the observed
series but we place almost no restrictions on their dynamics for moderate
values of the observed series. Results on the subgeometric ergodicity
of nonlinear autoregressions have previously appeared only in the
first-order case. We provide an extension to the higher-order case
and show that the autoregressions we consider are, under appropriate
conditions, subgeometrically ergodic. As useful implications we also
obtain stationarity and $\beta$-mixing with subgeometrically decaying
mixing coefficients.

\bigskip{}
\bigskip{}
\bigskip{}

\noindent\textbf{JEL classification:} C22

\bigskip{}

\noindent \textbf{Keywords:} Nonlinear autoregressive model, subgeometric
ergodicity, Markov chain, $\beta$-mixing.
\end{abstract}
\vfill{}

\pagebreak{}

\section{Introduction}

Markov chain theory and the notion of geometric ergodicity have become
standard tools in econometrics and statistics when analyzing the stationarity
and ergodicity of nonlinear autoregressions or other nonlinear time
series models. A detailed discussion of the relevant Markov chain
theory will be given in Section \ref{sec:markov}. For now, consider
a Markov chain $X_{t}$ ($t=0,1,2,\ldots$) on the state space $\mathsf{X}$
and initialized from $X_{0}$ following some initial distribution\textcolor{red}{{}
}(that is not necessarily the stationary distribution). Geometric
ergodicity of $X_{t}$ entails that the $n$-step probability measures
$P^{n}(x\,;\,\cdot)=\Pr\left(X_{n}\in\cdot\,|\,X_{0}=x\right)$ converge
in total variation norm $\left\Vert \cdot\right\Vert _{TV}$ to the
stationary probability measure $\pi$ at rate $r^{n}$ (for some $r>1$),
that is, 
\begin{equation}
\lim_{n\to\infty}r^{n}\lVert P^{n}(x\,;\,\cdot)-\pi\rVert_{TV}=0\label{eq:Geom-erg}
\end{equation}
(the definition of $\left\Vert \cdot\right\Vert _{TV}$ and a formulation
of (\ref{eq:Geom-erg}) using a more general norm are given in Section
\ref{sec:markov}). A common and convenient way to establish geometric
ergodicity involves the verification of a so-called drift condition.
Useful implications obtained with this approach include the existence
of a stationary probability distribution $\pi$ of $X_{t}$ as well
as the geometric $\beta$-mixing of $X_{t}$. (For a definition of
$\beta$-mixing, see \citet[Sec 1.1]{doukhan1994mixing} or \citet[Ch 3]{bradley2007introduction}.)\textcolor{red}{{}
}The authoritative and classic reference to Markov chain theory is
the monograph of \citet{meyn1993markov,meyn2009markov}. Recent papers
establishing geometric ergodicity of different nonlinear time series
models include \citet{francq2006mixing}, \citet{ling2007double},
\citet{meitz2008ergodicity}, and \citet*{fokianos2009poisson}, among
others.

In this paper we consider autoregressions that may exhibit rather
arbitrary (stationary, unit root, explosive, nonlinear, etc.) behavior
for moderate values of the observed series and that behave similarly
to a unit root process for large values of the observed series. What
this exactly means will be clarified shortly, but first we would like
to emphasize that the autoregressions we consider will not necessarily
be geometrically ergodic. Under appropriate conditions they will,
nevertheless, satisfy a weaker form of so-called subgeometric ergodicity.
A Markov chain is said to be subgeometrically ergodic when the convergence
in (\ref{eq:Geom-erg}) takes place at a rate $r(n)$ slower than
geometric, that is, 
\begin{equation}
\lim_{n\to\infty}r(n)\lVert P^{n}(x\,;\,\cdot)-\pi\rVert_{TV}=0.\label{eq:SubGeom-erg}
\end{equation}
In the geometric case $r(n)=r^{n}$ with $r>1$ or, equivalently,
$r(n)=e^{cn}$ with $c>0$. Examples of rates slower than geometric
include subexponential rates (say, $r(n)=e^{cn^{\gamma}}$ with $c>0$
and $\gamma\in(0,1)$) and polynomial rates (say, $r(n)=(1+n)^{\beta}$
with $\beta>0$). For an up-to-date treatment of subgeometric ergodicity
we refer to Chapters 16 and 17 of \citet*{douc2018markov} (further
references will be given below).

As will be discussed in Section 2, subgeometric ergodicity can conveniently
be established by verifying a suitably formulated drift condition
and useful implications analogous to those in the case of geometric
ergodicity again follow. In particular, the existence of a stationary
probability distribution $\pi$ of $X_{t}$ as well as the finiteness
of certain moments are obtained. Moreover, in a companion paper \citet{meitz2019subgemix}
we show that subgeometric ergodicity implies $\beta$-mixing with
subgeometrically decaying mixing coefficients. Subgeometric ergodicity
therefore allows one to use limit theorems developed for $\beta$-mixing
processes.

The main aims of this paper are to establish subgeometric ergodicity
of certain higher-order nonlinear autoregressions and to illustrate
the potential of the concept of subgeometric ergodicity for nonlinear
time series models. To facilitate discussion, first consider a simple
special case at an informal level. Specifically, consider the univariate
first-order nonlinear autoregressive model
\begin{equation}
y_{t}=g(y_{t-1})+\varepsilon_{t},\quad t=1,2,\ldots,\label{NLAR(1)}
\end{equation}
where the error term $\varepsilon_{t}$ is a sequence of independent
and identically distributed (IID) zero-mean random variables and $g$
is a real-valued function. For now, assume that $g$ is such that
\begin{equation}
\left|g(x)\right|\leq\left(1-r\left|x\right|^{-\rho}\right)\left|x\right|\quad\textrm{for }\left|x\right|\geq M_{0}\qquad\qquad[r>0,\,M_{0}>r^{1/\rho},\,0<\rho\leq2],\label{Ineq g(x)_p=00003D1}
\end{equation}
and that $g(x)$ is bounded for $\left|x\right|\leq M_{0}$. A concrete
example where (\ref{Ineq g(x)_p=00003D1}) can be easily verified
is
\begin{equation}
y_{t}=\left(1-\frac{r_{0}}{1+\left|y_{t-1}\right|^{\rho}}\right)y_{t-1}+\varepsilon_{t}\qquad\qquad[r_{0}>0,\,0<\rho\leq2].\label{Example_1}
\end{equation}
The model defined in (\ref{Example_1}) can be thought of as first-order
autoregression with a time-varying autoregressive coefficient. For
large values of $\left|y_{t-1}\right|$ the autoregressive coefficient
takes values that are close to one and the generation mechanism of
$y_{t}$ is close to a random walk while for small values of $\left|y_{t-1}\right|$
the autoregressive coefficient is close to $1-r_{0}$ and, for $r_{0}$
not very close to zero, $y_{t}$ is generated from a less persistent
stationary autoregressive process. Overall, the generation mechanism
of $y_{t}$ fluctuates between these two borderline cases. Simulated
examples in Section 5 demonstrate that processes of the type described
in (\ref{Example_1}) can exhibit behavior close to a random walk
for rather long times before returning to a less persistent regime.

When $\rho\geq1$ the model defined by equation (\ref{Example_1})
can be viewed as a special case of the model
\begin{equation}
y_{t}=y_{t-1}+\tilde{g}(y_{t-1})+\varepsilon_{t},\label{RW-type example_2}
\end{equation}
where the function $\tilde{g}$ is bounded (but not constant).\footnote{This model belongs to a class of models referred to as ``random-walk-type
Markov chains'' by \citet{jarner2003necessary}; see particularly
equation (3) of their paper.} In Section \ref{sec:model}, a higher-order version of equation (\ref{RW-type example_2})
(without assuming boundedness) is used as a starting point of the
formulation of our general model. Our main results in Section \ref{sec:results}
show that, depending on the assumptions made, either geometric, subexponential,
or polynomial ergodicity is obtained.

The preceding discussion illustrates what kind of behavior the autoregressions
we consider may exhibit for large values of the observed series. However,
it should be emphasized that inequality (\ref{Ineq g(x)_p=00003D1})
restricts the regression function $g$ only for large values of its
argument. As long as the assumed boundedness condition imposed on
the function $g$ is satisfied, no restrictions are required when
the process evolves in the vicinity of the origin. Allowing unit root
type behavior for large (absolute) values of the observed series is
the main feature which distinguishes the models we consider from most
previous nonlinear autoregressions where stationary behavior is related
to large (absolute) values of the process.\footnote{For instance, \citet{lu1998geometric}, \citet{gourieroux2006stochastic},
and \citet{bec2008acr}, among others, establish geometric ergodicity
(and thus the existence of a stationary distribution) for autoregressions
whose behavior approaches stationarity when the process moves away
from the origin while in the vicinity of the origin its behavior can
be rather arbitrary.} 

Previously results on subgeometric ergodicity of nonlinear autoregressions
have been obtained in the probability literature by \citet{tuominen1994subgeometric},
\citet{veretennikov2000polynomial}, \citet{fort2003polynomial},
\citet{douc2004practical}, \nocite{klokov2004sub,klokov2005subexponential}Klokov
and Veretennikov (2004, 2005), and \citet{klokov2007lower}, among
others (further discussion on these and some related papers will be
provided in Section 4). To our knowledge, all of the previous results
concern only first-order models. We contribute to this literature
by obtaining results for more general higher-order autoregressions.
This is achieved using techniques similar to those in the aforementioned
papers, especially in \citet{fort2003polynomial} and \citet{douc2004practical}.
Depending on the assumptions imposed on the moments of the error term,
the resulting rate of ergodicity is either geometric or subexponential
or polynomial. 

The rest of the paper is organized as follows. Section \ref{sec:markov}
contains basic concepts of Markov chains and summarizes existing results
on subgeometric ergodicity. Section \ref{sec:model} introduces the
nonlinear autoregressive model considered and states the assumptions
used to obtain the results of the paper. The main results on subexponential
and polynomial ergodicity are given in Section \ref{sec:results}.
In Section 5 we provide examples of our general model. Section 6 concludes.
All proofs are collected in an Appendix and a Supplementary Appendix. 

Finally, a few notational conventions are given. The minimum (maximum)
of the real numbers $x$ and $y$ is denoted by $x\wedge y$ ($x\lor y$),
and $L$ and $\Delta$ signify the lag operator and the difference
operator, respectively (so that $\Delta x{}_{t}=(1-L)x_{t}=x_{t}-x_{t-1}$).
The notation $\boldsymbol{1}_{S}(x)$ is used for the indicator function
which takes the value one when $x$ belongs to the set $S$ and zero
elsewhere, and $\left|\cdot\right|$ is used for both an absolute
value and Euclidean norm. Furthermore, $\text{\textbf{0}}{}_{k}$
denotes a $k\times1$ vector of zeros and $\boldsymbol{\iota}_{k}=(1,0,\ldots,0)$
($k\times1$).

\section{Markov chains and subgeometric ergodicity\label{sec:markov} }

In this section we discuss basic concepts of Markov chains needed
to obtain our results. More comprehensive discussions can be found
in \citet{meyn2009markov} and \citet{douc2018markov}. Let $X_{t}$
($t=0,1,2,\ldots$) be a Markov chain on a general measurable state
space $(\mathsf{X},\mathcal{B}(\mathsf{X}))$ (with $\mathcal{B}(\mathsf{X})$
the Borel $\sigma$-algebra) and let $P^{n}(x\,;\,A)=\Pr(X_{n}\in A\mid X_{0}=x)$
signify its $n$-step transition probability measure. As in \citet{fort2003polynomial}
and \citet{douc2004practical} our goal is to establish the convergence
\textemdash{} in a suitably defined norm and at rate $r(n)$ \textemdash{}
of the $n$-step probability measures $P^{n}(x\,;\,\cdot)$ to the
stationary distribution $\pi$. To this end, let $f:\mathsf{X}\rightarrow[1,\infty)$
be an arbitrary fixed measurable function and, for any signed measure
$\mu$, define the $f$-norm $\left\Vert \mu\right\Vert _{f}$ as
\begin{equation}
\left\Vert \mu\right\Vert _{f}=\sup_{f_{0}:\left|f_{0}\right|\leq f}\left|\mu(f_{0})\right|,\label{eq:f-norm}
\end{equation}
where $\mu(f_{0})=\int_{x\in\mathsf{X}}f_{0}(x)\mu(dx)$ (and the
supremum in (\ref{eq:f-norm}) runs over all measurable functions
$f_{0}:\mathsf{X}\to\mathbb{R}$ such that $\left|f_{0}(x)\right|\leq f(x)$
for all $x\in\mathsf{X}$). When $f\equiv1$, the $f$-norm $\left\Vert \mu\right\Vert _{f}$
reduces to the total variation norm $\left\Vert \mu\right\Vert _{TV}=\sup_{f_{0}:\left|f_{0}\right|\leq1}\left|\mu(f_{0})\right|$
used in (\ref{eq:Geom-erg}) and (\ref{eq:SubGeom-erg}). 

We aim to establish that the $n$-step probability measures $P^{n}(x\,;\,\cdot)$
converge in $f$-norm and at rate $r(n)$ to the stationary probability
measure $\pi$ satisfying $\pi(f)<\infty$, that is, that
\begin{equation}
\lim_{n\to\infty}r(n)\lVert P^{n}(x\,;\,\cdot)-\pi\rVert_{f}=0\qquad\text{for }\pi\text{-almost all }x\in\mathsf{X}.~\footnotemark\label{f-ergodicity}
\end{equation}
\textcolor{red}{\footnotetext{That is, the convergence in (\ref{f-ergodicity}) is required to hold for all $x\in\mathsf{X}$ except for those $x$ in a set that has probability zero with respect to the stationary measure $\pi$.}}If
(\ref{f-ergodicity}) holds we say that the Markov chain $X_{t}$
is ($f,r$)-ergodic; this implicitly entails the existence of $\pi$
as well as certain moments as $\pi(f)<\infty$. (For instance, if
$\mathsf{X}=\mathbb{R}$ and $f(x)=1+x^{2}$, then ($f,r$)-ergodicity
implies that the stationary distribution of $X_{t}$ has finite second
moments.)

In the probability literature the preceding definition of ($f,r$)-ergodicity
is standard. However, an equivalent and more transparent formulation
is obtained by replacing equation~(\ref{f-ergodicity})~with\hspace*{-10pt}

\[
\lim_{n\to\infty}r(n)\sup_{f_{0}:\left|f_{0}\right|\leq f}\left|E[f_{0}(X_{n})\mid X_{0}=x]-\pi(f_{0})\right|=0\qquad\text{for }\pi\text{-almost all }x\in\mathsf{X}
\]
(see \citet[p. 776]{tuominen1994subgeometric}). For instance, if
$f(x)=1+\left|x\right|$ the above equation shows that, for almost
any initial value $x,$ the conditional expectation $E[X_{n}\mid X_{0}=x]$
converges to $\int_{x\in\mathsf{X}}x\thinspace\pi(dx)$, the expectation
of the stationary distribution of $X_{t}$, and the rate of the convergence
is given by $r(n)$. 

Most of the recent ergodicity results obtained for nonlinear autoregressions
have established geometric ergodicity so that the rate of convergence
in (\ref{f-ergodicity}) is given by $r(n)=r^{n}$, $r>1$. The subgeometric
rate functions we consider are defined as follows (cf., e.g., \citet{nummelin1983rate}
and \citet{douc2004practical}). Let $\Lambda_{0}$ be the set of
positive nondecreasing functions $r_{0}\,:\,\mathbb{N}\rightarrow[1,\infty)$
such that $\ln[r_{0}(n)]/n$ decreases to zero as $n\rightarrow\infty$.
The class of subgeometric rate functions, denoted by $\Lambda$, consists
of positive functions $r\,:\,\mathbb{N}\rightarrow(0,\infty)$ for
which there exists some $r_{0}\in\Lambda_{0}$ such that
\[
0<\liminf_{n\rightarrow\infty}\frac{r(n)}{r_{0}(n)}\leq\limsup_{n\rightarrow\infty}\frac{r(n)}{r_{0}(n)}<\infty.
\]
Typical examples are obtained of rate functions $r$ for which these
inequalities hold with (for notational convenience, we set $\ln(0)=0$)
\[
r_{0}(n)=(1+\ln(n))^{\alpha}\,\cdot\,(1+n)^{\beta}\,\cdot\,e^{cn^{\gamma}},\qquad\alpha,\beta,c\geq0,\,\gamma\in(0,1).
\]
The rate function $r_{0}(n)$ is called subexponential when $c>0$,
polynomial when $c=0$ and $\beta>0$, and logarithmic when $\beta=c=0$
and $\alpha>0$. \citet[Sec 3.3]{douc2004practical} consider subexponential
convergence rates whereas \citet[Sec 2.2]{fort2003polynomial} consider
polynomial convergence rates in model (\ref{NLAR(1)}) (see also the
related references mentioned in these papers).

The proofs of our results make use of the following condition adapted
from \citet[Defn 16.1.7]{douc2018markov}.\footnote{A somewhat more general version which allows $V$ to be extended-real-valued
(i.e., $V\,:\,\mathsf{X}\rightarrow[1,\infty]$) is given in \citet{douc2004practical}.} 

\bigskip{}
\vspace*{\fill}
\pagebreak{}

\noindent \textbf{Condition D}. There exist a measurable function
$V\,:\,\mathsf{X}\rightarrow[1,\infty)$, a concave increasing continuously
differentiable function $\phi\,:\,[1,\infty)\rightarrow(0,\infty)$,
a measurable set $C$, and a finite constant $b$ such that 
\begin{equation}
E\left[V(X_{1})\,\left|\,X_{0}=x\right.\right]\leq V(x)-\phi\left(V(x)\right)+b\boldsymbol{1}_{C}(x),\qquad x\in\mathsf{X}.\label{Drift condition}
\end{equation}

\bigskip{}

Conditions of this kind are known as drift conditions; when $\phi(v)=\lambda v$
for some $\lambda>0$ the so-called Foster-Lyapunov drift condition
used to establish geometric ergodicity is obtained. For ease of discussion
and reference, the following theorem summarizes geometric, subexponential,
and polynomial ergodicity results that can be obtained using Condition
D. (For the definitions of irreducibility, aperiodicity, and petite
sets appearing in the theorem we refer the reader to \citet{meyn2009markov}.) 
\begin{thm}[(\citet{meyn2009markov}, \citet{douc2004practical})]
 Suppose $X_{t}$ is a $\psi$-irreducible and aperiodic Markov chain
on $(\mathsf{X},\mathcal{B}(\mathsf{X}))$ and that Condition D holds
with a petite set $C$ such that $\sup_{x\in C}V(x)<\infty$ and the
function $\phi$ being either\vspace*{-2pt}
\begin{lyxlist}{(ii; subexponential case)  }
\item [{(i;~geometric~case)}] $\phi(v)=\lambda v$ for some $\lambda>0$,\vspace*{-6pt}
\item [{(ii;~subexponential~case)}] $\phi(v)=c(v+v_{0})/[\ln(v+v_{0})]^{\alpha}$
for some $c,\alpha,v_{0}>0$, or\vspace*{-6pt}
\item [{(iii;~polynomial~case)}] $\phi(v)=cv^{\alpha}$ for some $\alpha\in[0,1)$
and $c\in(0,1]$.\vspace*{-2pt}
\end{lyxlist}
Then $X_{t}$ is $(f,r)$-ergodic with either\vspace*{-2pt}
\begin{lyxlist}{(iii)}
\item [{(i)}] $f=V$ and $r(n)=r^{n}$ for some $r>1$ (or, equivalently,
$r(n)=(e^{c})^{n}$ for some $c>0$),\vspace*{-6pt}
\item [{(ii)}] $f=V^{\delta}$ and $r(n)=(e^{d})^{n^{1/(1+\alpha)}}$ for
any $\delta\in(0,\negthinspace1)$ and any $d\in(0,(1\negthinspace-\negthinspace\delta)\negthinspace\left\{ c(1\negthinspace+\negthinspace\alpha)\right\} ^{1/(1+\alpha)})$,
or\hspace*{-4pt}\vspace*{-6pt}
\item [{(iii)}] $f=V^{1-\delta(1-\alpha)}$ and $r(n)=n^{\delta-1}$ for
any $\delta\in[1,1/(1-\alpha)]$.\bigskip{}
\end{lyxlist}
\end{thm}
In the geometric case the result of Theorem 1 is given in \citet{meyn2009markov},
and in the subexponential and polynomial cases the result can be obtained
from \citet{douc2004practical}; some further details are provided
in the proof of Theorem 1 in the Supplementary Appendix. Note that
in the subexponential case choosing $v_{0}$ sufficiently large ensures
the concavity of $\phi$ required in Condition D and also that (in
the subexponential case) results with a faster rate of convergence
and/or larger $f$-norm could be obtained at the expense of more complex
notation; see \citet[Sec 2.3]{douc2004practical}. 

An essential feature of the subgeometric ergodicity results in Theorem
1 is that there is a trade-off between the rate of convergence and
the size of the $f$-norm; in Theorem 1 the choice of $\delta$ reflects
this. If a fast rate of convergence is desired one has to accept a
small $f$-norm (recall from (\ref{f-ergodicity}) that the size of
the $f$-norm is directly proportional to the order of finite moments
the stationary distribution is guaranteed to have). For instance,
in the polynomial case choosing $\delta=1/(1-\alpha)$ gives the fastest
rate of convergence and with this choice the $f$-norm reduces to
the total variation norm (so that $f\equiv1$); the extreme case $\alpha=0$
results in $r(n)\equiv1$ and standard ergodicity. In the subexponential
case values of $\delta$ that are close to zero (one) correspond to
small (large) $f$-norms.

It is also worth noting that Condition D is only sufficient, not necessary,
for ($f,r$)-ergodicity. It is therefore possible that with another
drift condition (not necessarily a special case of Condition D) a
better rate function could be obtained, but presumably at the cost
of a smaller norm. Being able to obtain necessary conditions for particular
subgeometric ergodicity rates would be of interest but we will not
pursue this issue. Necessary conditions for geometric and polynomial
ergodicity in the context of random-walk-type Markov chains (see (\ref{RW-type example_2}))
are given in \citet{jarner2003necessary} (for an application of this
result to a threshold autoregressive model, see \citet{meitz2019subgemix}).

As already indicated in the Introduction, the ergodicity results of
Theorem 1 imply results on $\beta$-mixing or, more specifically,
on convergence rates of $\beta$-mixing coefficients $\beta(n)$ ($n=1,2,\ldots$)
(for a definition of $\beta(n)$ and properties of $\beta$-mixing,
see \citet[Sec 1.1]{doukhan1994mixing}, \citet[Ch 3]{bradley2007introduction},
or \citet{meitz2019subgemix}). To illustrate this point, let $\mu$
signify the distribution of $X_{0}$, the initial value of the Markov
chain $X_{t}$, and assume that$\int_{x\in\mathsf{X}}V(x)\mu(dx)<\infty$
(with $V$ as in Theorem 1). Then, using Theorems 1 and 2 of \citet{meitz2019subgemix}
the three cases in Theorem 1 imply the following convergence rates
for $\beta$-mixing coefficients (here $c$ and $\alpha$ are as in
Theorem 1):\footnote{See also the discussion following Theorem 2 of \citet{meitz2019subgemix},
and note that their Theorem 2(e) is also used to obtain the subexponential
rate shown here.}
\begin{lyxlist}{XXXXXXXXXXXXXXX}
\item [{(i)~geometric~case:}] $\lim_{n\rightarrow\infty}\tilde{r}^{n}\beta(n)=0$
for some $\tilde{r}>1$;
\item [{(ii)~subexponential~case:}] $\lim_{n\rightarrow\infty}(e^{\tilde{d}})^{n^{1/(1+\alpha)}}\beta(n)=0$
for any $\tilde{d}\in(0,\left\{ c(1\negthinspace+\negthinspace\alpha)/2\right\} ^{1/(1+\alpha)})$;
\item [{(iii)~polynomial~case:}] $\lim_{n\rightarrow\infty}n^{\alpha/(1-\alpha)}\beta(n)=0$. 
\end{lyxlist}
Thus, the convergence rates of the $\beta$-mixing coefficients are
qualitatively similar to the fastest convergence rates of ergodicity
obtained in Theorem 1 (as indicated above, a slight improvement can
be achieved in the subexponential case). These results, combined with
the fact that the $(f,r)$-ergodicity given in Theorem 1 implies finiteness
of moments, make possible to use limit theorems developed for $\beta$-mixing
processes (and also for $\alpha$-mixing processes because $\beta$-mixing
is known to imply $\alpha$-mixing).

\section{Model and assumptions\label{sec:model}}

We now introduce a higher-order generalization of the model discussed
in the Introduction. Suppose the process $y_{t}$ ($t=1,2,\ldots$)
is generated by
\begin{equation}
y_{t}=\varphi_{1}y_{t-1}+\cdots+\varphi_{p}y_{t-p}+\tilde{g}(y_{t-1},\ldots,y_{t-p})+\varepsilon_{t},\label{NLAR(p)_phi}
\end{equation}
where $\tilde{g}$ is a real-valued function, the error term $\varepsilon_{t}$
is a sequence of IID random variables, and exactly one of the roots
of the polynomial $\varphi(z)=1-\varphi_{1}z-\cdots-\varphi_{p}z^{p}$
is equal to unity and (when $p\geq2$) all others lie outside the
unit circle. Thus, the regression function of the model has a linear
part and a nonlinear part, and without the nonlinear part we have
a standard linear $p$th order autoregression with a single unit root
(cf. model (\ref{RW-type example_2})).

To express (\ref{NLAR(p)_phi}) in a different way, set $\pi_{j}=-\sum_{i=j+1}^{p}\varphi_{i}$
($j=1,\ldots,p-1$; when $p=1$, set $\pi_{1}=\cdots=\pi_{p-1}=0$)
so that we can express the polynomial $\varphi(z)$ as
\[
\varphi(z)=(1-z)(1-\pi_{1}z-\cdots-\pi_{p-1}z^{p-1}),
\]
where the roots of the polynomial $\varpi(z)=1-\pi_{1}z-\cdots-\pi_{p-1}z^{p-1}$
lie outside the unit circle. This shows that we can write equation
(\ref{NLAR(p)_phi}) alternatively as
\begin{equation}
y_{t}=y_{t-1}+\pi_{1}\Delta y_{t-1}+\cdots+\pi_{p-1}\Delta y_{t-p+1}+\tilde{g}(y_{t-1},\ldots,y_{t-p})+\varepsilon_{t}.\label{DNLAR(p)}
\end{equation}
Denoting $u_{t}=y_{t}-\pi_{1}y_{t-1}-\cdots-\pi_{p-1}y_{t-p+1}$ equation
(\ref{DNLAR(p)}) can be written as
\begin{equation}
y_{t}=\pi_{1}y_{t-1}+\cdots+\pi_{p-1}y_{t-p+1}+u_{t-1}+\tilde{g}(y_{t-1},\ldots,y_{t-p})+\varepsilon_{t}\label{NLAR(p)_pi}
\end{equation}
or as $u_{t}=u_{t-1}+\tilde{g}(y_{t-1},\ldots,y_{t-p})+\varepsilon_{t}$;
when $p=1$ we obtain $y_{t}=y_{t-1}+\tilde{g}(y_{t-1})+\varepsilon_{t}$
as in (\ref{RW-type example_2}). The formulation in (\ref{NLAR(p)_pi})
is convenient in our theoretical developments and will therefore be
used instead of (\ref{DNLAR(p)}). One reason for this convenience
is that in cases where the function $\tilde{g}$ depends on $y_{t-1},\ldots,y_{t-p}$
only through the linear combination $u_{t-1}=y_{t-1}-\pi_{1}y_{t-2}-\cdots-\pi_{p-1}y_{t-p}$
we can write equation (\ref{NLAR(p)_pi}) (with a slight abuse of
notation) in a more compact way as $u_{t}=u_{t-1}+\tilde{g}(u_{t-1})+\varepsilon_{t}$.
Then the process $u_{t}$ can be treated as the first-order model
(\ref{RW-type example_2}) and, as will be discussed shortly, with
a suitable assumption, we can make use of results in \citet[Sec 2.2]{fort2003polynomial}
and \citet[Sec 3.3]{douc2004practical} (this turns out to be the
case even when $\tilde{g}$ is not a function of the process $u_{t-1}$
only). 

Next we introduce the assumptions needed to prove our results. Our
first assumption restricts the dynamics in equation (\ref{NLAR(p)_pi}). 
\begin{assumption}
\noindent \label{assu:dynamics}Suppose the polynomial $\varpi(z)=1-\pi_{1}z-\cdots-\pi_{p-1}z^{p-1}$
and the function $\tilde{g}\,:\,\mathbb{R}^{p}\rightarrow\mathbb{R}$
in (\ref{NLAR(p)_pi}) satisfy the following conditions:\vspace*{-2pt}
\begin{lyxlist}{(ii)}
\item [{(i)}] \noindent The roots of $\varpi(z)$ lie outside the unit
circle.\vspace*{-2pt}
\item [{(ii)}] The function $\tilde{g}$ is measurable, bounded on compact
subsets of $\mathbb{R}^{p}$, and there exists a measurable function
$g\,:\,\mathbb{R}\rightarrow\mathbb{R}$ with the property $\left|g(x)\right|\rightarrow\infty$
as $\left|x\right|\rightarrow\infty$ such that the following two
conditions hold.\vspace*{-2pt}
\begin{lyxlist}{(ii.b)}
\item [{(ii.a)}] \noindent With $\boldsymbol{x}=(x_{1},\ldots,x_{p})$
and $u=x_{1}-\pi_{1}x_{2}-\cdots-\pi_{p-1}x_{p}$, the function $\tilde{g}$
satisfies
\begin{equation}
\left|u+\tilde{g}(\boldsymbol{x})-g(u)\right|\leq\left|\epsilon(\boldsymbol{x})\boldsymbol{x}\right|,\label{Inequality Ass 2a}
\end{equation}
where $\epsilon(\boldsymbol{x})$ is a real-valued function such that
$\left|\epsilon(\boldsymbol{x})\right|=o(\left|\boldsymbol{x}\right|^{-d})$
as $\left|\boldsymbol{x}\right|\rightarrow\infty$ for some $d>0$.\vspace*{-2pt}
\item [{(ii.b)}] There exist positive constants $r$, $M_{0}$, $K_{0}$,
and $0<\rho\leq2$ such that for all $u\in\mathbb{R}$ 
\begin{equation}
\left|g(u)\right|\leq\begin{cases}
(1-r\left|u\right|^{-\rho})\left|u\right| & \textrm{for }\left|u\right|\geq M_{0},\\
K_{0} & \textrm{for }\left|u\right|\leq M_{0}.
\end{cases}\label{Inequality_Ass 2}
\end{equation}
\end{lyxlist}
\end{lyxlist}
\end{assumption}
Assumption \ref{assu:dynamics}(i) corresponds to the conventional
stationarity condition of a linear autoregression in that it requires
the roots of the polynomial $\varpi(z)$ to lie outside the unit circle.
In the first-order case $p=1$, this condition becomes redundant because
then $\pi_{1}=\cdots=\pi_{p-1}=0$. 

Assumption \ref{assu:dynamics}(ii) requires the function $\tilde{g}$
to be bounded on compact subsets and links it to another function
$g$. Condition (ii.a) controls the difference between the functions
$u+\tilde{g}(\boldsymbol{x})$ and $g(u)$ or, in model (\ref{NLAR(p)_pi}),
the difference between the processes $u_{t-1}+\tilde{g}(y_{t-1},\ldots,y_{t-p})$
and $g(u_{t-1})$. In the special case where the function $\tilde{g}$
depends on $u$ only, condition (ii.a) becomes obvious because then
one can choose $u+\tilde{g}(\boldsymbol{x})=g(u)$ and $\epsilon(\boldsymbol{x})=0$,
and it suffices to check condition (ii.b) only. In this case we can
use results in \citet[Sec 2.2]{fort2003polynomial} and \citet[Sec 3.3]{douc2004practical}
directly in our proofs. However, we can do the same, albeit in a more
complicated way, also when the function $\tilde{g}$ depends on the
whole $p$-dimensional vector $\boldsymbol{x}$, but then the difference
between the functions $u+\tilde{g}(\boldsymbol{x})$ and $g(u)$ may
not increase ``too fast'' when $\left|\boldsymbol{x}\right|$ gets
large. What is ``too fast'' is controlled by the function $\epsilon(\boldsymbol{x})$,
and when $d\geq1$ the difference between $u+\tilde{g}(\boldsymbol{x})$
and $g(u)$ becomes negligible when $\left|\boldsymbol{x}\right|$
increases.

Condition (ii.a) implies that $\left|u+\tilde{g}(\boldsymbol{x})\right|\leq\left|g(u)\right|+\left|\epsilon(\boldsymbol{x})\boldsymbol{x}\right|$,
which combined with condition (ii.b) yields
\begin{equation}
\left|u+\tilde{g}(\boldsymbol{x})\right|\leq\left(1-r\left|u\right|^{-\rho}\right)\left|u\right|+o(\left|\boldsymbol{x}\right|^{-d})\left|\boldsymbol{x}\right|\quad\textrm{for }\left|u\right|\geq M_{0}.\label{Implication of Ass 2}
\end{equation}
This fact is used in our proofs. Note also that condition (ii.a) is
implied by the equality $u+\tilde{g}(\boldsymbol{x})=g(u)+\tilde{\epsilon}(\boldsymbol{x})\theta'\boldsymbol{x}$
where $\theta$ is a $p$-dimensional parameter vector and, if $\tilde{\epsilon}(\boldsymbol{x})=o(\left|\boldsymbol{x}\right|^{-d})$
is assumed, condition (ii.a) holds with $\epsilon(\boldsymbol{x})=\left|\theta\right|\tilde{\epsilon}(\boldsymbol{x})$.
This approach for checking condition (ii.a) is illustrated in Section
5. 

Condition (ii.b) is similar to its first-order counterpart (\ref{Ineq g(x)_p=00003D1})
to which it reduces when $p=1$. Note that apart from the boundedness
condition, no restrictions are placed on $g(u)$ for moderate values
of $u$. In the higher-order case this assumption concerns the filtered
process $u_{t}=\varpi(L)y_{t}$. In the first-order case we also have
$\boldsymbol{x}=u$ and the easiest way to verify Assumption 1(ii)
may then be to define the function $g$ as $g(x)=x+\tilde{g}(x)$
(and $\epsilon(\boldsymbol{x})=0$), and verify condition (ii.b) directly.
When $p\geq2,$ the fact that the domain of the function $\tilde{g}$
is larger than that of $g$ complicates the situation in that then
no simple connection between inequalities (\ref{Inequality Ass 2a})
and (\ref{Inequality_Ass 2}) can generally be found. An example of
this case is provided in Section 5.

Our second assumption gives conditions required of the error term
in equation (\ref{NLAR(p)_pi}).
\begin{assumption}
\label{assu:errors}$\{\varepsilon_{t},\,t=1,2,\ldots\}$ is a sequence
of IID random variables that is independent of $(y_{0},\ldots,y_{-p+1})$
(with $p$ as in Assumption 1), the distribution of $\varepsilon_{1}$
has a (Lebesgue) density that is bounded away from zero on compact
subsets of $\mathbb{R}$, and either
\begin{lyxlist}{00}
\item [{(a)}] $E\bigl[e^{\beta_{0}\left|\varepsilon_{1}\right|^{\kappa_{0}}}\bigr]<\infty$
for some $\ensuremath{\beta_{0}>0}$ and $\ensuremath{\kappa_{0}\in(0,1]}$,
and $E[\varepsilon_{1}]=0$; or
\item [{(b)}] $E\left[\left|\varepsilon_{1}\right|^{s_{0}}\right]<\infty$
for some $s_{0}>\rho$ (with $\rho$ as in Assumption 1), and $E[\varepsilon_{1}]=0$
holds if $\rho\geq1$.
\end{lyxlist}
\end{assumption}
Assumption 2(a) corresponds to Assumption 3.3 of \citet[Sec 3.3]{douc2004practical},
whereas Assumption 2(b) is a combination of the conditions imposed
in (NSS 1), (NSS 4), and Lemma 3 of \citet[Sec 2.2]{fort2003polynomial}.
The boundedness condition imposed in Assumption 2 on the density of
the error term is stronger than would be needed but is used for simplicity
(see Assumption 3.3 of \citet[Sec 3.3]{douc2004practical} for a more
general alternative). 

Note that finiteness of the first expectation in Assumption \ref{assu:errors}(a)
holds with $\kappa_{0}=1$ if the distribution of $\varepsilon_{1}$
has a moment generating function in some interval of the origin. Although
many widely used distributions satisfy this condition some heavy tailed
distributions are ruled out (this applies to distributions whose densities
cannot be bounded by a term of the form $c_{1}e^{-c_{2}\left|x\right|}$
with $c_{1}$ and $c_{2}$ positive constants). An example is Student's
t-distribution irrespective of the value of the degrees of freedom
parameter. The condition in Assumption \ref{assu:errors}(b) is used
to address this issue. In this condition the case $0<\rho<s_{0}<1$
is rather extreme in that not even the expectation $E[\varepsilon_{1}]$
is assumed to exist.

\section{Results\label{sec:results}}

We now present our ergodicity results which we base on model (\ref{NLAR(p)_pi}).
In Section 4.1 the rate of ergodicity established is subexponential
whereas a slower polynomial rate of ergodicity is obtained in Section
4.2. The difference between these two cases stems from the assumed
moment conditions: in Section 4.1 the condition in Assumption 2(a)
is assumed whereas in Section 4.2 the weaker condition in Assumption
2(b) is employed. First we have to present the companion form of model
(\ref{NLAR(p)_pi}) which applies to both of these cases and will
be needed in the proofs of our theorems.

To simplify notation, denote $\boldsymbol{y}_{t}=(y_{t},\ldots,y_{t-p+1})$
and define the function $\overline{g}\,:\,\mathbb{R}^{p}\rightarrow\mathbb{R}$
as 
\[
\overline{g}(\boldsymbol{x})=x_{1}-\pi_{1}x_{2}-\cdots-\pi_{p-1}x_{p}+\tilde{g}(\boldsymbol{x})=u+\tilde{g}(\boldsymbol{x})
\]
so that $\overline{g}(\boldsymbol{y}_{t-1})=u_{t-1}+\tilde{g}(\boldsymbol{y}_{t-1})$.
It is readily seen that the companion form related to equation (\ref{NLAR(p)_pi})
reads as
\[
\left[\begin{array}{c}
y_{t}\\
y_{t-1}\\
\vdots\\
\vdots\\
y_{t-p+1}
\end{array}\right]=\begin{bmatrix}\pi_{1} & \pi_{2} & \cdots & \pi_{p-1} & 0\\
1 & 0 & \cdots & 0 & 0\\
0 & \ddots & \ddots & \vdots & \vdots\\
\vdots & \ddots & \ddots & 0 & 0\\
0 & \cdots & 0 & 1 & 0
\end{bmatrix}\left[\begin{array}{c}
y_{t-1}\\
y_{t-2}\\
\vdots\\
\vdots\\
y_{t-p}
\end{array}\right]+\overline{g}(\boldsymbol{y}_{t-1})\left[\begin{array}{c}
1\\
0\\
\vdots\\
\vdots\\
0
\end{array}\right]+\varepsilon_{t}\left[\begin{array}{c}
1\\
0\\
\vdots\\
\vdots\\
0
\end{array}\right]
\]
or, with obvious matrix notation,
\begin{equation}
\boldsymbol{y}_{t}=\boldsymbol{\Phi}\boldsymbol{y}_{t-1}+\overline{g}(\boldsymbol{y}_{t-1})\boldsymbol{\iota}_{p}+\varepsilon_{t}\boldsymbol{\iota}_{p}\label{Companion form}
\end{equation}
(when $p=1$, $\boldsymbol{\Phi}=0$). Thus, Assumption \ref{assu:errors}
ensures that $\boldsymbol{y}_{t}$ is a Markov chain on $\mathbb{R}^{p}$.
For later purposes it is convenient to transform the companion form
(\ref{Companion form}). To this end, we define the matrices 
\begin{equation}
\negthickspace\negthickspace\negthickspace\mathbf{A}=\setlength{\arraycolsep}{2pt}
\global\long\def\arraystretch{0.9}%
\begin{bmatrix}1 & -\pi_{1} & -\pi_{2} & \cdots & -\pi_{p-1}\\
0 & 1 & 0 & \cdots & 0\\
\vdots & \:\:\:\:\ddots & \ddots & \ddots & \vdots\\
0 & \:\:\: & \ddots & \ddots & 0\\
0 & \cdots & \cdots & 0 & 1
\end{bmatrix}\quad\text{and}\quad\mathbf{\Pi}=\mathbf{A}\boldsymbol{\Phi}\mathbf{A}^{-1}=\begin{bmatrix}0 & 0 & 0 & \cdots & 0\\
1 & \pi_{1} & \pi_{2} & \cdots & \pi_{p-1}\\
0 & 1 & 0 & \cdots & 0\\
\vdots & \ddots & \ddots & \ddots & \vdots\\
0 & \cdots & 0 & 1 & 0
\end{bmatrix}=\begin{bmatrix}0 & \text{\textbf{0}}'_{p-1}\\
\boldsymbol{\iota}_{p-1} & \boldsymbol{\Pi}_{1}
\end{bmatrix},\negthickspace\negthickspace\negthickspace\negthickspace\label{Matrix Pi}
\end{equation}
where $\mathbf{A}$ is nonsingular and $\boldsymbol{\Pi}_{1}$ is
the $(p-1)\times(p-1)$ dimensional lower right hand corner of $\mathbf{\Pi}$
(when $p=1$, $\mathbf{A}=1$ and $\mathbf{\Pi}=0$). With these definitions
(\ref{Companion form}) can be transformed into
\begin{equation}
\mathbf{A}\boldsymbol{y}_{t}=\mathbf{\Pi}\mathbf{A}\boldsymbol{y}_{t-1}+\overline{g}(\boldsymbol{y}_{t-1})\boldsymbol{\iota}_{p}+\varepsilon_{t}\boldsymbol{\iota}_{p},\label{Companion form_A}
\end{equation}
where $\mathbf{A}\boldsymbol{y}_{t}=(u_{t},y_{t-1},\ldots,y_{t-p+1})$.
Now, for any $p$-dimensional vector $\boldsymbol{x}$, form the partition
$\boldsymbol{x}=(x_{1},\ldots,x_{p})=(x_{1},\boldsymbol{x}_{2})$
and define $\boldsymbol{z}(\boldsymbol{x})=(z_{1}(\boldsymbol{x}),\boldsymbol{z}_{2}(\boldsymbol{x}))=\boldsymbol{A}\boldsymbol{x}$,
where (due to (\ref{Matrix Pi})) $z_{1}(\boldsymbol{x})=x_{1}-\pi_{1}x_{2}-\cdots-\pi_{p-1}x_{p}$
and $\boldsymbol{z}_{2}(\boldsymbol{x})=\boldsymbol{x}_{2}$ (when
$p=1$, $\boldsymbol{x}_{2}$ and $\boldsymbol{z}_{2}(\boldsymbol{x})$
are dropped). With this notation equation (\ref{Companion form_A})
can be expressed as
\begin{equation}
\begin{bmatrix}z_{1}(\boldsymbol{y}_{t})\\
\boldsymbol{z}_{2}(\boldsymbol{y}_{t})
\end{bmatrix}=\setlength{\arraycolsep}{2pt}\begin{bmatrix}0 & \text{\textbf{0}}'_{p-1}\\
\boldsymbol{\iota}_{p-1} & \boldsymbol{\Pi}_{1}
\end{bmatrix}\begin{bmatrix}z_{1}(\boldsymbol{y}_{t-1})\\
\boldsymbol{z}_{2}(\boldsymbol{y}_{t-1})
\end{bmatrix}+\overline{g}(\boldsymbol{y}_{t-1})\boldsymbol{\iota}_{p}+\varepsilon_{t}\boldsymbol{\iota}_{p}=\begin{bmatrix}\overline{g}(\boldsymbol{y}_{t-1})+\varepsilon_{t}\\
\boldsymbol{\Pi}_{1}\boldsymbol{z}_{2}(\boldsymbol{y}_{t-1})+z_{1}(\boldsymbol{y}_{t-1})\boldsymbol{\iota}_{p-1}
\end{bmatrix}.\label{Companion form_A2}
\end{equation}
The first equation in (\ref{Companion form_A2}) is now in a form
that can be analyzed by using the results in \citet{fort2003polynomial}
and \citet{douc2004practical}. As for the second equation, by Assumption
\ref{assu:dynamics}(i) the roots of the polynomial $\varpi(z)$ lie
outside the unit circle, so that the eigenvalues of the matrix $\boldsymbol{\Pi}_{1}$
are smaller than one in absolute value. As is well known, this implies
the existence of a matrix norm $\left\Vert \cdot\right\Vert _{*}$
induced by a vector norm, also denoted by $\left\Vert \cdot\right\Vert _{*}$,
such that $\left\Vert \boldsymbol{\Pi}_{1}\right\Vert _{*}\leq\eta$
for some $\eta<1$ (see, e.g., Definition 5.6.1 and Lemma 5.6.10 in
\citet{horn2013matrix}). These facts will be useful in our proofs.

\subsection{Subexponential case}

Our results make use of Condition D which requires choosing the function
$V$. To this end, let $b_{1}$, $b_{2}$, and $b_{3}$ be positive
constants whose values (to be specified later) depend on the constants
$\beta_{0}$, $\kappa_{0}$, and $\rho$ introduced in Assumptions
1 and 2; for $b_{3}$ we already mention that it will always satisfy
$b_{3}\in(0,1]$. When $p\geq2$, we define the function $V$ as 
\begin{equation}
V(\boldsymbol{x})=\tfrac{1}{2}\exp\bigl\{ b_{1}\left|z_{1}(\boldsymbol{x})\right|^{b_{3}}\bigr\}+\tfrac{1}{2}\exp\bigl\{ b_{2}\left\Vert \boldsymbol{z}_{2}(\boldsymbol{x})\right\Vert _{*}^{b_{3}}\bigr\}\label{Def. V_1}
\end{equation}
and when $p=1$, we define $V(x)=\exp\{b_{1}\left|x\right|^{b_{3}}\}$. 

Now we can state the following theorem which makes use of the stronger
moment requirement in Assumption 2(a). (The proof is given in the
Supplementary Appendix.)
\begin{thm}
\noindent Suppose $p\geq2$ and consider the Markov chain $\boldsymbol{y}_{t}$
defined in equation (\ref{Companion form}). Let Assumptions 1 and
2(a) hold, suppose that in Assumption 1 the constants $\rho$ and
$d$ satisfy $0<\rho<2$ and $d=\rho/b_{3}$, and let $V(\boldsymbol{x})$
be as in (\ref{Def. V_1}). 
\begin{lyxlist}{000}
\item [{(i)}] \noindent If $\rho>\kappa_{0}$, then $\boldsymbol{y}_{t}$
is ($f,r$)-ergodic with the subexponential convergence rate $r(n)=e^{kn^{b_{3}/\rho}}$
and the function $f$ given by $f(\boldsymbol{x})=V(\boldsymbol{x})^{\delta}$;
this result holds for any choice of $\delta\in(0,1)$, for any $k$
such that $0<k<(1-\delta)\left\{ c\rho/b_{3}\right\} ^{b_{3}/\rho}$,
and for some (small) $b_{1},\,b_{2}\in(0,\beta_{0})$, $b_{3}=\kappa_{0}\wedge(2-\rho)\in(0,1)$,
and some (small) $c>0$. 
\item [{(ii)}] If $\rho=\kappa_{0}$, then $\boldsymbol{y}_{t}$ is geometrically
ergodic with the convergence rate $r(n)=e^{cn}$ and the function
$f$ given by $f(\boldsymbol{x})=V(\boldsymbol{x})$; this result
holds for some (small) $b_{1},\,b_{2}\in(0,\beta_{0})$, $b_{3}=\kappa_{0}\in(0,1]$,
and some $c>0$.
\end{lyxlist}
When $p=1$, consider the Markov chain $y_{t}$ defined by $y_{t}=y_{t-1}+\tilde{g}(y_{t-1})+\varepsilon_{t}$.
The above results hold for $y_{t}$ with the function $V$ defined
as $V(x)=\exp\{b_{1}\left|x\right|^{b_{3}}\}$ (and the constant $d$
becomes redundant). 
\end{thm}
In Theorem 2, the case $\rho=\kappa_{0}$ represents a qualitative
change in the ergodic behavior of the considered Markov chain: For
$\rho>\kappa_{0}$ a slower subexponential convergence rate is obtained
and $\rho=\kappa_{0}$ is the borderline case where a change to the
faster geometric rate occurs. For $\rho<\kappa_{0}$, geometric ergodicity
could also be established but we omit this case for brevity (in the
first-order case this result is an immediate consequence of Theorem
3.3 of \citet[Sec 3.3]{douc2004practical}). Note also that, by the
definition of the constant $b_{3}$, the rate of ergodicity in the
subexponential case decreases as the value of $\rho$ increases.

Previously, \citet[Sec 3.3]{douc2004practical} obtained the results
of Theorem 2 in the first-order case; our primary purpose here is
to provide higher-order analogs of their results. \citet[2005]{klokov2004sub}
and \citet{klokov2007lower} have also studied the first-order model
(\ref{NLAR(1)}) satisfying inequality (\ref{Ineq g(x)_p=00003D1})
with $1<\rho<2$ but otherwise their assumptions are rather different
from ours. They obtain subexponential bounds for ergodicity in total
variation norm (i.e., ($1,r$)-ergodicity) and for $\beta$-mixing
coefficients but they do not discuss general ($f,r$)-ergodicity.

As discussed in Section 2, we can also establish $\beta$-mixing and,
in contrast to \citet[2005]{klokov2004sub} and \citet{klokov2007lower},
we can permit all initial values with distribution $\mu$ such that
$\int_{x\in\mathbb{R}^{p}}V(x)\mu(dx)<\infty$ (and $V$ as in Theorem
2). Specifically, the discussion at the end of Section 2 and Theorem
2 imply $\beta$-mixing with the following rates: In case (i) the
rate is subexponential, i.e., $\lim_{n\rightarrow\infty}e^{\tilde{k}n^{b_{3}/\rho}}\beta(n)=0$
with any $\tilde{k}\in(0,\left\{ c\rho/2b_{3}\right\} ^{b_{3}/\rho})$,
and in case (ii) the rate is geometric, i.e., $\lim_{n\rightarrow\infty}\tilde{r}^{n}\beta(n)=0$
for some $\tilde{r}>1$ (or, equivalently, $\lim_{n\rightarrow\infty}e^{\tilde{c}n}\beta(n)=0$
for some $\tilde{c}>0$). 

The following corollary is an immediate consequence of the discussion
after equality (\ref{f-ergodicity}) (for a formal result, see Theorem
14.0.1 in \citet{meyn2009markov}).
\begin{cor*}[\textbf{to Theorem 2}]
 Let $\pi$ signify the stationary distribution of $\boldsymbol{y}_{t}$
in Theorem 2 and let the function $f$ be as in cases (i) and (ii)
of Theorem 2. Then $\pi(f)=\int_{\boldsymbol{x}\in\mathbb{R}^{p}}f(\boldsymbol{x})\pi(d\boldsymbol{x})<\infty$;
in particular, $\pi(\left|\boldsymbol{x}\right|^{s})<\infty$ for
all $s>0$ so that the stationary distribution has finite moments
of all orders.
\end{cor*}
As we remarked after Theorem 1, in the subexponential case of Theorem
2 results with a faster rate of convergence and/or larger $f$-norm
could be obtained at the expense of more complex notation. This means
that in the above corollary finiteness of slightly larger moments
could be obtained and the subexponential $\beta$-mixing rate discussed
above could similarly be slightly improved.

\subsection{Polynomial case}

Next we consider ergodicity results relying only on the weaker moment
requirement in Assumption 2(b). This will below lead to a slower polynomial
rate of ergodicity. The key result used to relax the moment requirement
is Lemma 3 of \citet[Sec 2.2]{fort2003polynomial} (which the authors
use in conjunction with their analog of Condition D; we depart from
their approach and use Condition D which corresponds to an analogous
condition described in Section 1.2 in \citet{fort2003polynomial}). 

The function $V$ employed is now different from the subexponential
case. When $p\geq2$, we define the function $V$ as 
\begin{equation}
V(\boldsymbol{x})=1+\left|z_{1}(\boldsymbol{x})\right|^{s_{0}}+s_{1}\left\Vert \boldsymbol{z}_{2}(\boldsymbol{x})\right\Vert _{*}^{\alpha s_{0}},\label{Def. V_2}
\end{equation}
where $s_{0}$ is as in Assumption 2(b), $\alpha=1-\rho/s_{0}$ with
$\rho$ as in Assumption 1(ii.b), and $s_{1}$ is a positive constant
(to be specified later); when $p=1$, we define $V(x)=1+\left|x\right|^{s_{0}}$.

The following theorem presents the ergodicity result obtained when
using the weaker moment condition in Assumption 2(b). (The proof is
given in the Supplementary Appendix.)
\begin{thm}
\noindent Suppose $p\geq2$ and consider the Markov chain $\boldsymbol{y}_{t}$
defined in equation (\ref{Companion form}). Let Assumptions 1 and
2(b) hold, suppose that in Assumption 1 the constants $\rho$ and
$d$ satisfy $0<\rho\leq2$ and $d=\rho/s_{0}$ when $s_{0}<1$ and
$d=\rho$ when $s_{0}\geq1$, and let $V(\boldsymbol{x})$ be as in
(\ref{Def. V_2}). Assume further that either \vspace*{-4pt}
\begin{lyxlist}{000000}
\item [{~~~(i)}] \noindent $0<\rho<1$ and $s_{0}>\rho$,\vspace*{-6pt}
\item [{~~~(ii)}] \noindent $1\leq\rho<2$ and either $s_{0}=2$ or
$s_{0}\geq4$, or\vspace*{-6pt}
\item [{~~~(iii)}] \noindent $\rho=2$ and $s_{0}\geq4$ with $s_{0}r-\frac{1}{2}s_{0}(s_{0}-1)E[\varepsilon_{1}^{2}]>0$.\vspace*{-4pt}
\end{lyxlist}
\noindent Then $\boldsymbol{y}_{t}$ is ($f,r$)-ergodic with the
polynomial convergence rate $r(n)=n^{\delta-1}$ and the function
$f$ given by $f(\boldsymbol{x})=V(\boldsymbol{x})^{1-\delta\rho/s_{0}}$;
this result holds for any choice of $\delta\in[1,s_{0}/\rho]$ and
for some (small) $s_{1}>0$. 

\medskip{}

\noindent When $p=1$, consider the Markov chain $y_{t}$ defined
by $y_{t}=y_{t-1}+\tilde{g}(y_{t-1})+\varepsilon_{t}$. The above
results hold for $y_{t}$ with the functions $V$ and $f$ defined
as $V(x)=1+\left|x\right|^{s_{0}}$ and $f(x)=1+\left|x\right|^{s_{0}-\delta\rho}$
(and the constant $d$ becomes redundant).
\end{thm}
\noindent Options (i)\textendash (iii) in Theorem 3 represent the
combinations of the values of $\rho$ and $s_{0}$ for which the result
can be obtained by relying on the corresponding cases (i)\textendash (iii)
in Lemma 3 of \citet[Sec 2.2]{fort2003polynomial}. Unlike in Theorem
2, the case $\rho=2$ is allowed, but then an additional and rather
intricate moment condition is required. A further departure from Theorem
2 is that the same polynomial rate of ergodicity is obtained in all
cases. However, similarly to the subexponential case in Theorem 2,
the rate of ergodicity decreases as the value of $\rho$ increases.
Also, from the discussion at the end of Section 2 we can conclude
that the rate of $\beta$-mixing implied by Theorem 3 is polynomial
and, specifically, $\lim_{n\rightarrow\infty}n^{s_{0}/\rho-1}\beta(n)=0$.

The first-order case of Theorem 3 was obtained by \citet[Sec 2.2]{fort2003polynomial}
(with slightly different assumptions). Polynomial ergodicity results
for first-order autoregressions similar to that in (\ref{NLAR(1)})
have previously appeared also in \citet[Sec 5.2]{tuominen1994subgeometric}
(in the case $0<\rho<1$), \citet{tanikawa2001markov} (in the case
$\rho=1$), and \textcolor{black}{\citet{veretennikov2000polynomial}
and \citet{klokov2007lower} (in the case $\rho=2$; these authors
also obtain polynomial bounds for }$\beta$-mixing coefficients\textcolor{black}{{}
}but do not consider general ($f,r$)-ergodicity\textcolor{black}{).}

The following corollary on the moments of the stationary distribution
is proved in the Supplementary Appendix. (In contrast to the subexponential
case, using the ($f,r$)-ergodicity result of Theorem 3 would here
yield a weaker moment result; hence, some extra steps are needed.)
\begin{cor*}[\textbf{to Theorem 3}]
 Let $\pi$ signify the stationary distribution of $\boldsymbol{y}_{t}$
in Theorem 3. Then $\pi(f)=\int_{\boldsymbol{x}\in\mathbb{R}^{p}}f(\boldsymbol{x})\pi(d\boldsymbol{x})<\infty$
with $f(\boldsymbol{x})=\left|\boldsymbol{x}\right|^{s_{0}-\rho}$
so that the stationary distribution has finite moments up to order
$s_{0}-\rho$.
\end{cor*}

\section{Illustrative examples}

In this section we discuss special cases of the general model introduced
in Section \ref{sec:model}. Using the three equivalent formulations
(\ref{NLAR(p)_phi})\textendash (\ref{NLAR(p)_pi}) in Section 3,
the model considered can be written as 
\[
y_{t}-\varphi_{1}y_{t-1}-\cdots-\varphi_{p}y_{t-p}=\Delta y_{t}-\pi_{1}\Delta y_{t-1}-\cdots-\pi_{p-1}\Delta y_{t-p+1}=u_{t}-u_{t-1}=\tilde{g}(y_{t-1},\ldots,y_{t-p})+\varepsilon_{t},
\]
where the polynomial $\varphi(z)$ has precisely one unit root, can
be decomposed as $\varphi(z)=(1-z)\varpi(z)$, and $u_{t}=\varpi(L)y_{t}$.
Further formal assumptions will be stated in Propositions 1 and 2
below. 

First-order subgeometrically ergodic autoregressions were already
exemplified, albeit at a rather general level, in \citet[Sec 2.2]{fort2003polynomial}
and \citet[Secs 3.3, 3.4]{douc2004practical}. In \citet[Sec 5]{meitz2019subgemix}
we study rates of subgeometric ergodicity and $\beta$-mixing in a
first-order multi-regime self-exciting threshold autoregressive (SETAR)
model; the proof of Theorem 3 in that paper illustrates how Theorems
2 and 3 of the present paper can be applied in a first-order case.
In what follows we focus on examples of higher-order subgeometrically
ergodic autoregressive models.

We consider three main examples. We first give a heuristic overview
of them and then present the formalities. The first example we consider
can be expressed as 
\begin{equation}
u_{t}=u_{t-1}+I(u_{t-1})+\varepsilon_{t},\label{eq:Ex1}
\end{equation}
where $I(u_{t-1})$ is not constant and will be interpreted as a time-varying
drift or intercept term (the case where $I(u_{t-1})$ is a constant
is not of interest here because then model (\ref{eq:Ex1}) reduces
to a nonstationary unit root model (with or without a drift)). We
consider the case where $I(u_{t-1})$ takes values in a bounded interval
and fluctuates suitably between increasing and decreasing drifts.
This ensures that model (\ref{eq:Ex1}) will be (subgeometrically)
ergodic and stationary under appropriate conditions (see Proposition
1 below) even though its dynamics involve a unit root component and
a (time-varying) drift.

The second example we consider is 
\begin{equation}
u_{t}-\nu=S(u_{t-1})(u_{t-1}-\nu)+\varepsilon_{t},\label{eq:Ex2}
\end{equation}
where $\nu\in\mathbb{R}$ is an intercept term and $S(u_{t-1})$ will
be interpreted as a time-varying slope coefficient. In the cases $S(u_{t-1})\equiv0$
and $S(u_{t-1})\equiv1$ model (\ref{eq:Ex2}) reduces to the (linear)
stationary model $\varpi(L)y_{t}=u_{t}=\varepsilon_{t}$ and to the
nonstationary unit root model $\varphi(L)y_{t}=\varepsilon_{t}$,
respectively. We consider the case of $S(u_{t-1})$ being time-varying,
taking values in some interval $[s,1)$ ($s<1$), and attaining values
arbitrarily close to 1 for values of $u_{t-1}$ large in absolute
value. Under appropriate conditions (see Proposition 2 below), model
(\ref{eq:Ex2}) will be (subgeometrically) ergodic and stationary
although exhibiting features similar to a unit root process.

Our third example generalizes the second one and illustrates that
one can allow nonlinear dependence not only on $u_{t-1}$ but on the
entire $\boldsymbol{y}_{t-1}$. Specifically, we consider model 
\begin{equation}
u_{t}=S(u_{t-1})u_{t-1}+F(\boldsymbol{y}_{t-1})+\varepsilon_{t},\label{eq:Ex3}
\end{equation}
where we have omitted the intercept for simplicity, $S(u_{t-1})$
again represents a time-varying slope coefficient, and the term $F(\boldsymbol{y}_{t-1})$
captures the nonlinear dependence on the entire $\boldsymbol{y}_{t-1}$. 

\subsection{Example with time-varying intercept term of LSTAR type}

We consider example (\ref{eq:Ex1}) with the time-varying intercept
term $I(u_{t-1})$ specified as in logistic smooth transition autoregressive
(LSTAR) models (see, e.g., \citet{vandijk2002smooth}). Specifically,
we choose 
\begin{equation}
I(u_{t-1})=\nu_{1}L(u_{t-1};b,a_{1})+\nu_{2}(1-L(u_{t-1};b,a_{2}))\label{Function I(u)}
\end{equation}
with $L(u;b,a)=1/(1+e^{-b(u-a)})$ denoting the logistic function.
The parameters $b,a_{1},a_{2}$ are assumed to satisfy $b>0$ and
$a_{1}\leq a_{2}$ as usual, and $\nu_{1},\nu_{2}$ are assumed to
satisfy $\nu_{1}<0<\nu_{2}$ to obtain ergodicity below. The time-varying
intercept term $I(u_{t-1})$ now takes values in the interval $(\nu_{1},\nu_{2})$.
Note that for large values of $b$ the logistic function $L(u;b,a)$
is close to the indicator function and then this model provides a
close approximation to the above-mentioned threshold autoregressive
model where $L(u_{t-1};b,a_{i})$ is replaced with an indicator function. 

The following proposition shows the ergodicity of this model. 
\begin{prop}
Consider the process $y_{t}$ defined by $u_{t}=u_{t-1}+I(u_{t-1})+\varepsilon_{t}$
as in (\ref{eq:Ex1}) (with $u_{t}=\varpi(L)y_{t}$ and the roots
of $\varpi(z)$ outside the unit circle), and with $I(u_{t-1})$ as
in (\ref{Function I(u)}) (with $\nu_{1}<0<\nu_{2}$). Assume further
that either\vspace*{-4pt}
\begin{lyxlist}{00}
\item [{\small\ \ (1)}] Assumption 2(a) is satisfied with $\kappa_{0}\in(0,1)$,\vspace*{-6pt}
\item [{\small\ \ (2)}] Assumption 2(a) is satisfied with $\kappa_{0}=1$,
or\vspace*{-6pt}
\item [{\small\ \ (3)}] Assumption 2(b) is satisfied with either $s_{0}=2$
or $s_{0}\geq4$.\vspace*{-4pt}
\end{lyxlist}
Then, under condition {\small (1)/(2)/(3)}, the process $\boldsymbol{y}_{t}=(y_{t},\ldots,y_{t-p+1})$
is either\vspace*{-4pt}
\begin{lyxlist}{00}
\item [{\small\ \ (1)}] subexponentially ergodic with convergence rate
$r(n)=(e^{k})^{n^{\kappa_{0}}}$ (for some $k>0$),\vspace*{-6pt}
\item [{\small\ \ (2)}] geometrically ergodic with convergence rate $r(n)=(e^{c})^{n}$
(for some $c>0$), or\vspace*{-6pt}
\item [{\small\ \ (3)}] polynomially ergodic with convergence rate $r(n)=n^{s_{0}-1}$.
\end{lyxlist}
\end{prop}
The proof of Proposition 1 is a straightforward application of Theorems
2 and 3 in the case $\rho=1$ (see the Appendix). Depending on moment
assumptions the rate of ergodicity is geometric, subexponential, or
polynomial, and similar rates also apply to $\beta$-mixing coefficients
(see the discussions following Theorems 2 and 3 as well as the Corollaries
to these theorems for existence of finite moments of the stationary
distribution).

\begin{figure}[p]
\begin{minipage}[t][1\totalheight][c]{0.33\textwidth}%
\includegraphics[width=1.1\textwidth]{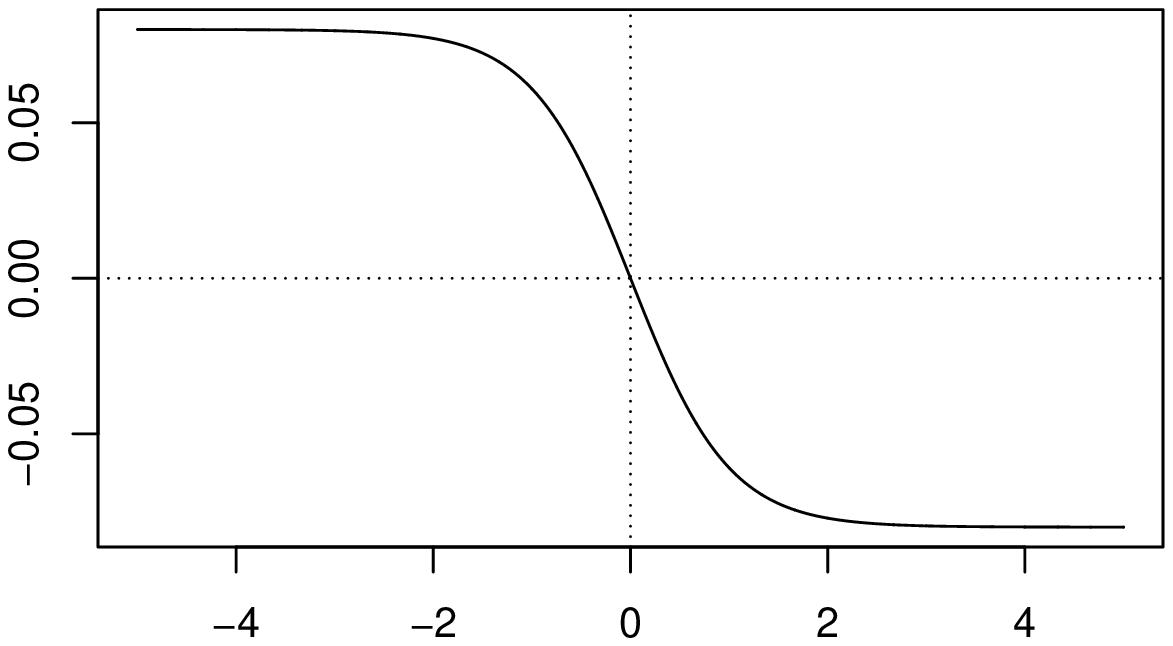}%
\end{minipage}\hfill{}%
\begin{minipage}[t][1\totalheight][c]{0.33\textwidth}%
\includegraphics[width=1.1\textwidth]{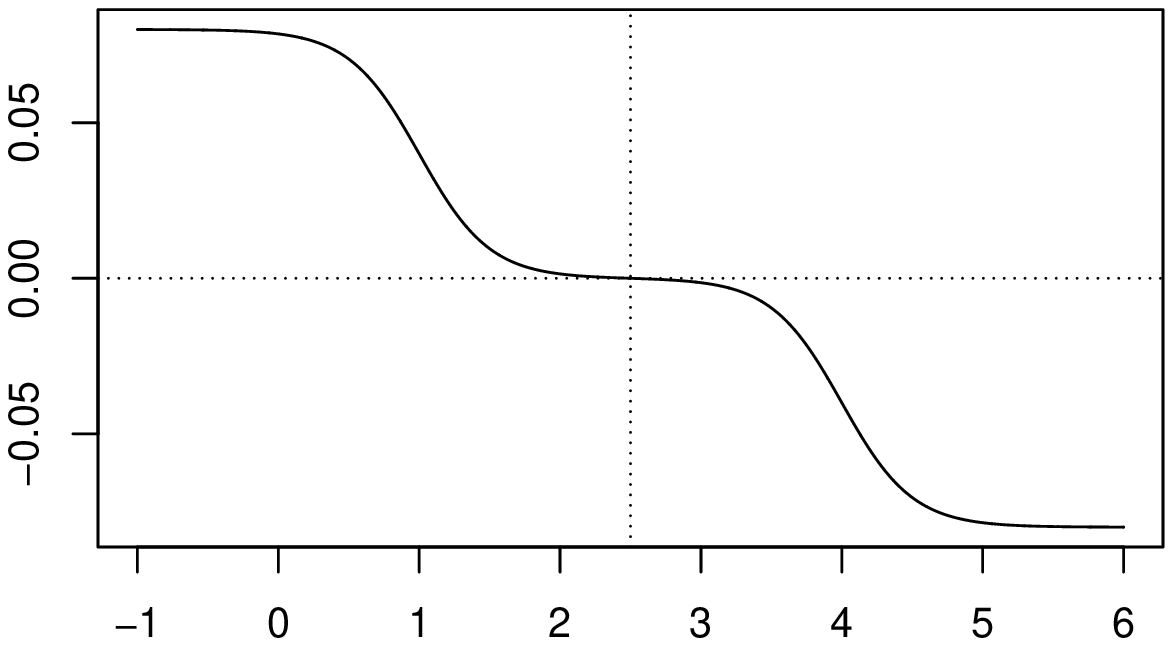}%
\end{minipage}\hfill{}%
\begin{minipage}[t][1\totalheight][c]{0.33\textwidth}%
\includegraphics[width=1.1\textwidth]{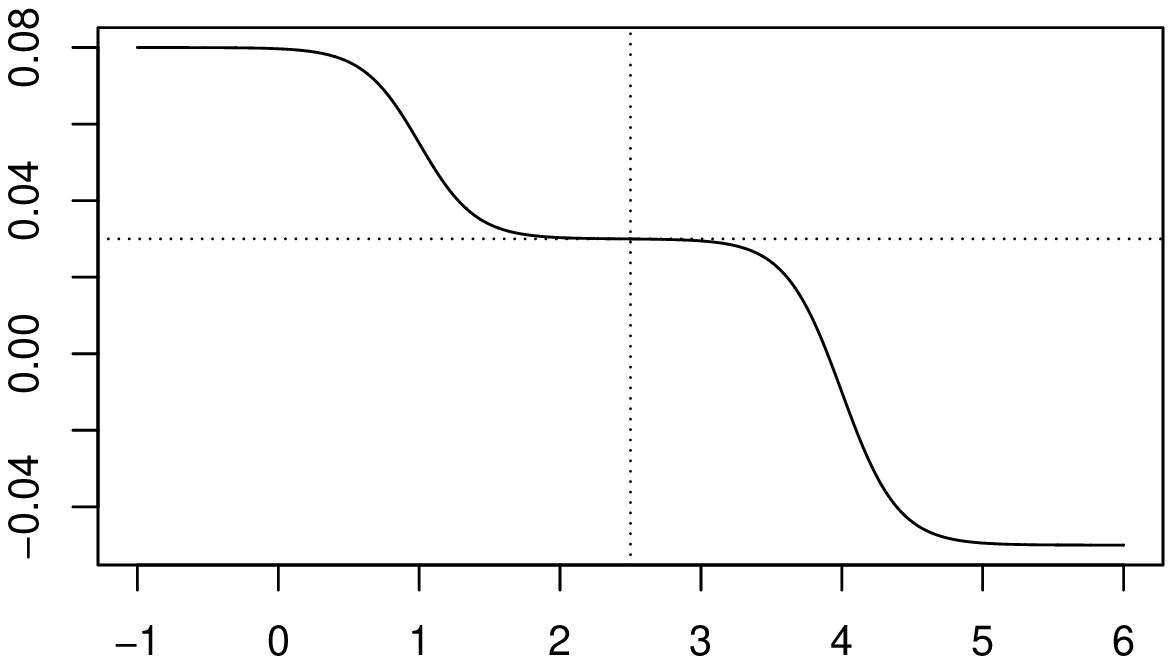}%
\end{minipage}\vspace*{-25pt}

\begin{minipage}[t][1\totalheight][c]{0.33\textwidth}%
\includegraphics[width=1.1\textwidth]{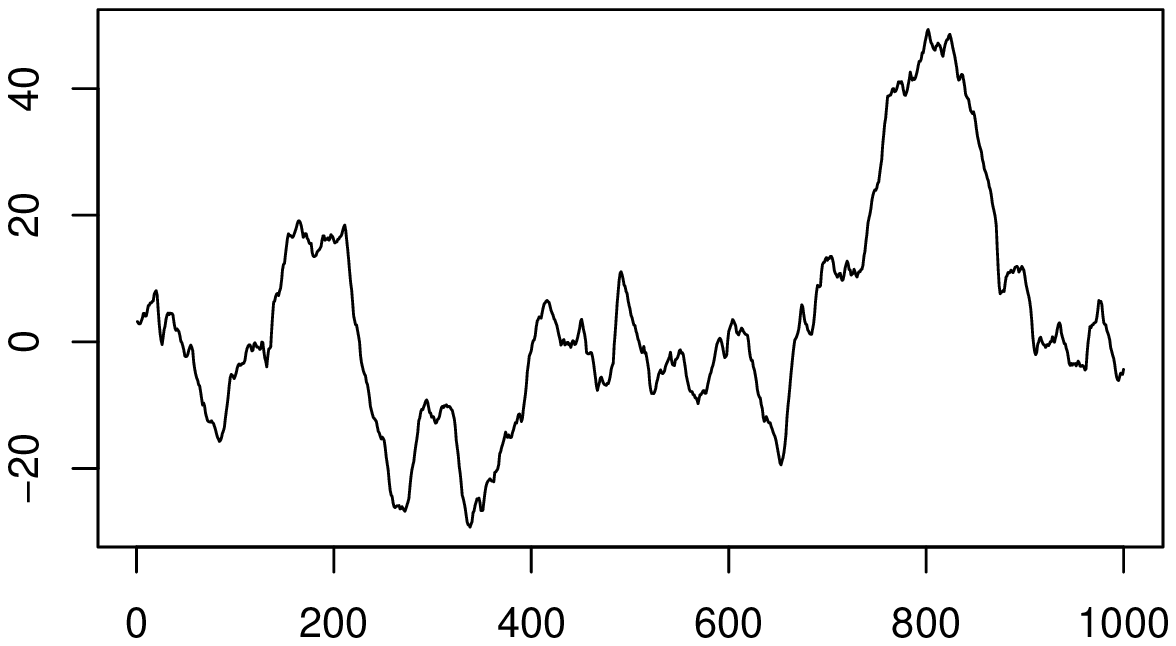}%
\end{minipage}\hfill{}%
\begin{minipage}[t][1\totalheight][c]{0.33\textwidth}%
\includegraphics[width=1.1\textwidth]{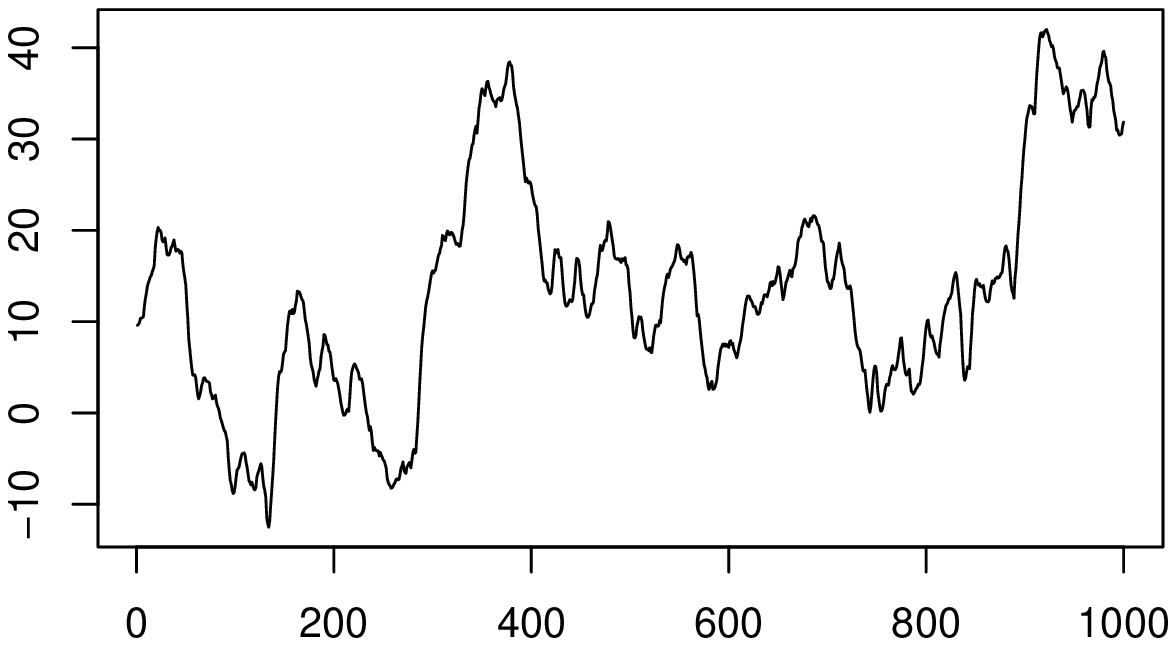}%
\end{minipage}\hfill{}%
\begin{minipage}[t][1\totalheight][c]{0.33\textwidth}%
\includegraphics[width=1.1\textwidth]{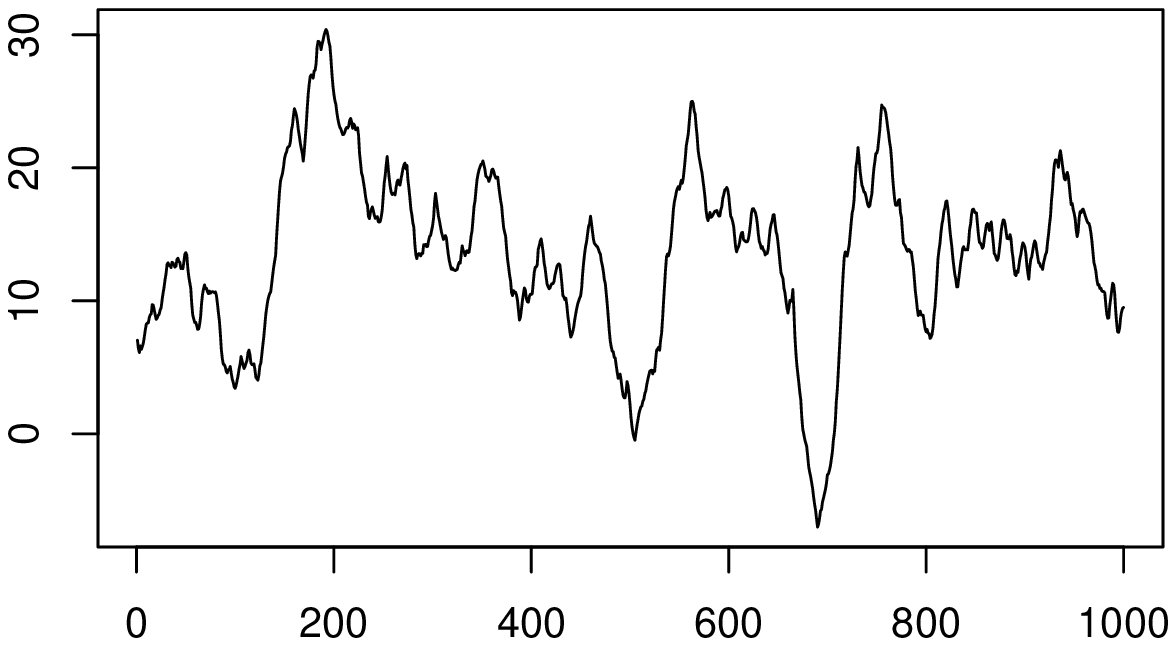}%
\end{minipage}\vspace*{-25pt}

\begin{minipage}[t][1\totalheight][c]{0.33\textwidth}%
\includegraphics[width=1.1\textwidth]{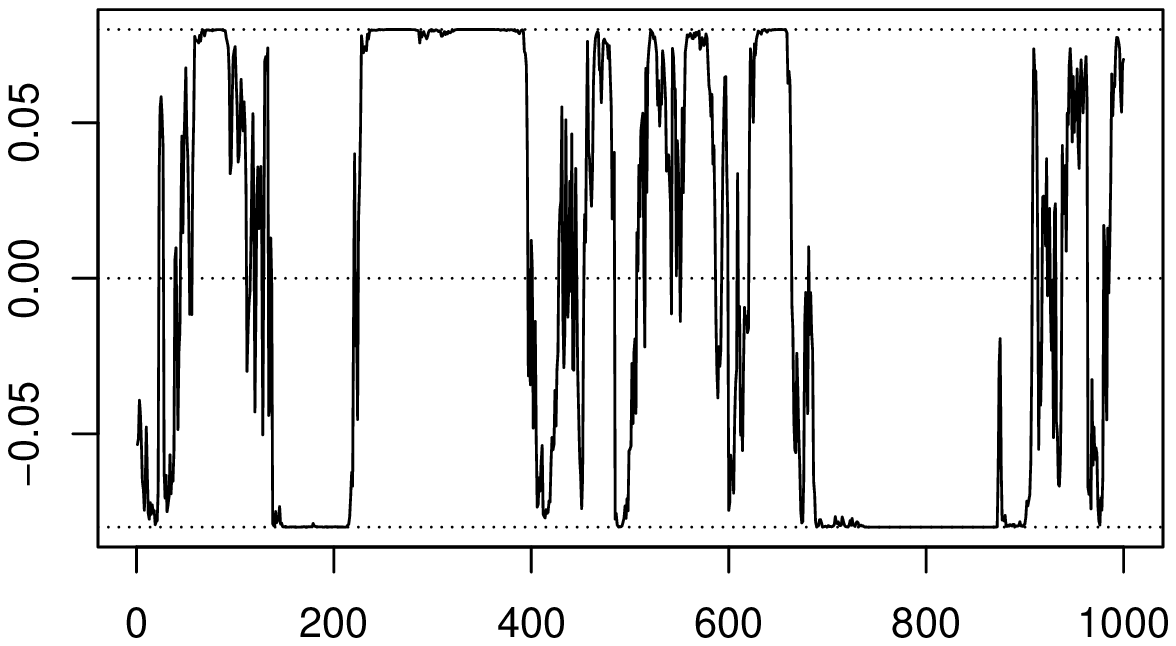}%
\end{minipage}\hfill{}%
\begin{minipage}[t][1\totalheight][c]{0.33\textwidth}%
\includegraphics[width=1.1\textwidth]{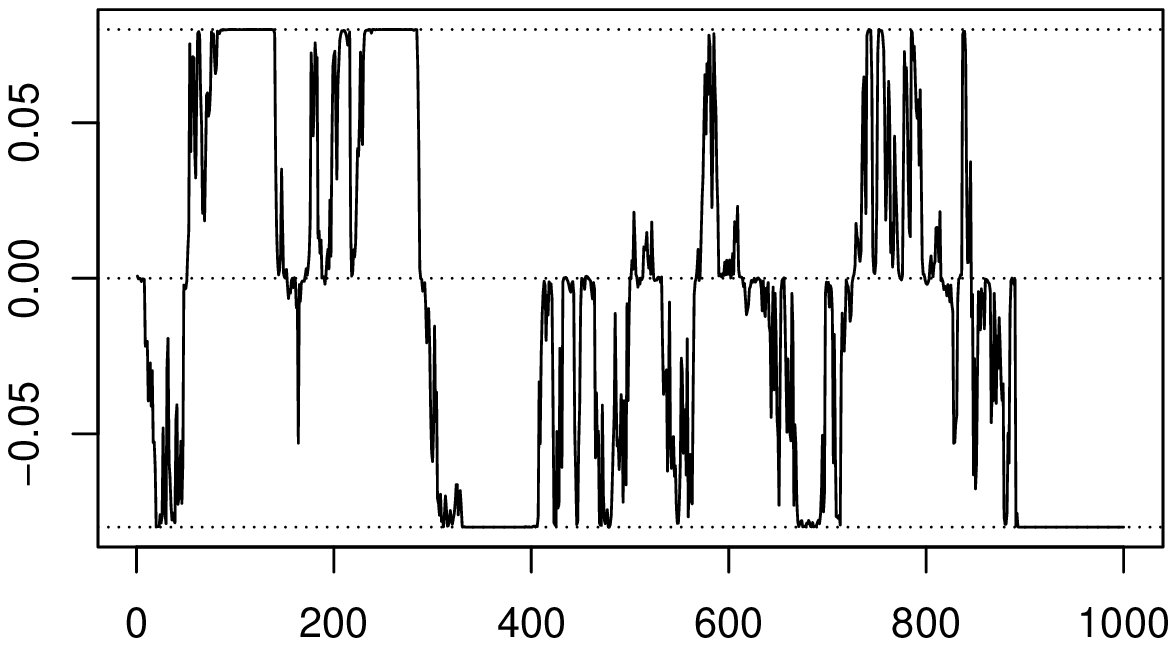}%
\end{minipage}\hfill{}%
\begin{minipage}[t][1\totalheight][c]{0.33\textwidth}%
\includegraphics[width=1.1\textwidth]{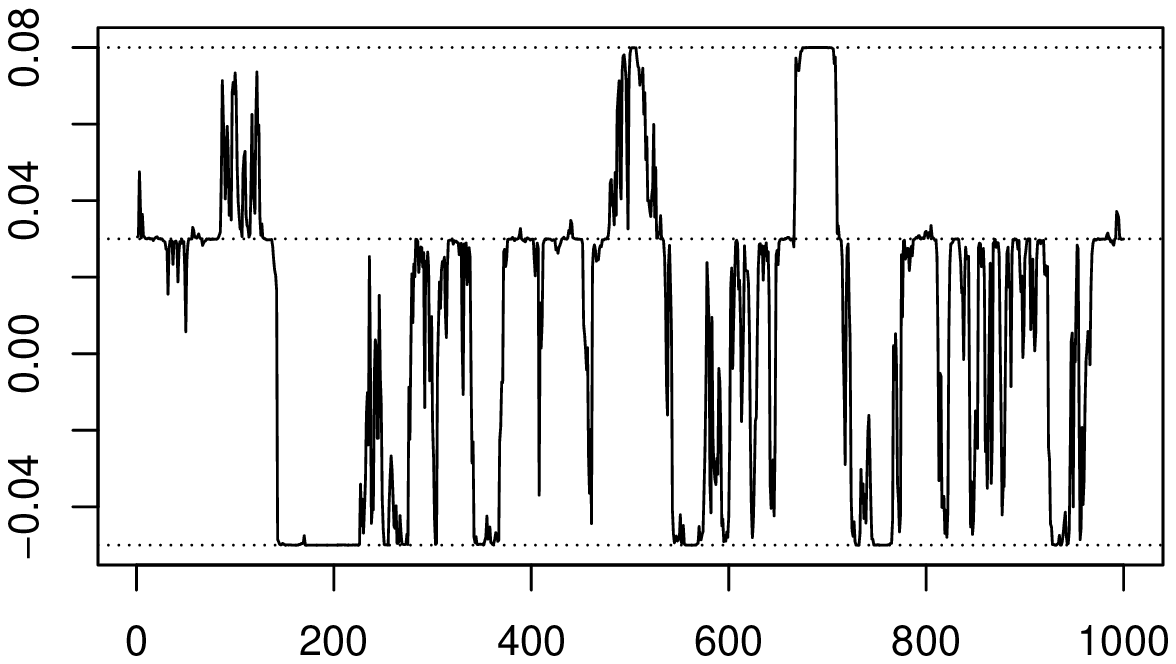}%
\end{minipage}\vspace*{-25pt}

\begin{minipage}[t][1\totalheight][c]{0.33\textwidth}%
\includegraphics[width=1.1\textwidth]{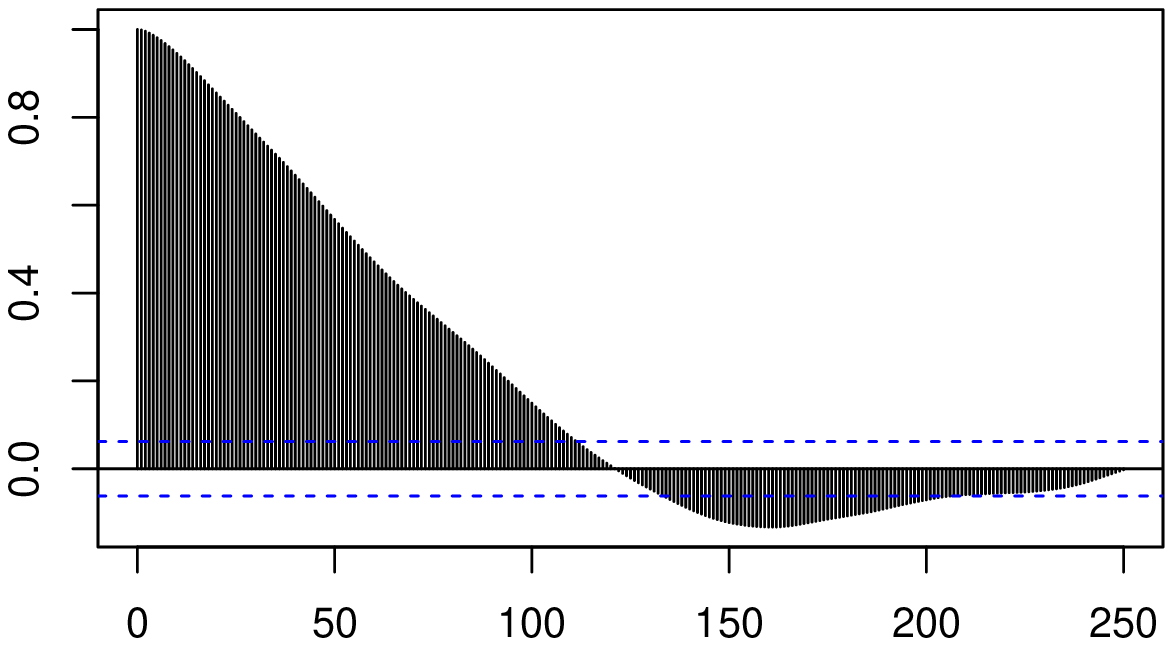}%
\end{minipage}\hfill{}%
\begin{minipage}[t][1\totalheight][c]{0.33\textwidth}%
\includegraphics[width=1.1\textwidth]{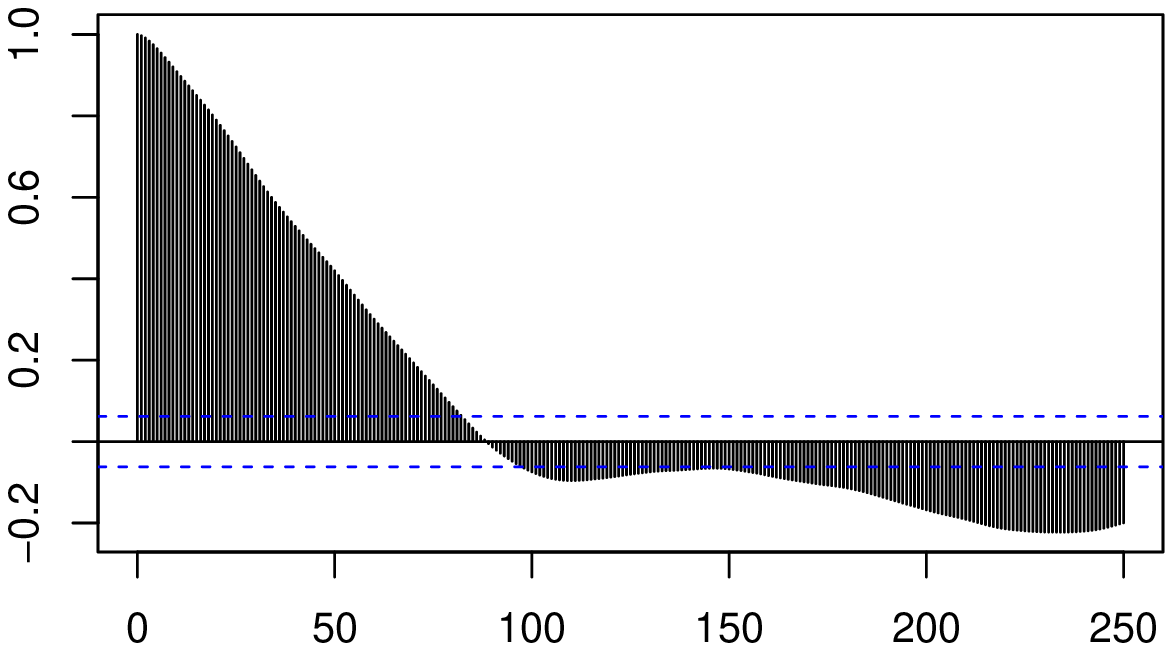}%
\end{minipage}\hfill{}%
\begin{minipage}[t][1\totalheight][c]{0.33\textwidth}%
\includegraphics[width=1.1\textwidth]{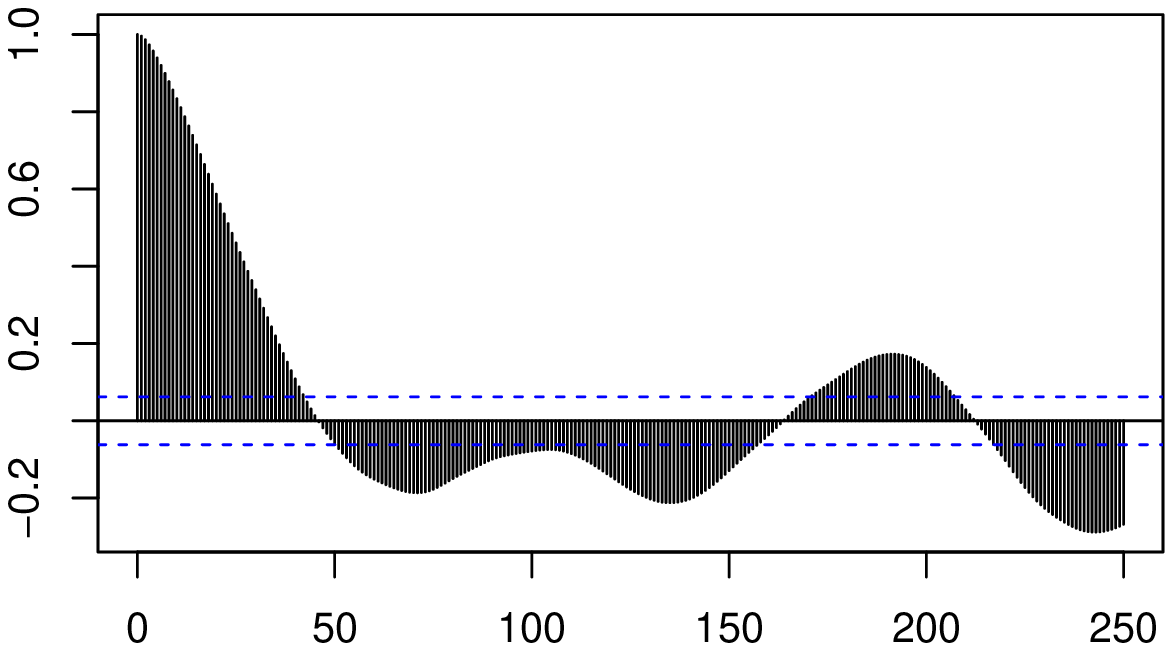}%
\end{minipage}\vspace*{-15pt}
\caption{\label{FigEx1} \uline{Top row.} Graphs of the function $I(u)=\nu_{1}L(u;b,a_{1})+\nu_{2}(1-L(u;b,a_{2}))$
in (\ref{Function I(u)}). Left: $b=2,$ $\nu_{1}=-0.08$, $\nu_{2}=0.08$,
and $a_{1}=a_{2}=0$. Middle: $b=4$, $\nu_{1}=-0.08$, $\nu_{2}=0.08$,
$a_{1}=1$, $a_{2}=4$. Right: $b=5$, $\nu_{1}=-0.05$, $\nu_{2}=0.08$,
$a_{1}=1$, and $a_{2}=4$ (the horizontal dotted line at $0.03$).
\uline{Second row.} Simulated time series of $y_{t}$ corresponding
to the time-varying intercept functions $I(u_{t})$ in the top row.
The 1000 observations are generated from model (\ref{eq:Ex1}) with
$p=2$, $\pi_{1}=0.75$, and $\varepsilon_{t}$ distributed as rescaled
Student's $t$ distribution with five degrees of freedom (and $E[\varepsilon_{1}^{2}]$
equal to $0.300$ on the left, $0.250$ in the middle, and $0.125$
on the right). \uline{Third row.} The corresponding time series
graphs of $I(u_{t})$. The dotted lines show the minimum and maximum,
and the value zero in the first and second columns and $0.03$ in
the third column. \uline{Bottom row.} The corresponding autocorrelation
functions of the simulated series. The first three autocorrelations
are 0.999, 0.996, 0.992 (left); 0.997, 0.991, 0.984 (middle); and
0.996, 0.987, 0.973 (right). }
\end{figure}

We now provide intuitive and graphical illustrations of the behavior
processes covered by Proposition 1 can exhibit. An informal description
captures the main features concisely: For `values of $u_{t-1}$ in
the extreme left tail', the intercept term $I(u_{t-1})$ is close
to $\nu_{2}>0$ resulting in increasing drift towards `central values
of $u_{t-1}$'; conversely, for `values of $u_{t-1}$ in the extreme
right tail', $I(u_{t-1})$ is close to $\nu_{1}<0$ and decreasing
drift towards `central values of $u_{t-1}$' takes place; finally,
for such `central values of $u_{t-1}$', unit root behavior without
drift occurs. This informal description is illustrated in the top
row of Figure \ref{FigEx1} using three different cases of the function
$I(\cdot)$ in (\ref{Function I(u)}); the precise parameter values
used can be found in the caption of Figure~\ref{FigEx1}.

The graph on the left illustrates the simplest possible case. When
$u$ takes negative values the function $I(u)$ takes positive values;
for $u<-2$ the values of $I(u)$ are very close to its supremum,
in this example $0.08$. As $u_{t}=I(u_{t-1})+u_{t-1}+\varepsilon_{t}$,
the behavior of the process $u_{t}$ is for $u_{t-1}<0$ (and particularly
for $u_{t-1}<-2$) close to that of a unit root process with an increasing
drift. A similar behavior occurs when $u_{t-1}$ takes positive values
(and particularly when $u_{t-1}>2$) but then the drift is decreasing.
Thus, large absolute values of $u_{t-1}$ induce a drift towards the
origin which provides intuition why the process $u_{t}$ can be ergodic
and stationary even though its behavior shows resemblance to a unit
root process with a (negligible) drift near the origin where $I(u)\approx0$.
The other two graphs in the top row of Figure \ref{FigEx1} illustrate
similar but somewhat more involved possibilities for the function
$I(u)$. The middle graph illustrates that the `unit root regime'
can take place over a wider range of values of $u$ and these values
need not be centered at the origin; in this graph $I(u)\approx0$
for $u$ roughly between $1.5$ and $3.5$. The graph on the right
illustrates a possibility in which also the `middle regime' corresponds
to a unit root with positive drift.

The second row of Figure \ref{FigEx1} presents three examples of
simulated time series of $y_{t}$. These are generated from second-order
versions of model (\ref{eq:Ex1}) using the three time-varying intercept
terms $I(u_{t-1})$ depicted in the first row. In all cases the autoregressive
coefficient $\pi_{1}$ is equal to 0.75 and the error terms $\varepsilon_{t}$
are generated from rescaled Student's $t$-distribution with five
degrees of freedom (and different variances, see the caption for details).
The third and fourth rows of Figure \ref{FigEx1} show the time series
graphs of the time-varying intercept terms $I(u_{t})$ and the autocorrelation
functions of $y_{t}$, respectively, in these three examples. The
time series graphs of $y_{t}$ in the second row bear a resemblance
to those of unit root processes but nevertheless exhibit mean-reverting
behavior. However, the mean reversion takes place slower than it would
for a geometrically ergodic process. The autocorrelation functions
also show very strong persistence. These features are related to the
fact that the considered models are only polynomially ergodic with
rate $r(n)=n^{4-\delta}$ for some $\delta>0$ (this follows from
Proposition 1 as the error terms used only have moments of order smaller
than five).

We notice from Figure \ref{FigEx1} that the periods when $u_{t}$
takes large absolute values can be rather long and they can contain
both increasing and decreasing periods for $y_{t}$. An example is
the time series in the first column around $t\approx800$ corresponding
to the largest peak of the time series of $y_{t}$; similar features
occur also in the time series in the second and third columns, often
around peaks or troughs of the series. We also note from the second
and third columns of Figure \ref{FigEx1} that the time series of
$I(u_{t})$ stays close to 0 or 0.03, respectively, for some time
(corresponding to behavior of $u_{t}$ close to that of a unit root
process without a drift or with a drift). An example is the rather
long decreasing period in the time series of $y_{t}$ in the third
column, roughly between $t\approx200$ and $t\approx500$, that is
to large extent due to unit root type behavior.

\subsection{Example with time-varying slope term of ESTAR type}

Next we consider example (\ref{eq:Ex2}) with the time-varying slope
term $S(u_{t-1})$ being either 
\begin{equation}
S_{1}(u_{t-1})=1-\frac{r_{0}}{h(u_{t-1})}\qquad\text{or}\qquad S_{2}(u_{t-1})=\exp\Bigl\{-\frac{r_{0}}{h(u_{t-1})}\Bigr\},\label{Functions S}
\end{equation}
where $r_{0}>0$ and the positive-valued function $h$ is such that
$h(u)$ is large whenever $u$ is large in absolute value (formal
requirements for $h$ are given in Proposition 2 below). Then the
time-varying slope term $S(u_{t-1})$ takes values in some interval
$[s,1)$ ($s<1$) and attains values arbitrarily close to 1 for values
of $u_{t-1}$ large in absolute value. The shapes of $S_{1}(u)$ and
$S_{2}(u)$ as functions of $u$ resemble the `inverted bell curve
form' commonly employed in exponential STAR (ESTAR) models (see, e.g.,
\citet{vandijk2002smooth}). The main difference between the two functions
in (\ref{Functions S}) is that $S_{2}(u)$ takes values in the unit
interval $(0,1)$ whereas $S_{1}(u)$ can also take negative values.

Before providing concrete examples we state the following proposition
which imposes conditions on the function $h$ above to ensure that
the results of Theorems 2 and 3 hold for model (\ref{eq:Ex2}) with
$S(\cdot)$ as in (\ref{Functions S}). The proof is straightforward
and available in the Appendix.
\begin{prop}
Consider the process $y_{t}$ defined by $u_{t}-\nu=S(u_{t-1})(u_{t-1}-\nu)+\varepsilon_{t}$
as in (\ref{eq:Ex2}) (with $u_{t}=\varpi(L)y_{t}$ and the roots
of $\varpi(z)$ outside the unit circle) with $S(u_{t-1})$ being
either $S_{1}(u_{t-1})$ or $S_{2}(u_{t-1})$ in (\ref{Functions S})
(with $r_{0}>0$) and with the function $h$ therein satisfying\vspace*{-4pt}
\begin{lyxlist}{000}
\item [{(h)}] $h:\mathbb{R}\rightarrow(0,\infty)$ is measurable, bounded
on compact sets, satisfies $h(u)\rightarrow\infty$ as $\left|u\right|\rightarrow\infty$,
and is such that $c_{1}h(u)\leq\left|u\right|^{\rho}$ and $\left|u\right|^{\rho+c_{2}}\leq c_{3}h^{2}(u)$
for $\left|u\right|\geq M_{0}$, some $c_{1},c_{2},c_{3},M_{0}>0$,
and $0<\rho\leq2$.\vspace*{-4pt}
\end{lyxlist}
Assume further that either Assumption\vspace*{-4pt}
\begin{lyxlist}{00}
\item [{\small\ \ (1)}] 2(a) is satisfied with $\kappa_{0}<\rho$,\vspace*{-6pt}
\item [{\small\ \ (2)}] 2(a) is satisfied with $\kappa_{0}=\rho$, or\vspace*{-6pt}
\item [{\small\ \ (3)}] 2(b) is satisfied with an $s_{0}$ such that one
of the conditions (i)\textendash (iii) of Theorem 3 holds.\vspace*{-4pt}
\end{lyxlist}
Then, under condition {\small (1)/(2)/(3)}, the process $\boldsymbol{y}_{t}=(y_{t},\ldots,y_{t-p+1})$
is either\vspace*{-4pt}
\begin{lyxlist}{00}
\item [{\small\ \ (1)}] subexponentially ergodic,\vspace*{-6pt}
\item [{\small\ \ (2)}] geometrically ergodic, or\vspace*{-6pt}
\item [{\small\ \ (3)}] polynomially ergodic.\vspace*{-4pt}
\end{lyxlist}
Moreover, condition (h) above is satisfied for i) $h(u)=1+\lvert u-a\rvert^{\rho}$,
ii) $h(u)=(1+\lvert u-a\rvert)^{\rho}$, iii) $h(u)=(1+(u-a)^{2})^{\rho/2}$,
iv) $h(u)=1+\lvert u-a_{1}\rvert^{\rho_{1}}+\lvert u-a_{2}\rvert^{\rho_{2}}$,
v) $h(u)=1+(1+\lvert u-a_{1}\rvert)^{\rho_{1}}+(1+\lvert u-a_{2}\rvert)^{\rho_{2}}$,
or vi) $h(u)=1+(1+(u-a_{1})^{2})^{\rho_{1}/2}+(1+(u-a_{2})^{2})^{\rho_{2}/2}$
($\rho,\rho_{1},\rho_{2}\in(0,2]$; $a,a_{1},a_{2}\in\mathbb{R}$).
\end{prop}
The obtained rate of ergodicity is again either geometric, subexponential,
or polynomial depending on the moment assumptions made (the precise
convergence rates can be obtained from Theorems 2 and 3; for existence
of moments, see the corollaries to Theorems 2 and 3). The last part
of the proposition lists several potential concrete choices for the
function $h$, of which cases (iii) and (vi) may be convenient if
differentiability of the function $h$ is desired. 

To illustrate the type of behavior processes covered by Proposition
2 may exhibit, consider as a simple example model (\ref{eq:Ex2})
with $\nu=0$, the slope term $S_{1}(u_{t-1})$ in (\ref{Functions S}),
and $h(u)=1+\lvert u\rvert^{\rho}$ (cf.~(\ref{Example_1})). The
considered model then becomes 
\begin{equation}
u_{t}=\biggl(1-\frac{r_{0}}{1+\lvert u_{t-1}\rvert^{\rho}}\biggr)u_{t-1}+\varepsilon_{t}.\label{Specific example_1}
\end{equation}
The shape of the slope coefficient $S_{1}(u_{t-1})$ as a function
of $u_{t-1}$ is now similar to that of an inverted bell curve increasing
monotonically to unity as $\lvert u_{t-1}\rvert$ increases, and $S_{1}(u_{t-1})$
takes values within the interval $[1-r_{0},1)$. Note that depending
on the value of $r_{0}$, the slope coefficient takes values potentially
only near the unity (say, in the interval $[0.95,1)$) or even in
rather extreme ranges (say, in the interval $[-100,1)$) so that very
different behaviors can be accommodated. 

To explain why such processes can still be ergodic and stationary,
note that for large values of $\lvert u_{t-1}\rvert$ the slope $S_{1}(u_{t-1})$
always takes values within $(-1,1)$ (or a much smaller subset of
it near unity). This prevents the process $u_{t}$ from exploding
and ensures mean-reverting behavior. Note that for larger values of
$\rho$ (within the permitted range $0<\rho\leq2$), the slope $S_{1}(u_{t-1})$
approaches unity faster as $\lvert u_{t-1}\rvert$ increases, intuitively
corresponding to more `wandering' behavior of the observed series.
This is reflected in Proposition 2 as slower rates of ergodicity being
related to larger values of the parameter $\rho$ (see also Theorems
2 and 3).

\begin{figure}[p]
\begin{minipage}[t][1\totalheight][c]{0.33\textwidth}%
\includegraphics[width=1.1\textwidth]{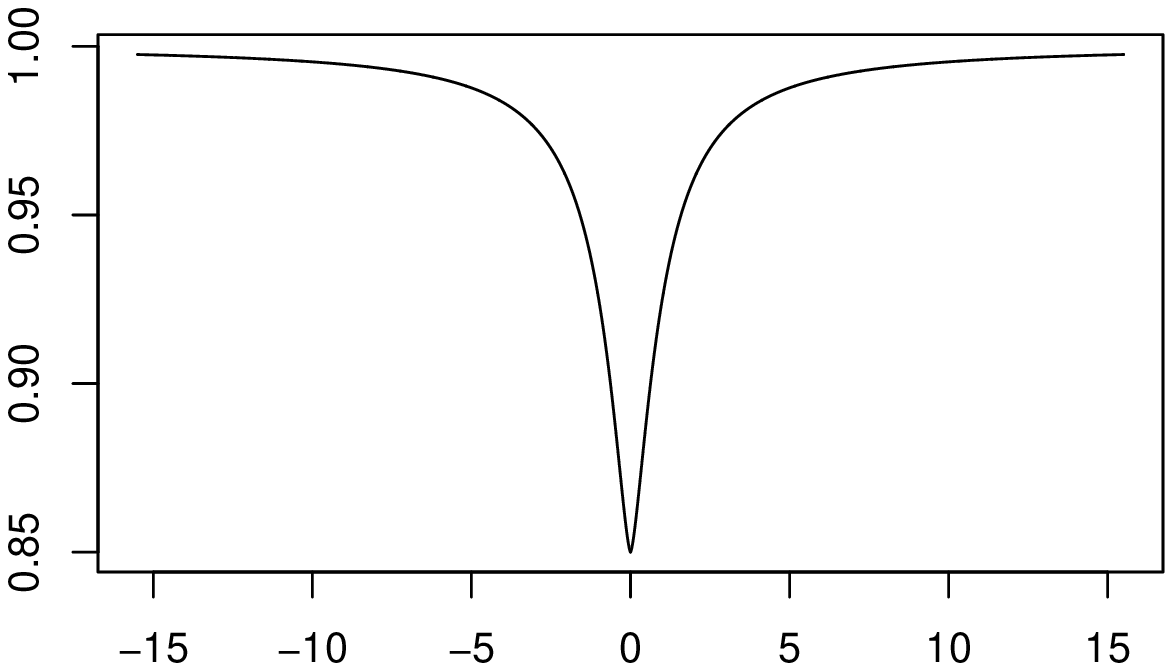}%
\end{minipage}\hfill{}%
\begin{minipage}[t][1\totalheight][c]{0.33\textwidth}%
\includegraphics[width=1.1\textwidth]{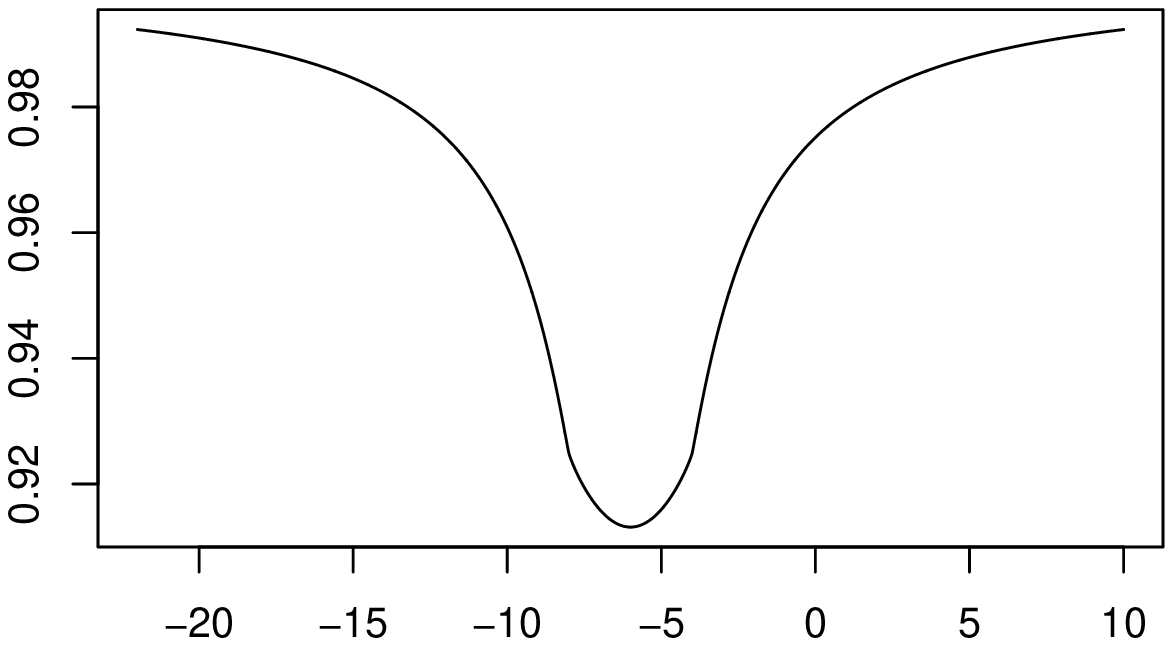}%
\end{minipage}\hfill{}%
\begin{minipage}[t][1\totalheight][c]{0.33\textwidth}%
\includegraphics[clip,width=1.1\textwidth]{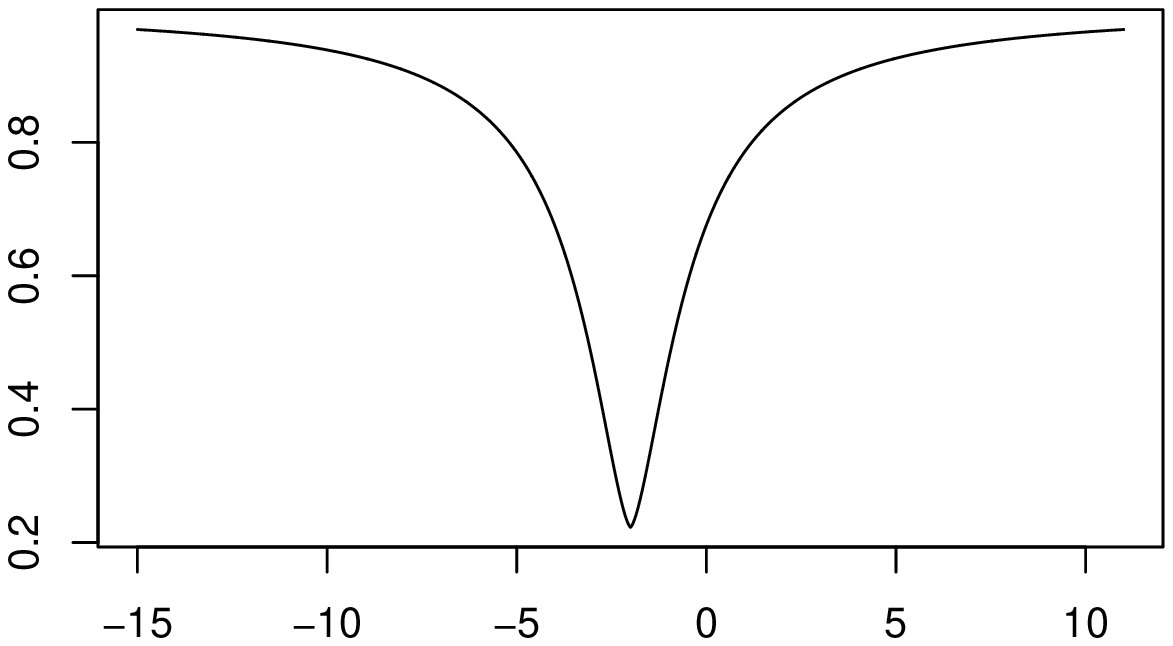}%
\end{minipage}\vspace*{-25pt}

\begin{minipage}[t][1\totalheight][c]{0.33\textwidth}%
\includegraphics[width=1.1\textwidth]{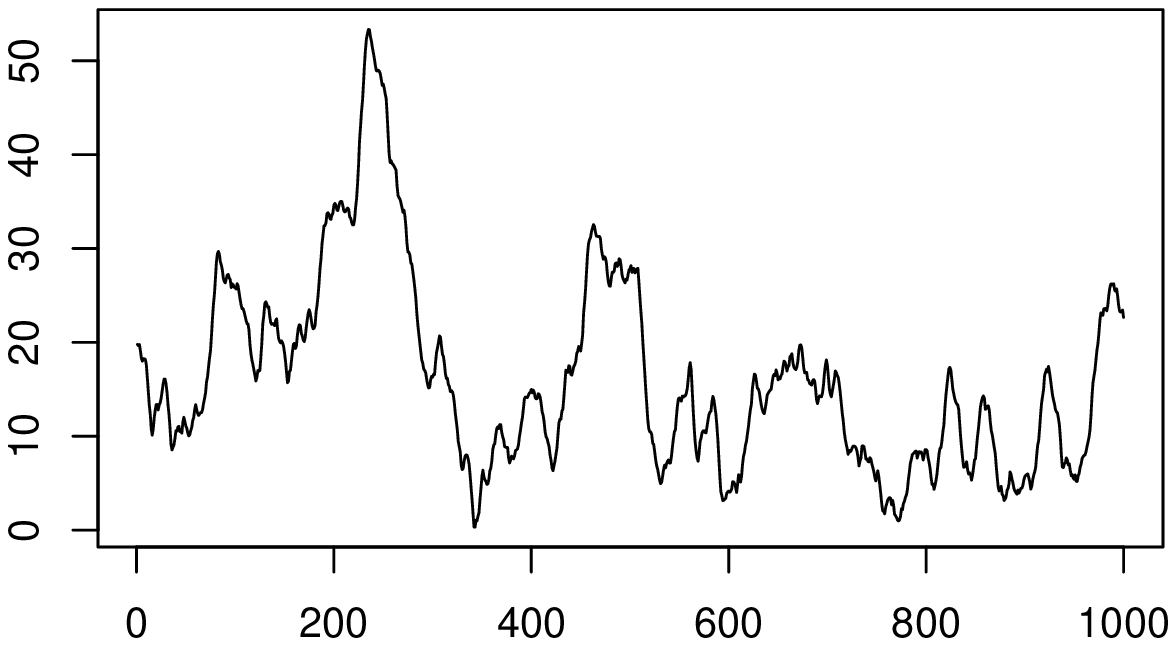}%
\end{minipage}\hfill{}%
\begin{minipage}[t][1\totalheight][c]{0.33\textwidth}%
\includegraphics[width=1.1\textwidth]{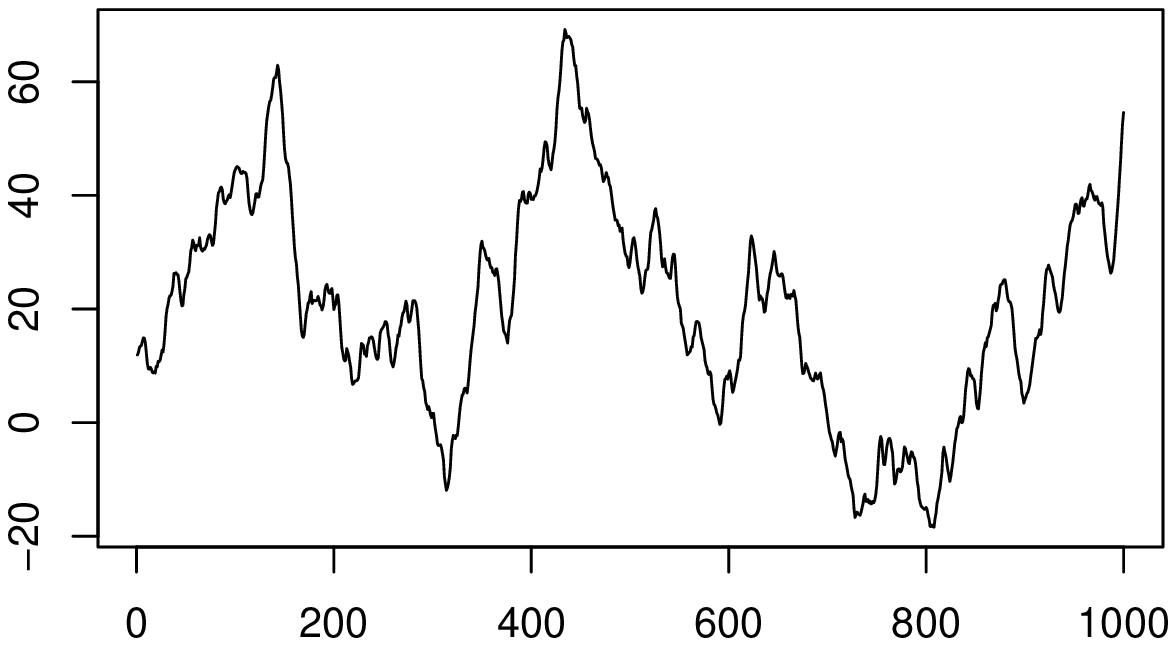}%
\end{minipage}\hfill{}%
\begin{minipage}[t][1\totalheight][c]{0.33\textwidth}%
\includegraphics[bb=0bp 0bp 396bp 287bp,width=1.1\textwidth]{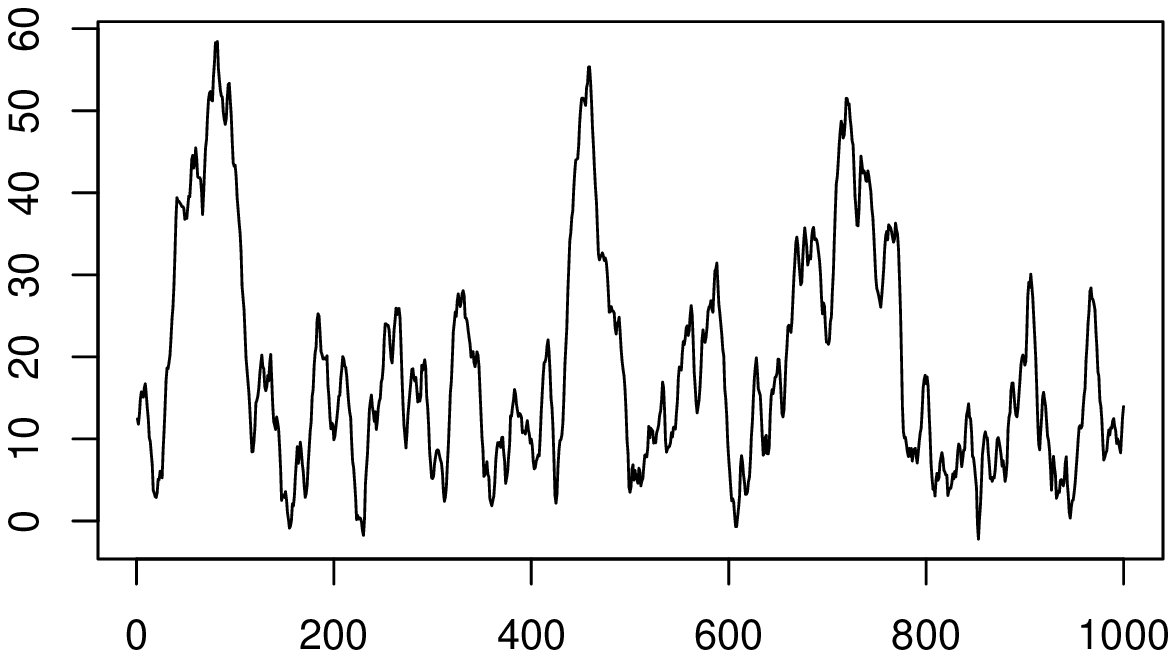}%
\end{minipage}\vspace*{-25pt}

\begin{minipage}[t][1\totalheight][c]{0.33\textwidth}%
\includegraphics[width=1.1\textwidth]{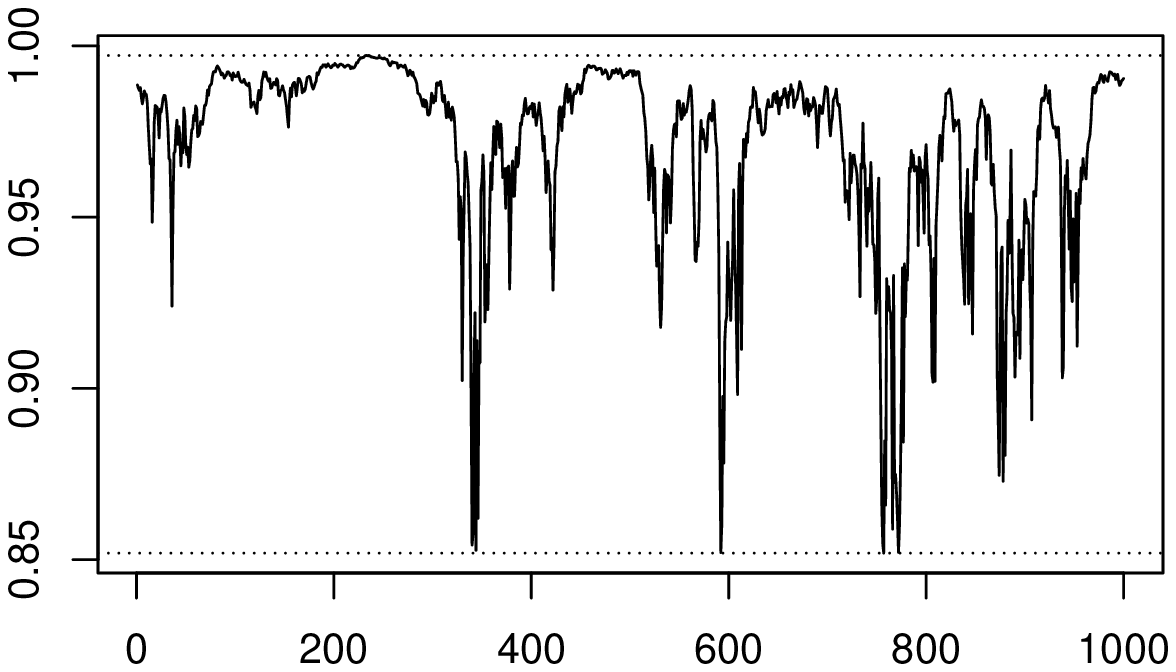}%
\end{minipage}\hfill{}%
\begin{minipage}[t][1\totalheight][c]{0.33\textwidth}%
\includegraphics[width=1.1\textwidth]{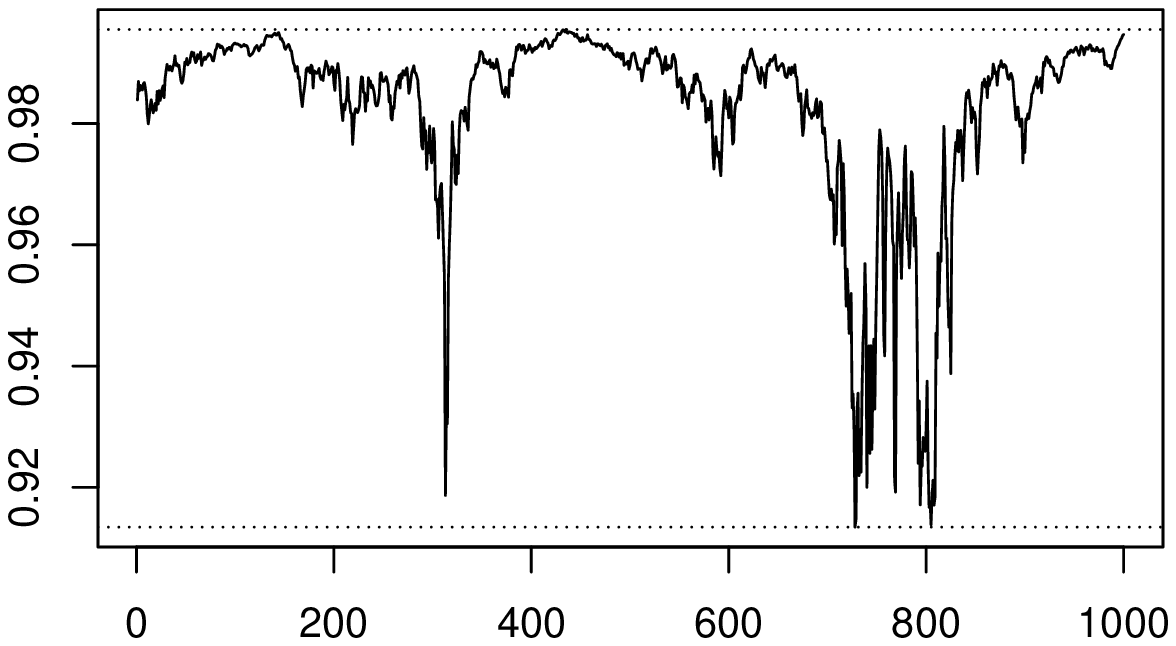}%
\end{minipage}\hfill{}%
\begin{minipage}[t][1\totalheight][c]{0.33\textwidth}%
\includegraphics[width=1.1\textwidth]{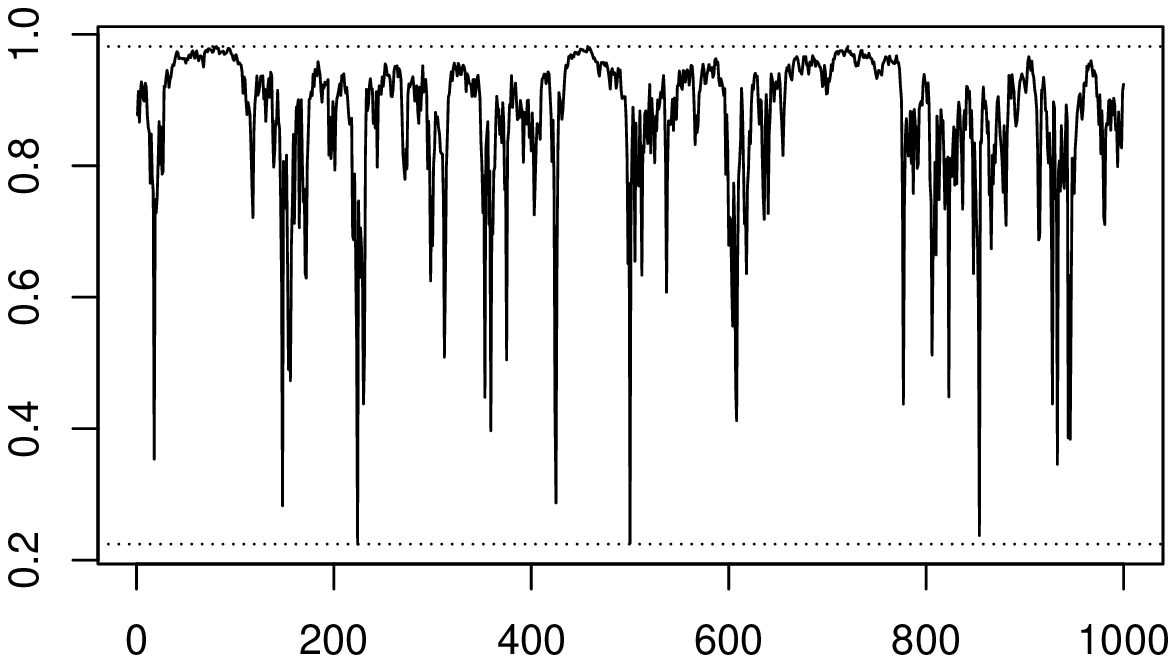}%
\end{minipage}\vspace*{-25pt}

\begin{minipage}[t][1\totalheight][c]{0.33\textwidth}%
\includegraphics[width=1.1\textwidth]{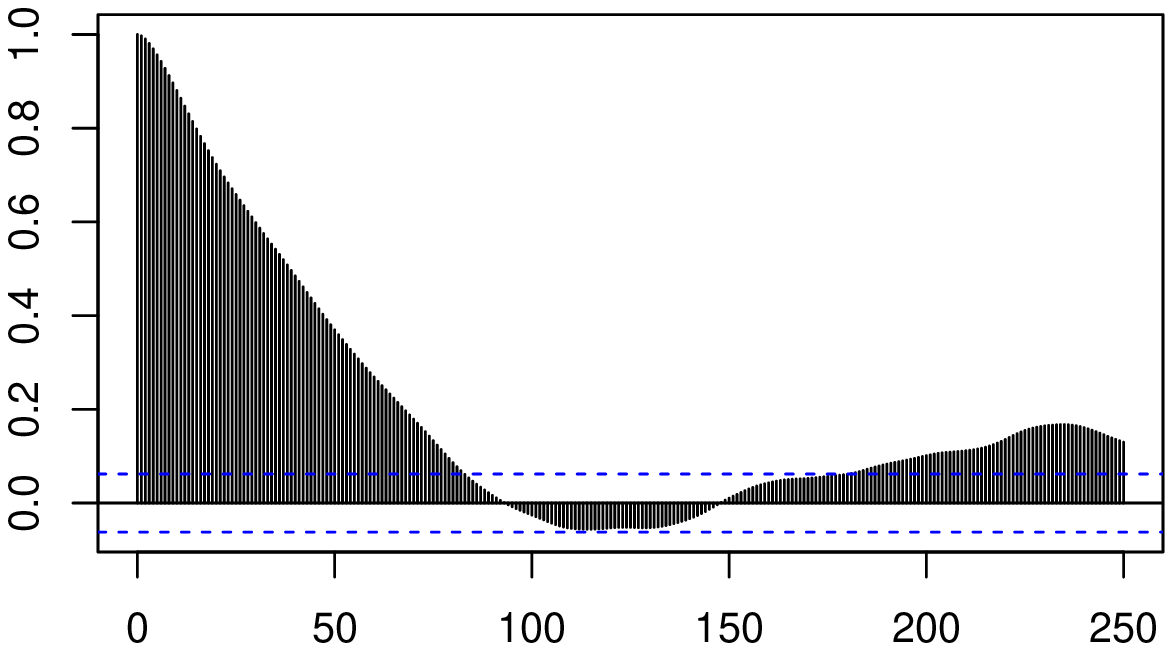}%
\end{minipage}\hfill{}%
\begin{minipage}[t][1\totalheight][c]{0.33\textwidth}%
\includegraphics[width=1.1\textwidth]{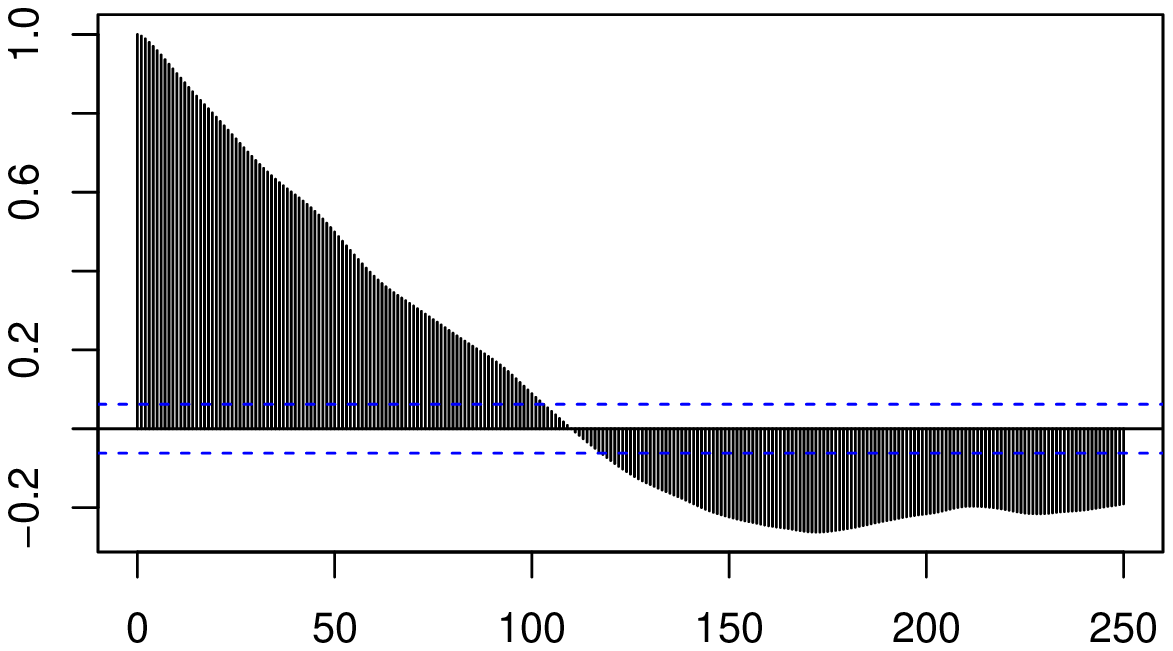}%
\end{minipage}\hfill{}%
\begin{minipage}[t][1\totalheight][c]{0.33\textwidth}%
\includegraphics[width=1.1\textwidth]{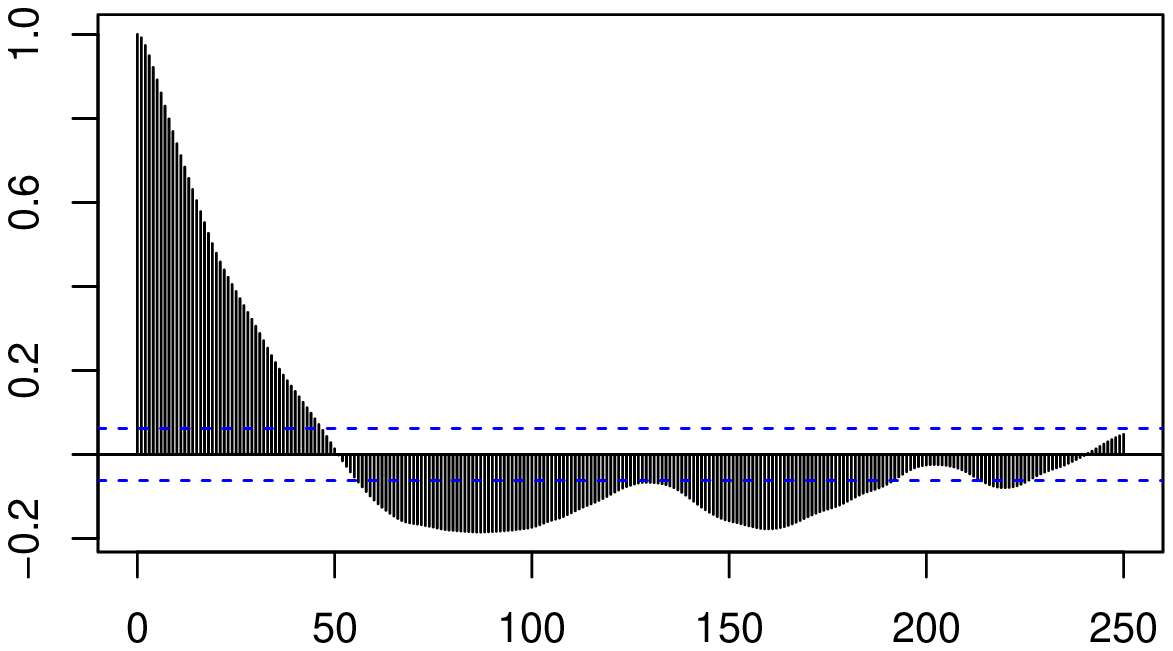}%
\end{minipage}\vspace*{-15pt}
\caption{\label{FigEx2} \uline{Top row.} Graphs of $S(u)$ in (\ref{Functions S}).
Left: $S_{1}(u)$ with $r_{0}=0.15$ and $h(u)=1+\left|u\right|^{1.5}$.
Middle: $S_{1}(u)$ with $r_{0}=0.5$ and $h(u)=1+\lvert u+4\rvert^{1.25}+\lvert u+8\rvert^{1.25}$.
Right: $S_{2}(u)$\textbf{ }with $r_{0}=1.5$ and $h(u)=1+\left|u+2\right|^{1.5}$.
\uline{Second row.} Simulated time series of $y_{t}$ corresponding
to the time-varying slope functions $S(u_{t})$ in the top row. The
1000 observations are generated from model (\ref{eq:Ex2}) with $p=2$,
$\pi_{1}=0.75$, $\nu=2$, and Gaussian $\varepsilon_{t}$ ($\varepsilon_{t}\sim N(0,0.25)$
on the left; $\varepsilon_{t}\sim N(0,0.75)$ in the middle; $\varepsilon_{t}\sim N(0,1.5)$
on the right). \uline{Third row.} The corresponding time series
graphs of $S_{1}(u_{t-1})=1-r_{0}/h(u_{t-1})$ on the left and in
the middle, and $S_{2}(u_{t-1})=\exp\{-r_{0}/h(u_{t-1})\}$ on the
right. The dotted lines show the minimum and maximum. \uline{Bottom
row.} The corresponding autocorrelation functions of the simulated
series. The first three autocorrelations are 0.997, 0.990, 0.981 (left);
0.996, 0.989, 0.980 (middle); and 0.992, 0.974, 0.950 (right). }
\end{figure}

Figure \ref{FigEx2} illustrates the preceding discussion. The top
row depicts examples of the function $S(u)$ with some particular
choices of the function $h$ (see the caption of Figure 2 for the
details). In the figures on the left and in the middle $S(u)=S_{1}(u)$
whereas in the figure on the right $S(u)=S_{2}(u)$. In each example
the shape of the function $S(u)$ is similar to that of an inverted
bell curve increasing monotonically to unity as $\lvert u\rvert$
increases.

Rows 2\textendash 4 of Figure \ref{FigEx2} present simulated time
series of $y_{t}$, time series graphs of the time-varying slope coefficients
$S(u_{t-1})$, and the autocorrelation functions of $y_{t}$, respectively,
in these three examples. In the first and second columns the time
series graphs of $y_{t}$ and the related autocorrelation functions
indicate unit root type behavior and strong persistence; in these
cases also the slope coefficient $S(u_{t-1})$ takes values mostly
larger than $0.90$ and often very close to one. Compared to these
to cases, the time series graph of $y_{t}$ in the third column appears
less \textquoteleft wandering\textquoteright{} and this feature is
also reflected in the related autocorrelation function which decays
faster and in a slope coefficient $S(u_{t-1})$ taking values further
away from unity. Despite the three examples exhibiting somewhat different
behaviours, all of them exhibit mean-reverting behavior and are subexponentially
ergodic (this is due to Proposition 2 because $\rho>1$ and the error
terms used are normally distributed and hence satisfy Assumption 2(a)
with $\kappa_{0}=1$). The mean reversion again takes place slower
than would be the case for geometrically ergodic processes.

\subsection{Example of a more general formulation}

Finally, we briefly consider example (\ref{eq:Ex3}) and for simplicity
set $p=2$. The time-varying slope term $S(u_{t-1})$ can be either
one of the two options in (\ref{Functions S}). As for $F(\bm{y}_{t-1})$,
we set $F(\bm{y}_{t-1})=\exp\{-\gamma\lvert\boldsymbol{y}_{t-1}\rvert^{2}\}(\theta_{1}y_{t-1}+\theta_{2}y_{t-2})$,
where $\boldsymbol{y}_{t-1}=(y_{t-1},y_{t-2})$, $\gamma>0$, and
$\theta=(\theta_{1},\theta_{2})$ can take any values in $\mathbb{R}^{2}$.
That is, the considered model reads as
\begin{equation}
y_{t}=\pi_{1}y_{t-1}+S(u_{t-1})(y_{t-1}-\pi_{1}y_{t-2})+\exp\{-\gamma\lvert\boldsymbol{y}_{t-1}\rvert^{2}\}(\theta_{1}y_{t-1}+\theta_{2}y_{t-2})+\varepsilon_{t}.\label{General example_2}
\end{equation}
On the right hand side of (\ref{General example_2}) the term $\exp\{-\gamma\lvert\boldsymbol{y}_{t-1}\rvert^{2}\}$
has a bell-shaped form (as a function of $\lvert\boldsymbol{y}_{t-1}\rvert$)
with maximum at the origin while for choices such as $h(u_{t-1})=1+\lvert u\rvert^{\rho}$
the shape of $S(u_{t-1})$ is that of an inverted bell curve. Thus,
given the shape of the terms $S(u_{t-1})$ and $\exp\{-\gamma\lvert\boldsymbol{y}_{t-1}\rvert^{2}\}$,
model (\ref{General example_2}) can be viewed as a certain type of
three-regime ESTAR model (see, e.g., \citet{vandijk2002smooth}).

It is straightforward to check that model (\ref{General example_2})
satisfies Assumption 1(ii) for both of the two options of $S$ and
with $h$ as in Proposition 2 (details are available in the Appendix;
note that in this example the function $\tilde{g}$ in equation (\ref{NLAR(p)_phi})
does not depend on $u_{t-1}$ only). It may be worth noting that Assumption
1(ii.a) does not hold if we replace the norm $\lvert\boldsymbol{y}_{t-1}\rvert$
in (\ref{General example_2}) with a linear function of $\boldsymbol{y}_{t-1}$
such as $u_{t-1}$. Another point worth noting is that Assumption
1(ii) does not rule out the possibility of setting $\pi_{1}=0$ (and
$u_{t-1}=y_{t-1}$) in (\ref{General example_2}), and similarly for
its higher-order counterparts where some or even all the coefficients
$\pi_{1},\ldots,\pi_{p-1}$ may be equal to zero.

Allowing the parameters $\theta_{1}$ and $\theta_{2}$ in model (\ref{General example_2})
to be totally unrestricted highlights the fact that the autoregressions
we consider may exhibit rather arbitrary behavior for moderate values
of the observed series. As indicated in the Introduction, geometrically
ergodic nonlinear autoregressions with features of this kind have
previously been considered by \citet{lu1998geometric}, \citet{gourieroux2006stochastic},
\citet{bec2008acr}, and others. However, in most of these previous
models stationarity is approached the further away the process moves
from the origin whereas in our model unit root type behavior prevails
for large absolute values of the process.

\section{Conclusions}

In this paper we examined the subgeometric ergodicity of certain higher-order
nonlinear autoregressive models. Generalizing existing first-order
results, we provided conditions that ensure subexponential and polynomial
ergodicity of the considered autoregressions. These results were established
by utilizing suitably formulated drift conditions. Relying on results
in a companion paper \citet{meitz2019subgemix}, useful conclusions
on the convergence rates of $\beta$-mixing coefficients were also
obtained. 

After obtaining theoretical results for rather general models we considered
concrete examples and illustrated them with simulation. However, further
work is needed to judge the usefulness of these models in practical
applications. Several extensions could also be envisioned. For instance,
subgeometric ergodicity of multivariate higher-order autoregressions
or of models with conditional heteroskedasticity are interesting topics
left for future work.

\section*{Appendix}

This Appendix provides the technical details for the results in Section
5. Proofs of Theorems 1\textendash 3 and of Corollary to Theorem 3
are available in the Supplementary Appendix.
\begin{proof}[\textbf{\emph{Proof of Proposition 1}}]
\noindent  We first verify that Assumption 1 holds with $\rho=1$.
Condition (i) holds by assumption. As $u_{t}=u_{t-1}+I(u_{t-1})+\varepsilon_{t}$,
the function $\tilde{g}$ in condition (ii) depends on $u$ only and
equals $I(u)$ (cf.~eqn (\ref{NLAR(p)_pi})) so that condition (ii.a)
holds with $g(u)=u+I(u)$ and $\epsilon(\boldsymbol{x})=0$ (the condition
$\left|g(u)\right|\rightarrow\infty$ as $\left|u\right|\rightarrow\infty$
clearly holds). Thus is suffices to check condition (ii.b) with $g(u)=u+I(u)$
and $I(u)$ as in (\ref{Function I(u)}). Clearly $g(u)$ is bounded
on bounded subsets so the latter part of (\ref{Inequality_Ass 2})
holds. We now show that the former part of (\ref{Inequality_Ass 2})
holds with $\rho=1$, that is, that $\left|g(u)\right|\leq\left|u\right|-r$
for some $r,M_{0}>0$ and $\left|u\right|\geq M_{0}$. Note that the
logistic function $L(u;b,a)$ satisfies $L(u;b,a)\to0$ as $u\rightarrow-\infty$
and $L(u;b,a)\to1$ as $u\rightarrow\infty$. As $I(u)=\nu_{1}L(u;b,a_{1})+\nu_{2}(1-L(u;b,a_{2}))$
with $\nu_{1}<0<\nu_{2}$, we can find a $M_{0}>0$ such that $I(u)>\nu_{2}/2>0$
for $u<-M_{0}$ and $I(u)<\nu_{1}/2<0$ for $u>M_{0}$. Setting $r=\min\{-\nu_{1}/2,\nu_{2}/2\}$
(and choosing a larger $M_{0}$ if necessary) we then have $\left|g(u)\right|=u+I(u)<u-(-\nu_{1}/2)\leq\left|u\right|-r$
for $u>M_{0}$ and $\left|g(u)\right|=-u-I(u)<-u-\nu_{2}/2\leq\left|u\right|-r$
for $u<-M_{0}$, establishing the former part of (\ref{Inequality_Ass 2}).
Thus Assumption 1 holds with $\rho=1$. 

To complete the proof, note that when condition (i) of the Proposition
holds, Theorem 2(i) applies (with $b_{3}=\kappa_{0}$) and process
$\boldsymbol{y}_{t}$ is subexponentially ergodic with convergence
rate $r(n)=(e^{k})^{n^{\kappa_{0}}}$ (for some $k>0$). Similarly,
when condition (ii) of the Proposition holds, Theorem 2(ii) applies
and process $\boldsymbol{y}_{t}$ is geometrically ergodic with convergence
rate $r(n)=(e^{c})^{n}$ (for some $c>0$). Finally, when condition
(iii) of the Proposition holds, Theorem 3 applies and process $\boldsymbol{y}_{t}$
is polynomially ergodic with convergence rate $r(n)=n^{s_{0}-1}$.
\end{proof}
\smallskip{}

\begin{proof}[\textbf{\emph{Proof of Proposition 2}}]
\noindent Assumption 1(i) again holds by assumption. For Assumption
1(ii), note that (\ref{eq:Ex2}) can be written as $u_{t}=u_{t-1}+[1-S(u_{t-1})]\nu+[S(u_{t-1})-1]u_{t-1}+\varepsilon_{t}$
(cf.~eqn (\ref{NLAR(p)_pi})) and choosing $\tilde{g}(\boldsymbol{x})=[1-S(u)]\nu+[S(u)-1]u$
and $g(u)=[1-S(u)]\nu+S(u)u$ it is seen that Assumption 1(ii.a) holds
with $\epsilon(\boldsymbol{x})=0$. For Assumption 1(ii.b), it again
suffices to consider the former part of (\ref{Inequality_Ass 2}).
The assumptions $c_{1}h(u)\leq\left|u\right|^{\rho}$ and $\left|u\right|^{\rho+c_{2}}\leq c_{3}h^{2}(u)$
made of function $h$ imply that, for large enough $\left|u\right|$,
\begin{equation}
\hspace*{-10pt}0\leq1-\frac{r_{0}}{h(u)}\leq1-\frac{c_{1}r_{0}}{\left|u\right|^{\rho}},\quad\frac{1}{h(u)\left|u\right|}\leq\frac{c_{3}^{1/2}}{\left|u\right|^{1-\rho/2+c_{2}/2}\left|u\right|^{\rho}},\quad\text{and}\quad0<\frac{r_{0}^{2}}{2h^{2}(u)}\leq\frac{c_{3}r_{0}^{2}}{2}\frac{1}{\left|u\right|^{\rho+c_{2}}},\hspace*{-7pt}\label{Prop2Proof1}
\end{equation}
where (in the middle inequality) $1-\rho/2+c_{2}/2>0$ as $\rho\leq2$
and $c_{2}>0$. Thus for the $S_{1}(u)$ in (\ref{Functions S}) and
large enough $\left|u\right|$, 
\[
\left|g(u)\right|\leq\frac{r_{0}\left|\nu\right|}{h(u)\left|u\right|}\left|u\right|+\left|1-\frac{r_{0}}{h(u)}\right|\left|u\right|\leq\biggl[1+\biggl(\frac{r_{0}\left|\nu\right|c_{3}^{1/2}}{\left|u\right|^{1-\rho/2+c_{2}/2}}-c_{1}r_{0}\biggr)\frac{1}{\left|u\right|^{\rho}}\biggr]\left|u\right|
\]
where the upper bound is dominated by $\bigl(1-\frac{c_{1}r_{0}/2}{\left|u\right|^{\rho}}\bigr)\left|u\right|$
for sufficiently large $\left|u\right|$ (so that Assumption 1(ii.b)
holds with $r=c_{1}r_{0}/2$). For $S_{2}(u)=\exp\{-r_{0}/h(u)\}$
in (\ref{Functions S}), note that the inequality $1+x\leq e^{x}\leq1+x+x^{2}/2$
(for $x\leq0$) yields $1-S_{2}(u)\leq r_{0}/h(u)$ and $S_{2}(u)\leq1-r_{0}/h(u)+r_{0}^{2}/2h^{2}(u)$
(for all $u$). These inequalities, together with the fact that $0<S_{2}(u)<1$,
imply that 
\[
\left|g(u)\right|\leq\left|1-S_{2}(u)\right|\left|\nu\right|+\left|S_{2}(u)\right|\left|u\right|\leq\frac{r_{0}\left|\nu\right|}{h(u)\left|u\right|}\left|u\right|+\left(1-\frac{r_{0}}{h(u)}+\frac{r_{0}^{2}}{2h^{2}(u)}\right)\left|u\right|.
\]
Using the inequalities in (\ref{Prop2Proof1}) implies that, for large
enough $\left|u\right|$, the above upper bound is dominated by
\[
\biggl[1+\biggl(\frac{r_{0}\left|\nu\right|c_{3}^{1/2}}{\left|u\right|^{1-\rho/2+c_{2}/2}}-c_{1}r_{0}+\frac{c_{3}r_{0}^{2}}{2\left|u\right|^{c_{2}}}\biggr)\frac{1}{\left|u\right|^{\rho}}\biggr]\left|u\right|
\]
which in turn is again dominated by $\bigl(1-\frac{c_{1}r_{0}/2}{\left|u\right|^{\rho}}\bigr)\left|u\right|$
for sufficiently large $\left|u\right|$ (so that Assumption 1(ii.b)
holds with $r=c_{1}r_{0}/2$).

The statements (1)\textendash (3) now follow from Theorems 2(i), 2(ii),
and 3, respectively.

Finally we show that condition (h) in the Proposition is satisfied
for the six choices of $h(u)$ given. For case (i) note that, for
$\left|u\right|>a$,
\begin{align*}
h(u) & =\left(1+\frac{1}{\lvert u-a\rvert^{\rho}}\right)\cdot\left|1-\frac{a}{\left|u\right|}\right|^{\rho}\cdot\lvert u\rvert^{\rho},
\end{align*}
where the product of the first two terms on the right hand side converges
to one as $\left|u\right|\rightarrow\infty$, implying that, for some
$c_{1}\in(0,1)$ and large enough $\left|u\right|$, $h(u)\leq\left|u\right|^{\rho}/c_{1}$
or, equivalently, $c_{1}h(u)\leq\left|u\right|^{\rho}$. Furthermore,
let $c_{2}\in(0,\rho)$ and write $\lvert u\rvert^{2\rho}=\lvert u\rvert^{\rho-c_{2}}\lvert u\rvert^{\rho+c_{2}}$.
Then the preceding discussion implies that, for some $c_{3}\in(0,1)$
and large enough $\left|u\right|$, $h^{2}(u)\geq\left|u\right|^{\rho+c_{2}}/c_{3}$
or, equivalently, $\left|u\right|^{\rho+c_{2}}\leq c_{3}h^{2}(u)$
. Thus, condition (h) holds in case (i).

Regarding cases (ii) and (iii), as $\rho\in(0,2]$, Loève's $c_{r}$-inequality
shows that $(1+\lvert u-a\rvert)^{\rho}\leq2(1+\lvert u-a\rvert^{\rho})$,
and the arguments used in case (i) above yield, for $\left|u\right|$
large enough, $(c_{1}/2)h(u)\leq\left|u\right|^{\rho}$ with $c_{1}$
as in case (i). Similarly in case (iii), $(1+(u_{t-1}-a)^{2})^{\rho/2}\leq1+\lvert u_{t-1}-a\rvert^{\rho}$
so that $c_{1}h(u)\leq\left|u\right|^{\rho}$ holds with $c_{1}$
as in case (i). As for the inequality concerning $h^{2}(u)$, consider
case (ii) and note that, for large enough $\left|u\right|$ and some
$c_{2}\in(0,\rho)$, 
\[
h^{2}(u)=\left(\frac{1+\lvert u-a\rvert}{\left|u\right|}\right)^{2\rho}\cdot\lvert u\rvert^{\rho-c_{2}}\cdot\lvert u\rvert^{\rho+c_{2}},
\]
where the product of the first two terms on the right hand side tends
to infinity as $\left|u\right|\rightarrow\infty$. Thus, as in case
(i) we obtain $\left|u\right|^{\rho+c_{2}}\leq c_{3}h^{2}(u)$, and
similar arguments can be used by replacing the definition of $h$
in the above equality with the one in case (iii).

Now consider case (iv) and let $\rho=\rho_{1}\lor\rho_{2}$. We then
have, for large enough $\left|u\right|$, $h(u)\leq(1+\lvert u-a_{1}\rvert^{\rho})+(1+\lvert u-a_{2}\rvert^{\rho})$,
and arguments used in case (i) above show that we can find $c_{1}\in(0,1)$
such that $c_{1}h(u)\leq\left|u\right|^{\rho}$ holds. Verifying the
desired inequality for $h^{2}(u)$ can be established by using arguments
used in case (ii) above. Cases (v) and (vi) can be handled with arguments
similar to those already used; we omit the details.
\end{proof}
\smallskip{}

\begin{proof}[\textbf{\emph{Checking Assumption 1(ii) for model (\ref{General example_2}).}}]
\textbf{} Suppose that $S(u_{t-1})$ is either one of the two options
in (\ref{Functions S}) and that the function $h$ satisfies condition
(h) in Proposition 2. Clearly, we can write model (\ref{General example_2})
as $u_{t}=u_{t-1}+[S(u_{t-1})-1]u_{t-1}+\exp\{-\gamma\lvert\boldsymbol{y}_{t-1}\rvert^{2}\}(\theta_{1}y_{t-1}+\theta_{2}y_{t-2})+\varepsilon_{t}$.
Choosing $\tilde{g}(\boldsymbol{x})=[S(u)-1]u+\exp\{-\gamma\lvert\boldsymbol{x}\rvert^{2}\}\theta'\boldsymbol{x}$
($\boldsymbol{x}=(x_{1},x_{2})$) and $g(u)=S(u)u$ yields $u+\tilde{g}(\boldsymbol{x})-g(u)=\exp\{-\gamma\lvert\boldsymbol{x}\rvert^{2}\}\theta'\boldsymbol{x}$,
implying that Assumption 1(ii.a) holds \textcolor{black}{with any
}positive $d$ (see inequality (\ref{Implication of Ass 2}) and the
following discussion). As $g(u)=S(u)u$, validity of Assumption 1(ii.b)
can be checked as in the proof of Proposition 2 by setting the intercept
term $\nu$ therein to zero. 
\end{proof}
\bigskip{}

\bibliographystyle{chicago}
\bibliography{SubGE}

\begin{thebibliography}{}

\bibitem[\protect\citeauthoryear{Bec, Rahbek, and Shephard}{Bec
  et~al.}{2008}]{bec2008acr}
Bec, F., A.~Rahbek, and N.~Shephard (2008).
\newblock The {ACR} model: a multivariate dynamic mixture autoregression.
\newblock {\em Oxford Bulletin of Economics and Statistics\/}~{\em 70\/}(5),
  583--618.

\bibitem[\protect\citeauthoryear{Bradley}{Bradley}{2007}]{bradley2007introduction}
Bradley, R.~C. (2007).
\newblock {\em Introduction to Strong Mixing Conditions}, Volume
  {\hspace*{-3pt}}s 1--3.
\newblock Kendrick Press.

\bibitem[\protect\citeauthoryear{Cline and Pu}{Cline and
  Pu}{1998}]{cline1998verifying}
Cline, D. B.~H. and H.~H. Pu (1998).
\newblock Verifying irreducibility and continuity of a nonlinear time series.
\newblock {\em Statistics and Probability Letters\/}~{\em 40\/}(2), 139--148.

\bibitem[\protect\citeauthoryear{Douc, Fort, Moulines, and Soulier}{Douc
  et~al.}{2004}]{douc2004practical}
Douc, R., G.~Fort, E.~Moulines, and P.~Soulier (2004).
\newblock Practical drift conditions for subgeometric rates of convergence.
\newblock {\em Annals of Applied Probability\/}~{\em 14\/}(3), 1353--1377.

\bibitem[\protect\citeauthoryear{Douc, Guillin, and Moulines}{Douc
  et~al.}{2008}]{douc2008bounds}
Douc, R., A.~Guillin, and E.~Moulines (2008).
\newblock Bounds on regeneration times and limit theorems for subgeometric
  {M}arkov chains.
\newblock {\em Annales de l'Institut Henri Poincar{\'e}, Probabilit{\'e}s et
  Statistiques\/}~{\em 44\/}(2), 239--257.

\bibitem[\protect\citeauthoryear{Douc, Moulines, Priouret, and Soulier}{Douc
  et~al.}{2018}]{douc2018markov}
Douc, R., E.~Moulines, P.~Priouret, and P.~Soulier (2018).
\newblock {\em Markov Chains}.
\newblock Springer.

\bibitem[\protect\citeauthoryear{Doukhan}{Doukhan}{1994}]{doukhan1994mixing}
Doukhan, P. (1994).
\newblock {\em Mixing: Properties and Examples}.
\newblock Springer.

\bibitem[\protect\citeauthoryear{Fokianos, Rahbek, and Tj{\o}stheim}{Fokianos
  et~al.}{2009}]{fokianos2009poisson}
Fokianos, K., A.~Rahbek, and D.~Tj{\o}stheim (2009).
\newblock Poisson autoregression.
\newblock {\em Journal of the American Statistical Association\/}~{\em 104},
  1430--1439.

\bibitem[\protect\citeauthoryear{Fort and Moulines}{Fort and
  Moulines}{2003}]{fort2003polynomial}
Fort, G. and E.~Moulines (2003).
\newblock Polynomial ergodicity of {M}arkov transition kernels.
\newblock {\em Stochastic Processes and their Applications\/}~{\em 103\/}(1),
  57--99.

\bibitem[\protect\citeauthoryear{Francq and Zako{\"\i}an}{Francq and
  Zako{\"\i}an}{2006}]{francq2006mixing}
Francq, C. and J.-M. Zako{\"\i}an (2006).
\newblock Mixing properties of a general class of {GARCH}(1,1) models without
  moment assumptions on the observed process.
\newblock {\em Econometric Theory\/}~{\em 22\/}(5), 815--834.

\bibitem[\protect\citeauthoryear{Gouri{\'e}roux and Robert}{Gouri{\'e}roux and
  Robert}{2006}]{gourieroux2006stochastic}
Gouri{\'e}roux, C. and C.~Y. Robert (2006).
\newblock Stochastic unit root models.
\newblock {\em Econometric Theory\/}~{\em 22\/}(6), 1052--1090.

\bibitem[\protect\citeauthoryear{Horn and Johnson}{Horn and
  Johnson}{2013}]{horn2013matrix}
Horn, R.~A. and C.~R. Johnson (2013).
\newblock {\em Matrix Analysis\/} (2nd ed.).
\newblock Cambridge University Press.

\bibitem[\protect\citeauthoryear{Jarner and Roberts}{Jarner and
  Roberts}{2002}]{jarner2002polynomial}
Jarner, S.~F. and G.~O. Roberts (2002).
\newblock Polynomial convergence rates of {M}arkov chains.
\newblock {\em Annals of Applied Probability\/}~{\em 12\/}(1), 224--247.

\bibitem[\protect\citeauthoryear{Jarner and Tweedie}{Jarner and
  Tweedie}{2003}]{jarner2003necessary}
Jarner, S.~F. and R.~L. Tweedie (2003).
\newblock Necessary conditions for geometric and polynomial ergodicity of
  random-walk-type {M}arkov chains.
\newblock {\em Bernoulli\/}~{\em 9\/}(4), 559--578.

\bibitem[\protect\citeauthoryear{Klokov}{Klokov}{2007}]{klokov2007lower}
Klokov, S.~A. (2007).
\newblock Lower bounds of mixing rate for a class of {M}arkov processes.
\newblock {\em Theory of Probability \& Its Applications\/}~{\em 51\/}(3),
  528--535.

\bibitem[\protect\citeauthoryear{Klokov and Veretennikov}{Klokov and
  Veretennikov}{2004}]{klokov2004sub}
Klokov, S.~A. and A.~Y. Veretennikov (2004).
\newblock Sub-exponential mixing rate for a class of {M}arkov chains.
\newblock {\em Mathematical Communications\/}~{\em 9}, 9--26.

\bibitem[\protect\citeauthoryear{Klokov and Veretennikov}{Klokov and
  Veretennikov}{2005}]{klokov2005subexponential}
Klokov, S.~A. and A.~Y. Veretennikov (2005).
\newblock On subexponential mixing rate for {M}arkov processes.
\newblock {\em Theory of Probability \& Its Applications\/}~{\em 49\/}(1),
  110--122.

\bibitem[\protect\citeauthoryear{Ling}{Ling}{2007}]{ling2007double}
Ling, S. (2007).
\newblock A double {AR}($p$) model:\ structure and estimation.
\newblock {\em Statistica Sinica\/}~{\em 17\/}(1), 161.

\bibitem[\protect\citeauthoryear{Lu}{Lu}{1998}]{lu1998geometric}
Lu, Z. (1998).
\newblock On the geometric ergodicity of a non-linear autoregressive model with
  an autoregressive conditional heteroscedastic term.
\newblock {\em Statistica Sinica\/}~{\em 8}, 1205--1217.

\bibitem[\protect\citeauthoryear{Meitz and Saikkonen}{Meitz and
  Saikkonen}{2008}]{meitz2008ergodicity}
Meitz, M. and P.~Saikkonen (2008).
\newblock Ergodicity, mixing, and existence of moments of a class of {M}arkov
  models with applications to {GARCH} and {ACD} models.
\newblock {\em Econometric Theory\/}~{\em 24\/}(5), 1291--1320.

\bibitem[\protect\citeauthoryear{Meitz and Saikkonen}{Meitz and
  Saikkonen}{2019}]{meitz2019subgemix}
Meitz, M. and P.~Saikkonen (2019).
\newblock Subgeometric ergodicity and $\beta$-mixing.
\newblock Available as arXiv:1904.07103.

\bibitem[\protect\citeauthoryear{Meyn and Tweedie}{Meyn and
  Tweedie}{1993}]{meyn1993markov}
Meyn, S.~P. and R.~L. Tweedie (1993).
\newblock {\em Markov Chains and Stochastic Stability}.
\newblock Springer.

\bibitem[\protect\citeauthoryear{Meyn and Tweedie}{Meyn and
  Tweedie}{2009}]{meyn2009markov}
Meyn, S.~P. and R.~L. Tweedie (2009).
\newblock {\em Markov Chains and Stochastic Stability\/} (2nd ed.).
\newblock Cambridge University Press.

\bibitem[\protect\citeauthoryear{Nummelin and Tuominen}{Nummelin and
  Tuominen}{1983}]{nummelin1983rate}
Nummelin, E. and P.~Tuominen (1983).
\newblock The rate of convergence in {O}rey's theorem for {H}arris recurrent
  {M}arkov chains with applications to renewal theory.
\newblock {\em Stochastic Processes and their Applications\/}~{\em 15\/}(3),
  295--311.

\bibitem[\protect\citeauthoryear{Tanikawa}{Tanikawa}{2001}]{tanikawa2001markov}
Tanikawa, A. (2001).
\newblock Markov chains satisfying simple drift conditions for subgeometric
  ergodicity.
\newblock {\em Stochastic Models\/}~{\em 17\/}(2), 109--120.

\bibitem[\protect\citeauthoryear{Tuominen and Tweedie}{Tuominen and
  Tweedie}{1994}]{tuominen1994subgeometric}
Tuominen, P. and R.~L. Tweedie (1994).
\newblock Subgeometric rates of convergence of $f$-ergodic {M}arkov chains.
\newblock {\em Advances in Applied Probability\/}~{\em 26\/}(3), 775--798.

\bibitem[\protect\citeauthoryear{van Dijk, Ter{\"a}svirta, and Franses}{van
  Dijk et~al.}{2002}]{vandijk2002smooth}
van Dijk, D., T.~Ter{\"a}svirta, and P.~H. Franses (2002).
\newblock Smooth transition autoregressive models --- a survey of recent
  developments.
\newblock {\em Econometric Reviews\/}~{\em 21\/}(1), 1--47.

\bibitem[\protect\citeauthoryear{Veretennikov}{Veretennikov}{2000}]{veretennikov2000polynomial}
Veretennikov, A.~Y. (2000).
\newblock On polynomial mixing and convergence rate for stochastic difference
  and differential equations.
\newblock {\em Theory of Probability \& Its Applications\/}~{\em 44\/}(2),
  361--374.

\end{thebibliography}

\pagebreak{}

\section*{Supplementary Appendix to \textquoteleft Subgeometrically ergodic
autoregressions\textquoteright{} by Meitz and Saikkonen}

\bigskip{}

\begin{proof}[\textbf{\emph{Proof of Theorem 1}}]
 In the geometric case, the result of Theorem 1 is given in \citet[Thm 15.0.1]{meyn2009markov}.
In the polynomial case, the result can be obtained by combining Theorem
2.8 of \citet{douc2004practical} with the discussion in their Section
2.3 (see also \citet{jarner2002polynomial}). In the subexponential
case, the function $\phi$ is concave and increasing as long as $v_{0}$
is chosen large enough (cf.~\citet[p. 243, the paragraph following Assumption 2]{douc2008bounds}).
Again, the result can be obtained by combining Theorem 2.8 of \citet{douc2004practical}
with the discussion in their Section 2.3; note also that the two functions
$\phi(v)=c(v+v_{0})/[\ln(v+v_{0})]^{\alpha}$ and $\phi_{0}(v)=cv/[\ln(v)]^{\alpha}$
both lead to the same rate function $r_{\phi}(n)$ given in \citet[p. 1365, line 6]{douc2004practical}.
\end{proof}
\bigskip{}

\noindent \textbf{Proof of Theorem 2}. First note that from equation
(\ref{Companion form}), Assumptions 1 and 2(a), and Theorem 2.2(ii)
of \citet{cline1998verifying} (see also Example 2.1 of that paper)
it follows that the Markov chain $\boldsymbol{y}_{t}$ is a $\psi$-irreducible
and aperiodic $T$-chain. Moreover, as in the proof of Lemma 1 of
\citet{lu1998geometric} it can be seen that $\psi$ is the Lebesque
measure and using Theorem 6.2.5 of \citet{meyn2009markov} we can
conclude that all compact sets of $\mathcal{B}(\mathbb{R}^{p})$ are
petite (and in this case small, as shown by Theorem 5.5.7 of Meyn
and Tweedie (2009)). The same also holds for the Markov chain $y_{t}$
in the case $p=1$.

In what follows, we first consider the case $p\geq2$ and consider
the case $p=1$ at the end of the proof.

\bigskip{}

\noindent \textbf{Part (i)}: In this case we have \textbf{$\rho>\kappa_{0}$}
and $b_{3}=\kappa_{0}\land(2-\rho)\in(0,1)$; for brevity, the notation
$b_{3}$ will be used. The choice of $b_{1}$ and $b_{2}$ will be
discussed later. We can make use of results in the proof of Theorem
3.3, part (i), in Douc et al. (2004, Sec. 3.3). Write the function
$V(\boldsymbol{x})$ as
\begin{equation}
V(\boldsymbol{x})=\tfrac{1}{2}\exp\bigl\{ b_{1}\left|z_{1}(\boldsymbol{x})\right|^{b_{3}}\bigr\}+\tfrac{1}{2}\exp\bigl\{ b_{2}\left\Vert \boldsymbol{z}_{2}(\boldsymbol{x})\right\Vert _{*}^{b_{3}}\bigr\}\overset{def}{=}V_{1}(z_{1}(\boldsymbol{x}))+V_{2}(\boldsymbol{z}_{2}(\boldsymbol{x})),\label{Drift_1}
\end{equation}
and consider $E\left[V(\boldsymbol{y}_{1})\mid\boldsymbol{y}_{0}=\boldsymbol{x}\right]$,
the conditional expectation in (\ref{Drift condition}). Note that
$\boldsymbol{z}(\boldsymbol{y}_{1})$ appearing in $V(\boldsymbol{y}_{1})$
can be expressed as (see (\ref{Companion form_A2}))
\[
\boldsymbol{z}(\boldsymbol{y}_{1})=\begin{bmatrix}z_{1}(\boldsymbol{y}_{1})\\
\boldsymbol{z}_{2}(\boldsymbol{y}_{1})
\end{bmatrix}=\boldsymbol{A}\boldsymbol{y}_{1}=\begin{bmatrix}0 & \text{\textbf{0}}'_{p-1}\\
\boldsymbol{\iota}_{p-1} & \boldsymbol{\Pi}_{1}
\end{bmatrix}\begin{bmatrix}z_{1}(\boldsymbol{y}_{0})\\
\boldsymbol{z}_{2}(\boldsymbol{y}_{0})
\end{bmatrix}+\overline{g}(\boldsymbol{y}_{0})\boldsymbol{\iota}_{p}+\varepsilon_{1}\boldsymbol{\iota}_{p}=\begin{bmatrix}\overline{g}(\boldsymbol{y}_{0})+\varepsilon_{1}\\
\boldsymbol{\Pi}_{1}\boldsymbol{z}_{2}(\boldsymbol{y}_{0})+z_{1}(\boldsymbol{y}_{0})\boldsymbol{\iota}_{p-1}
\end{bmatrix}.
\]
In what follows, we usually drop the argument from $\boldsymbol{z}(\boldsymbol{x})$
and its components and write, for example, $\boldsymbol{z}_{2}$ instead
of $\boldsymbol{z}_{2}(\boldsymbol{x})$. Now (dropping the argument
from $\boldsymbol{z}(\boldsymbol{x})$) 
\begin{align*}
E\left[V(\boldsymbol{y}_{1})\mid\boldsymbol{y}_{0}=\boldsymbol{x}\right] & =E\bigl[\tfrac{1}{2}\exp\bigl\{ b_{1}\left|\overline{g}(\boldsymbol{x})+\varepsilon_{1}\right|^{b_{3}}\bigr\}\bigr]+\tfrac{1}{2}\exp\bigl\{ b_{2}\left\Vert \boldsymbol{\Pi}_{1}\boldsymbol{z}_{2}+z_{1}\boldsymbol{\iota}_{p-1}\right\Vert _{*}^{b_{3}}\bigr\}\\
 & =E[V_{1}(\overline{g}(\boldsymbol{x})+\varepsilon_{1})]+V_{2}(\boldsymbol{\Pi}_{1}\boldsymbol{z}_{2}+z_{1}\boldsymbol{\iota}_{p-1}).
\end{align*}
Defining $V_{\epsilon}(\boldsymbol{x})=\exp\{b_{1}\left|\epsilon(\boldsymbol{x})\boldsymbol{x}\right|^{b_{3}}\}$
we bound the expectation on the right hand side as follows:
\begin{align*}
E\left[V_{1}\left(\overline{g}(\boldsymbol{x})+\varepsilon_{1}\right)\right] & =E\bigl[\tfrac{1}{2}\exp\bigl\{ b_{1}\left|\overline{g}(\boldsymbol{x})-g(z_{1})+g(z_{1})+\varepsilon_{1}\right|^{b_{3}}\bigr\}\bigr]\\
 & \leq\exp\bigl\{ b_{1}\left|\overline{g}(\boldsymbol{x})-g(z_{1})\right|^{b_{3}}\bigr\} E\bigl[\tfrac{1}{2}\exp\bigl\{ b_{1}\left|g(z_{1})+\varepsilon_{1}\right|^{b_{3}}\bigr\}\bigr]\\
 & \leq V_{\epsilon}(\boldsymbol{x})E\left[V_{1}\left(g(z_{1})+\varepsilon_{1}\right)\right],
\end{align*}
where the first inequality is due to the triangle inequality and the
fact that $b_{3}\in(0,1)$, and the second inequality follows from
the definition of the function $\overline{g}$ and inequality (\ref{Inequality Ass 2a})
in Assumption 1(ii). Thus, we can bound the conditional expectation
$E\left[V(\boldsymbol{y}_{1})\mid\boldsymbol{y}_{0}=\boldsymbol{x}\right]$
as
\begin{equation}
E\left[V(\boldsymbol{y}_{1})\mid\boldsymbol{y}_{0}=\boldsymbol{x}\right]\leq V_{\epsilon}(\boldsymbol{x})E\left[V_{1}\left(g(z_{1})+\varepsilon_{1}\right)\right]+V_{2}\left(\boldsymbol{\Pi}_{1}\boldsymbol{z}_{2}+z_{1}\boldsymbol{\iota}_{p-1}\right).\label{Drift_3}
\end{equation}

\noindent \textbf{Step 1: Bounding $E\left[V_{1}\left(g(z_{1})+\varepsilon_{1}\right)\right]$
in (\ref{Drift_3})}. We first note that the arguments used by \citet{douc2004practical}
to obtain their inequality (3.14) can be used to justify that, for
$\left|z_{1}\right|\geq M_{0}$, 
\[
V_{1}\left(g(z_{1})\right)-V_{1}\left(z_{1}\right)\leq(-b_{1}rb_{3}\left|z_{1}\right|^{b_{3}-\rho}+\tfrac{1}{2}b_{1}^{2}r^{2}b_{3}^{2}\left|z_{1}\right|^{2(b_{3}-\rho)})V_{1}(z_{1}).
\]
Moreover, repeating the arguments in \citet{douc2004practical} between
their (3.15)\textendash (3.19) it can be shown that, for $\left|z_{1}\right|$
large (which, due to our Assumption \ref{assu:dynamics}(ii), also
implies that $\left|g(z_{1})\right|$ is large) and some $c>0$, 
\[
E\left[V_{1}\left(g(z_{1})+\varepsilon_{1}\right)\right]-V_{1}\left(g(z_{1})\right)\leq\{\tfrac{1}{2}b_{1}^{2}b_{3}^{2}+c\left|z_{1}\right|^{-b_{3}}\}E[\varepsilon_{1}^{2}V_{1}(\varepsilon_{1})]\left|z_{1}\right|^{2b_{3}-2}V_{1}(z_{1}),
\]
so that, for $\left|z_{1}\right|$ large,
\[
E\left[V_{1}\left(g(z_{1})+\varepsilon_{1}\right)\right]-V_{1}\left(g(z_{1})\right)\leq b_{1}^{2}b_{3}^{2}E[\varepsilon_{1}^{2}V_{1}(\varepsilon_{1})]\left|z_{1}\right|^{2b_{3}-2}V_{1}(z_{1});
\]
note that due to Assumption \ref{assu:errors}(a) and the choice of
$b_{3}$, the condition $E\bigl[\left|\varepsilon_{1}\right|^{2}V_{1}(\varepsilon_{1})\bigr]<\infty$
\textcolor{black}{can be achieved by choosing the value of $b_{1}$
small enough.} From the above inequalities it follows that, for $\left|z_{1}\right|$
large, 
\begin{align}
E\left[V_{1}\left(g(z_{1})+\varepsilon_{1}\right)\right] & \leq V_{1}(z_{1})+k(z_{1})V_{1}(z_{1}),\label{eq:Th1(iii)Pr_E(V1)}
\end{align}
where
\begin{equation}
k(z_{1})=-b_{1}rb_{3}\left|z_{1}\right|^{b_{3}-\rho}+\tfrac{1}{2}b_{1}^{2}r^{2}b_{3}^{2}\left|z_{1}\right|^{2(b_{3}-\rho)}+b_{1}^{2}b_{3}^{2}E[\varepsilon_{1}^{2}V_{1}(\varepsilon_{1})]\left|z_{1}\right|^{2b_{3}-2}.\label{eq:Th1(iii)Pr_k(z1)}
\end{equation}

Next we obtain an upper bound for $k(z_{1})$. Note that we necessarily
have $b_{3}-\rho<0$ and $2b_{3}-2\leq b_{3}-\rho$ with equality
if and only if $b_{3}=2-\rho$ (these follow from $\rho>\kappa_{0}$
and $b_{3}=\kappa_{0}\land(2-\rho)$). First consider the case $2b_{3}-2=b_{3}-\rho$
so that $b_{3}=2-\rho$ and
\[
k(z_{1})=-\left(r-b_{1}b_{3}E[\varepsilon_{1}^{2}V_{1}(\varepsilon_{1})]\right)b_{1}b_{3}\left|z_{1}\right|^{b_{3}-\rho}+\tfrac{1}{2}b_{1}^{2}r^{2}b_{3}^{2}\left|z_{1}\right|^{2(b_{3}-\rho)}.
\]
As $b_{3}-\rho<0$, the inequality $\frac{1}{2}b_{1}r^{2}b_{3}\left|z_{1}\right|^{b_{3}-\rho}\leq\frac{1}{2}\epsilon_{1}$
holds for all large enough $\left|z_{1}\right|$ and with $\epsilon_{1}>0$
which can be chosen as close to zero as desired. Moreover, as $b_{3}$
and $E\left[\varepsilon_{1}^{2}V_{1}(\varepsilon_{1})\right]$ are
positive and here fixed, we can \textcolor{black}{choose the value
of $b_{1}$ small enough so that $b_{1}b_{3}E[\varepsilon_{1}^{2}V_{1}(\varepsilon_{1})]\leq\frac{1}{2}\epsilon_{1}$
holds}. Hence, for $\left|z_{1}\right|$ large, $k(z_{1})\leq-\left[r-\tfrac{1}{2}\epsilon_{1}-\tfrac{1}{2}\epsilon_{1}\right]b_{1}b_{3}\left|z_{1}\right|^{b_{3}-\rho}$
and here $\epsilon_{1}$ can be chosen small enough so that $r-\epsilon_{1}>0$
holds. Now consider the case $2b_{3}-2<b_{3}-\rho$ (so that $b_{3}<2-\rho$).
Write $k(z_{1})$ as 
\[
k(z_{1})=-(r-\tfrac{1}{2}b_{1}r^{2}b_{3}\left|z_{1}\right|^{b_{3}-\rho}-b_{1}b_{3}E[\varepsilon_{1}^{2}V_{1}(\varepsilon_{1})]\left|z_{1}\right|^{(2b_{3}-2)-(b_{3}-\rho)})b_{1}b_{3}\left|z_{1}\right|^{b_{3}-\rho}
\]
and note that $\tfrac{1}{2}b_{1}r^{2}b_{3}\left|z_{1}\right|^{b_{3}-\rho}+b_{1}b_{3}E[\varepsilon_{1}^{2}V_{1}(\varepsilon_{1})]\left|z_{1}\right|^{(2b_{3}-2)-(b_{3}-\rho)}\leq\epsilon_{2}$
holds with $0<\epsilon_{2}<r$ for all large enough $\left|z_{1}\right|$
so that the bound $k(z_{1})\leq-(r-\epsilon_{2})b_{1}b_{3}\left|z_{1}\right|^{b_{3}-\rho}$
is obtained. To combine the two cases, note that the arguments above
hold if $\epsilon_{1}$ and $\epsilon_{2}$ are replaced with $\epsilon_{3}=\epsilon_{1}\land\epsilon_{2}$.
Thus, defining the positive constant $\omega_{1}$ as $\omega_{1}=r-\epsilon_{3}$
we obtain, for $\left|z_{1}\right|$ large,
\[
k(z_{1})\leq-\omega_{1}b_{1}b_{3}\left|z_{1}\right|^{b_{3}-\rho}
\]
(cf.~\citet[top of p. 1373]{douc2004practical}). Combining this
with the inequality (\ref{eq:Th1(iii)Pr_E(V1)}) we obtain 
\[
E\left[V_{1}\left(g(z_{1})+\varepsilon_{1}\right)\right]\leq(1-\omega_{1}b_{1}b_{3}\left|z_{1}\right|^{b_{3}-\rho})V_{1}(z_{1}).
\]

\noindent \textbf{Step 2: Bounding $V_{\epsilon}(\boldsymbol{x})E\left[V_{1}\left(g(z_{1})+\varepsilon_{1}\right)\right]$
in (\ref{Drift_3})}. Using the bound just obtained, bound the first
term on the right hand side of (\ref{Drift_3}) as
\begin{align*}
V_{\epsilon}(\boldsymbol{x})E\left[V_{1}\left(g(z_{1})+\varepsilon_{1}\right)\right] & \leq\exp\bigl\{ b_{1}\left|\epsilon(\boldsymbol{x})\boldsymbol{x}\right|^{b_{3}}\bigr\}(1-\omega_{1}b_{1}b_{3}\left|z_{1}\right|^{b_{3}-\rho})V_{1}(z_{1})\\
 & =\tfrac{1}{2}(1-\omega_{1}b_{1}b_{3}\left|z_{1}\right|^{b_{3}-\rho})\exp\bigl\{ b_{1}\left|z_{1}\right|^{b_{3}}+b_{1}\left|\epsilon(\boldsymbol{x})\boldsymbol{x}\right|^{b_{3}}\bigr\}.
\end{align*}
For all $\left|z_{1}\right|$ large enough, $1-\omega_{1}b_{1}b_{3}\left|z_{1}\right|^{b_{3}-\rho}\in(0,1)$
and the same holds true for $k_{1}(z_{1})\,\overset{def}{=}\,1-\tfrac{1}{2}\omega_{1}b_{1}b_{3}\left|z_{1}\right|^{b_{3}-\rho}$.
Using the inequality $(1-u)^{\alpha}\leq1-\alpha u$ ($0\leq u,\alpha\leq1$)
we thus have
\[
1-\omega_{1}b_{1}b_{3}\left|z_{1}\right|^{b_{3}-\rho}=\bigl(1-\omega_{1}b_{1}b_{3}\left|z_{1}\right|^{b_{3}-\rho}\bigr)^{1/2}\bigl(1-\omega_{1}b_{1}b_{3}\left|z_{1}\right|^{b_{3}-\rho}\bigr)^{1/2}\leq k_{1}(z_{1})^{2}.
\]
Furthermore, as $\ln\left(k_{1}(z_{1})\right)=\ln(1-\tfrac{1}{2}\omega_{1}b_{1}b_{3}\left|z_{1}\right|^{b_{3}-\rho})\leq-\tfrac{1}{2}\omega_{1}b_{1}b_{3}\left|z_{1}\right|^{b_{3}-\rho}$
it follows that $k_{1}(z_{1})=\exp\left\{ \ln\left(k_{1}(z_{1})\right)\right\} \leq\exp\bigl\{-\tfrac{1}{2}\omega_{1}b_{1}b_{3}\left|z_{1}\right|^{b_{3}-\rho}\bigr\}$
and we can write
\[
V_{\epsilon}(\boldsymbol{x})E\left[V_{1}\left(g(z_{1})+\varepsilon_{1}\right)\right]\leq\tfrac{1}{2}k_{1}(z_{1})\exp\{\left(1-\tfrac{1}{2}\omega_{1}b_{3}\left|z_{1}\right|^{-\rho}\right)b_{1}\left|z_{1}\right|^{b_{3}}+b_{1}\left|\epsilon(\boldsymbol{x})\boldsymbol{x}\right|^{b_{3}}\}.
\]

Consider the argument of the exponential function on the right hand
side of the above inequality. As $\boldsymbol{z}=(z_{1},\boldsymbol{z}_{2})=\boldsymbol{A}\boldsymbol{x}$,
the equivalence of vector norms in $\mathbb{R}^{p}$ and straightforward
calculations show that, for some $c_{*}>0$, 
\[
\left|\epsilon(\boldsymbol{x})\boldsymbol{x}\right|=\left|\epsilon(\boldsymbol{x})\boldsymbol{A}^{-1}\boldsymbol{z}\right|\leq c_{*}\left|\epsilon(\boldsymbol{x})\right|\left|z_{1}\right|+c_{*}\left|\epsilon(\boldsymbol{x})\right|\left\Vert \boldsymbol{z}_{2}\right\Vert _{*}=\left|\epsilon_{1}(\boldsymbol{x})\right|\left|z_{1}\right|+\left|\epsilon_{1}(\boldsymbol{x})\right|\left\Vert \boldsymbol{z}_{2}\right\Vert _{*},
\]
where $\epsilon_{1}(\boldsymbol{x})=c_{*}\epsilon(\boldsymbol{x})$.
Hence, as Assumption 1(ii) holds with $d=\rho/b_{3}$, we have $\left|\epsilon_{1}(\boldsymbol{x})\right|=o(\left|\boldsymbol{x}\right|^{-\rho/b_{3}})$
and
\[
\left|\epsilon(\boldsymbol{x})\boldsymbol{x}\right|^{b_{3}}\leq o(\left|\boldsymbol{x}\right|^{-\rho})\left|z_{1}\right|^{b_{3}}+o(\left|\boldsymbol{x}\right|^{-\rho})\left\Vert \boldsymbol{z}_{2}\right\Vert _{*}^{b_{3}},
\]
so that, for all $\left|z_{1}\right|$ large (implying that $\left|\boldsymbol{z}\right|$
and hence that $\left|\boldsymbol{x}\right|$ is large\footnote{\label{fn:x_big_iff_z_big}Due to the nonsingularity of the matrix
$\boldsymbol{A}$, there exists a positive constant $c$ such that
$c^{-1}\left|\boldsymbol{z}\right|\leq\left|\boldsymbol{x}\right|\leq c\left|\boldsymbol{z}\right|$,
so that $\left|\boldsymbol{x}\right|\rightarrow\infty$ if and only
if $\left|\boldsymbol{z}\right|\rightarrow\infty$.}; see the discussion above Theorem 1),
\begin{align*}
\left(1-\tfrac{1}{2}\omega_{1}b_{3}\left|z_{1}\right|^{-\rho}\right)b_{1}\left|z_{1}\right|^{b_{3}}+b_{1}\left|\epsilon(\boldsymbol{x})\boldsymbol{x}\right|^{b_{3}} & \leq\left(1-\tfrac{1}{2}\omega_{1}b_{3}\left|z_{1}\right|^{-\rho}+o(\left|\boldsymbol{x}\right|^{-\rho})\right)b_{1}\left|z_{1}\right|^{b_{3}}+o(\left|\boldsymbol{x}\right|^{-\rho})b_{1}\left\Vert \boldsymbol{z}_{2}\right\Vert _{*}^{b_{3}}\\
 & \leq\left(1-\omega_{2}b_{3}\left|z_{1}\right|^{-\rho}\right)b_{1}\left|z_{1}\right|^{b_{3}}+o(\left|\boldsymbol{x}\right|^{-\rho})b_{1}\left\Vert \boldsymbol{z}_{2}\right\Vert _{*}^{b_{3}},
\end{align*}
where $0<\omega_{2}<\tfrac{1}{2}\omega_{1}$. Thus, we can conclude
that, for all $\left|z_{1}\right|$ large,
\[
V_{\epsilon}\left(\boldsymbol{x}\right)E\left[V_{1}\left(g(z_{1})+\varepsilon_{1}\right)\right]\leq\tfrac{1}{2}k_{1}(z_{1})\exp\bigl\{(1-\omega_{2}b_{3}\left|z_{1}\right|^{-\rho})b_{1}\left|z_{1}\right|^{b_{3}}+o(\left|\boldsymbol{x}\right|^{-\rho})b_{1}\left\Vert \boldsymbol{z}_{2}\right\Vert _{*}^{b_{3}}\bigr\}.
\]

Next define $\tau_{1}(z_{1})=1-\omega_{2}b_{3}\left|z_{1}\right|^{-\rho}$
and $\tau_{2}(z_{1})=1-\tau_{1}(z_{1})$, and note that $\tau_{1}(z_{1})\in(0,1)$
for any $\left|z_{1}\right|$ large. By the preceding discussion,
we then have, for all $\left|z_{1}\right|$ large,
\begin{align*}
V_{\epsilon}(\boldsymbol{x})E\left[V_{1}\left(g(z_{1})+\varepsilon_{1}\right)\right] & \leq\tfrac{1}{2}k_{1}(z_{1})\exp\bigl\{\tau_{1}(z_{1})b_{1}\left|z_{1}\right|^{b_{3}}+\tau_{2}(z_{1})\tau_{2}(z_{1})^{-1}o(\left|\boldsymbol{x}\right|^{-\rho})b_{1}\left\Vert \boldsymbol{z}_{2}\right\Vert _{*}^{b_{3}}\bigr\}\\
 & \leq\frac{\tau_{1}(z_{1})}{2}k_{1}(z_{1})\exp\bigl\{ b_{1}\left|z_{1}\right|^{b_{3}}\bigr\}+\frac{\tau_{2}(z_{1})}{2}k_{1}(z_{1})\exp\bigl\{\tau_{2}(z_{1})^{-1}o(\left|\boldsymbol{x}\right|^{-\rho})b_{1}\left\Vert \boldsymbol{z}_{2}\right\Vert _{*}^{b_{3}}\bigr\}\\
 & \leq\tfrac{1}{2}k_{1}(z_{1})\exp\bigl\{ b_{1}\left|z_{1}\right|^{b_{3}}\bigr\}+\tfrac{1}{2}\exp\bigl\{\tau_{2}(z_{1})^{-1}o(\left|\boldsymbol{x}\right|^{-\rho})\left\Vert \boldsymbol{z}_{2}\right\Vert _{*}^{b_{3}}\bigr\}\\
 & =k_{1}(z_{1})V_{1}(z_{1})+\tfrac{1}{2}\exp\bigl\{ o(1)\left\Vert \boldsymbol{z}_{2}\right\Vert _{*}^{b_{3}}\bigr\}.
\end{align*}
Here the second inequality is justified by the convexity of the exponential
function and the third one follows because $\tau_{1}(z_{1})\in(0,1)$
and $k_{1}(z_{1})\in(0,1)$ can be assumed. The last equality is due
to the definition of $V_{1}$ and the definition of $\tau_{2}(z_{1})$
which implies 
\[
\tau_{2}(z_{1})^{-1}o(\left|\boldsymbol{x}\right|^{-\rho})=\left(\omega_{2}b_{3}\right)^{-1}\left|z_{1}\right|^{\rho}o(\left|\boldsymbol{x}\right|^{-\rho})\leq c^{\rho}\left(\omega_{2}b_{3}\right)^{-1}\left|\boldsymbol{x}\right|^{\rho}o(\left|\boldsymbol{x}\right|^{-\rho})=o(1),
\]
where the inequality holds because $\left|z_{1}\right|\leq\left|\boldsymbol{z}\right|\leq c\left|\boldsymbol{x}\right|$
(see footnote \ref{fn:x_big_iff_z_big}) and where $o(1)\rightarrow0$
as $\left|\boldsymbol{x}\right|\rightarrow\infty$.

It will be convenient to modify the preceding upper bound of $V_{\epsilon}(\boldsymbol{x})E\left[V_{1}\left(g(z_{1})+\varepsilon_{1}\right)\right]$.
To this end, denote $\alpha=\rho/b_{3}-1$ ($>0$) and write $b_{1}\left|z_{1}\right|^{b_{3}-\rho}=b_{1}^{\rho/b_{3}}\bigl(b_{1}\left|z_{1}\right|^{b_{3}}\bigr)^{-\alpha}\geq b_{1}^{\rho/b_{3}}\left(1+\ln V_{1}(z_{1})\right)^{-\alpha}$
where the inequality is based on the definition of $V_{1}(z_{1})$
(also note that $\ln(\frac{1}{2})\approx-0.6931$). Thus, by the definition
of $k_{1}(z_{1})$ we have,
\[
k_{1}(z_{1})\leq1-\tfrac{1}{2}\omega_{1}b_{3}b_{1}^{\rho/b_{3}}\left(1+\ln V_{1}(z_{1})\right)^{-\alpha}.
\]
Using this upper bound and the definition
\begin{equation}
\phi_{1}\left(V_{1}(z_{1})\right)=\tfrac{1}{2}\omega_{1}b_{3}b_{1}^{\rho/b_{3}}\left(1+\ln V_{1}(z_{1})\right)^{-\alpha}V_{1}(z_{1})\quad(>0),\label{eq:Th1(iii)Pr_phi2}
\end{equation}
yields, for $\left|z_{1}\right|$ large \textcolor{black}{and for
a small enough choice of $b_{1}$}, the following bound for the first
term on the right hand side of (\ref{Drift_3}):
\[
V_{\epsilon}(\boldsymbol{x})E\left[V_{1}\left(g(z_{1})+\varepsilon_{1}\right)\right]\leq V_{1}(z_{1})-\phi_{1}\left(V_{1}(z_{1})\right)+\tfrac{1}{2}\exp\bigl\{ o(1)\left\Vert \boldsymbol{z}_{2}\right\Vert _{*}^{b_{3}}\bigr\}.
\]
To state this more formally,\textcolor{green}{{} }\textcolor{black}{we
can find $b_{1}=\tilde{b}_{1}<\beta_{0}$, }and $M_{1}\geq M_{0}$
such that the above inequality holds for $\left|z_{1}\right|>M_{1}$.
Moreover, as in Douc et al. (2004, p. 1373) these choices can be done
in such a way that, for some (finite) constant $\overline{M}_{1}$,
and for all $z_{1}$,
\begin{equation}
V_{\epsilon}(\boldsymbol{x})E\left[V_{1}\left(g(z_{1})+\varepsilon_{1}\right)\right]\leq V_{1}(z_{1})-\phi_{1}\left(V_{1}(z_{1})\right)+\tfrac{1}{2}\exp\bigl\{ o(1)\left\Vert \boldsymbol{z}_{2}\right\Vert _{*}^{b_{3}}\bigr\}+\overline{M}_{1}\boldsymbol{1}_{C_{1}}(z_{1}),\label{eq:Th1(i)PR_E(V1)_2}
\end{equation}
where $C_{1}=\left\{ z_{1}\in\mathbb{R}\,:\,\left|z_{1}\right|\leq M_{1}\right\} $.\bigskip{}

\noindent \textbf{Step 3: Bounding $V_{2}\left(\boldsymbol{\Pi}_{1}\boldsymbol{z}_{2}+z_{1}\boldsymbol{\iota}_{p-1}\right)$
in (\ref{Drift_3})}. Here we assume that the choice of $b_{1}$ is
fixed to the value $\tilde{b}_{1}$ specified above. Recall that $V_{2}(\boldsymbol{\Pi}_{1}\boldsymbol{z}_{2}+z_{1}\boldsymbol{\iota}_{p-1})=\tfrac{1}{2}\exp\bigl\{ b_{2}\left\Vert \boldsymbol{\Pi}_{1}\boldsymbol{z}_{2}+z_{1}\boldsymbol{\iota}_{p-1}\right\Vert _{*}^{b_{3}}\bigr\}$
and note that 
\[
b_{2}\left\Vert \boldsymbol{\Pi}_{1}\boldsymbol{z}_{2}+z_{1}\boldsymbol{\iota}_{p-1}\right\Vert _{*}^{b_{3}}\leq b_{2}\left(\left\Vert \boldsymbol{\Pi}_{1}\boldsymbol{z}_{2}\right\Vert _{*}+\left\Vert z_{1}\boldsymbol{\iota}_{p-1}\right\Vert _{*}\right)^{b_{3}}\leq b_{2}\eta^{b_{3}}\left\Vert \boldsymbol{z}_{2}\right\Vert _{*}^{b_{3}}+b_{2}\left\Vert \boldsymbol{\iota}_{p-1}\right\Vert _{*}^{b_{3}}\left|z_{1}\right|^{b_{3}},
\]
where we have made use of the fact $b_{3}\in(0,1)$ and Assumption
1(i) which implies that $\left\Vert \boldsymbol{\Pi}_{1}\right\Vert _{*}\leq\eta$
for some $\eta<1$ (see the discussion following equation (\ref{Companion form_A2})). 

Let $\tau_{1}\in(0,1)$ and $\tau_{2}=1-\tau_{1}$ be such that $\tau_{2}\in(\eta^{b_{3}},1)$,
and denote $b_{2,1}=b_{2}\left\Vert \boldsymbol{\iota}_{p-1}\right\Vert _{*}^{b_{3}}/\tau_{1}$
and $b_{2,2}=b_{2}/\tau_{2}$. Then, 
\begin{align*}
V_{2}\bigl(\boldsymbol{\Pi}_{1}\boldsymbol{z}_{2}+z_{1}\boldsymbol{\iota}_{p-1}\bigr) & \leq\tfrac{1}{2}\exp\bigl\{ b_{2}\eta^{b_{3}}\left\Vert \boldsymbol{z}_{2}\right\Vert _{*}^{b_{3}}+b_{2}\left\Vert \boldsymbol{\iota}_{p-1}\right\Vert _{*}^{b_{3}}\left|z_{1}\right|^{b_{3}}\bigr\}\\
 & =\tfrac{1}{2}\exp\bigl\{\tau_{2}b_{2,2}\eta^{b_{3}}\left\Vert \boldsymbol{z}_{2}\right\Vert _{*}^{b_{3}}+\tau_{1}b_{2,1}\left|z_{1}\right|^{b_{3}}\bigr\}\\
 & \leq\tfrac{\tau_{1}}{2}\exp\bigl\{ b_{2,1}\left|z_{1}\right|^{b_{3}}\bigr\}+\tfrac{\tau_{2}}{2}\exp\bigl\{ b_{2,2}\eta^{b_{3}}\left\Vert \boldsymbol{z}_{2}\right\Vert _{*}^{b_{3}}\bigr\}\\
 & \leq\tfrac{1}{2}\exp\bigl\{ b_{2,1}\left|z_{1}\right|^{b_{3}}\bigr\}+\tfrac{1}{2}\exp\bigl\{ b_{2,2}\eta^{b_{3}}\left\Vert \boldsymbol{z}_{2}\right\Vert _{*}^{b_{3}}\bigr\}\\
 & \overset{def}{=}V_{2,1}(z_{1})+V_{2,2}(\boldsymbol{z}_{2}),
\end{align*}
where the second inequality is justified by the convexity of the exponential
function. Now, as $\tau_{2}\in(\eta^{b_{3}},1)$, we have $b_{2,2}\eta^{b_{3}}=b_{2}\eta^{b_{3}}/\tau_{2}<b_{2}$,
and\textcolor{black}{{} we choose the value of $b_{2}$ so small that
$b_{2,1}=b_{2}\left\Vert \boldsymbol{\iota}_{p-1}\right\Vert _{*}^{b_{3}}/\tau_{1}<b_{1}=\tilde{b}_{1}$
with $\tilde{b}_{1}$ as fixed above.}

We next bound $V_{2,1}(z_{1})$ and $V_{2,2}(\boldsymbol{z}_{2})$.
For the former, write $V_{2,1}(z_{1})=\exp\bigl\{-\left(b_{1}-b_{2,1}\right)\left|z_{1}\right|^{b_{3}}\bigr\} V_{1}(z_{1})$
and use the facts $\ln V_{1}(z_{1})=\ln(\frac{1}{2})+b_{1}\left|z_{1}\right|^{b_{3}}$,
$\alpha=\rho/b_{3}-1>0$, and $b_{1}-b_{2,1}>0$ to obtain
\[
V_{2,1}(z_{1})=\frac{\left(1+\ln V_{1}(z_{1})\right)^{\alpha}b_{3}b_{1}^{\rho/b_{3}}}{b_{3}b_{1}^{\rho/b_{3}}\exp\bigl\{\left(b_{1}-b_{2,1}\right)\left|z_{1}\right|^{b_{3}}\bigr\}}\left(1+\ln V_{1}(z_{1})\right)^{-\alpha}V_{1}(z_{1})\leq\tfrac{1}{2}\epsilon_{4}b_{3}b_{1}^{\rho/b_{3}}\left(1+\ln V_{1}(z_{1})\right)^{-\alpha}V_{1}(z_{1}),
\]
where the inequality holds for any $\epsilon_{4}>0$ as long as $\left|z_{1}\right|$
is large enough. Using the definition of $\phi_{1}\left(V_{1}(z_{1})\right)$
in (\ref{eq:Th1(iii)Pr_phi2}) this implies a bound for $-\phi_{1}\left(V_{1}(z_{1})\right)+V_{2,1}(z_{1})$
which will be needed later:
\begin{align}
-\phi_{1}\left(V_{1}(z_{1})\right)+V_{2,1}(z_{1}) & \leq-\tfrac{1}{2}\omega_{1}b_{3}b_{1}^{\rho/b_{3}}\left(1+\ln V_{1}(z_{1})\right)^{-\alpha}V_{1}(z_{1})\nonumber \\
 & \qquad+\tfrac{1}{2}\epsilon_{4}b_{3}b_{1}^{\rho/b_{3}}\left(1+\ln V_{1}(z_{1})\right)^{-\alpha}V_{1}(z_{1})+\overline{M}_{1}\boldsymbol{1}_{C_{1}}(z_{1})\nonumber \\
 & =-\omega b_{3}b_{1}^{\rho/b_{3}}\left(1+\ln V_{1}(z_{1})\right)^{-\alpha}V_{1}(z_{1})+\overline{M}_{1}\boldsymbol{1}_{C_{1}}(z_{1}),\label{Bound_-phi2+V_2,1}
\end{align}
where $\omega=\tfrac{1}{2}(\omega_{1}-\epsilon_{4})$ and, as $\omega_{1}>0$
holds for (fixed) $b_{1}=\tilde{b}_{1}$, we can choose $\epsilon_{4}$
so small that $\omega>0$ holds. Note that here the last expression
provides a bound for $-\phi_{1}\left(V_{1}(z_{1})\right)+V_{2,1}(z_{1})$
that holds for all $z_{1}$ (although this may require redefining
the set $C_{1}$ and the value of the constant $\overline{M}_{1}$
which appear also in the upper bound obtained earlier for $E\left[V_{1}\left(g(z_{1})+\varepsilon_{1}\right)\right]$).
Denoting $\epsilon=\epsilon_{3}+\epsilon_{4}$ and using the definition
of $\omega_{1}$ (given at the end of Step 1) we therefore have $\omega=\tfrac{1}{2}(r-\epsilon)$.

Now consider $V_{2,2}(\boldsymbol{z}_{2})=\frac{1}{2}\exp\{b_{2,2}\eta^{b_{3}}\left\Vert \boldsymbol{z}_{2}\right\Vert _{*}^{b_{3}}\}$
and recall that $b_{2,2}\eta^{b_{3}}<b_{2}$. Using the definition
$V_{2}(\boldsymbol{z}_{2})=\frac{1}{2}\exp\{b_{2}\left\Vert \boldsymbol{z}_{2}\right\Vert _{*}^{b_{3}}\}$
we have, for some $\eta_{2}\in(0,1)$ and $\left\Vert \boldsymbol{z}_{2}\right\Vert _{*}$
bounded away from zero,
\[
V_{2,2}(\boldsymbol{z}_{2})=V_{2}(\boldsymbol{z}_{2})\frac{\exp\{b_{2,2}\eta^{b_{3}}\left\Vert \boldsymbol{z}_{2}\right\Vert _{*}^{b_{3}}\}}{\exp\{b_{2}\left\Vert \boldsymbol{z}_{2}\right\Vert _{*}^{b_{3}}\}}=V_{2}(\boldsymbol{z}_{2})\exp\{-\left(b_{2}-b_{2,2}\eta^{b_{3}}\right)\left\Vert \boldsymbol{z}_{2}\right\Vert _{*}^{b_{3}}\}\leq\eta_{2}V_{2}(\boldsymbol{z}_{2}),
\]
and furthermore
\[
V_{2}\bigl(\boldsymbol{\Pi}_{1}\boldsymbol{z}_{2}+z_{1}\boldsymbol{\iota}_{p-1}\bigr)\leq V_{2,1}(z_{1})+\eta_{2}V_{2}(\boldsymbol{z}_{2}),
\]
where the bound obtained above for $V_{2,1}(z_{1})$ has been omitted
but it will be used below.

\noindent \textbf{Step 4: Bounding $E\left[V(\boldsymbol{y}_{1})\,\left|\,\boldsymbol{y}_{0}=\boldsymbol{x}\right.\right]$
in (\ref{Drift_3})}. Using (\ref{eq:Th1(i)PR_E(V1)_2}) and the preceding
inequality obtained for $V_{2}\bigl(\boldsymbol{\Pi}_{1}\boldsymbol{z}_{2}+z_{1}\boldsymbol{\iota}_{p-1}\bigr)$
we can now write
\begin{align*}
E\left[V(\boldsymbol{y}_{1})\,\left|\,\boldsymbol{y}_{0}=\boldsymbol{x}\right.\right] & \leq V_{1}(z_{1})-\phi_{1}\left(V_{1}(z_{1})\right)+\tfrac{1}{2}\exp\bigl\{ o(1)\left\Vert \boldsymbol{z}_{2}\right\Vert _{*}^{b_{3}}\bigr\}+\overline{M}_{1}\boldsymbol{1}_{C_{1}}(z_{1})\\
 & \quad\quad\quad\quad\quad\quad\quad\quad+V_{2,1}(z_{1})+V_{2}(\boldsymbol{z}_{2})-(1-\eta_{2})V_{2}(\boldsymbol{z}_{2}).
\end{align*}
As $\left|\boldsymbol{z}_{2}\right|\leq\left|\boldsymbol{z}\right|\leq c\left|\boldsymbol{x}\right|$
(see footnote \ref{fn:x_big_iff_z_big}), the term $o(1)$ on the
right hand side converges to zero as $\left|\boldsymbol{z}_{2}\right|\rightarrow\infty$.
Thus, as $V_{2}(\boldsymbol{z}_{2})=\frac{1}{2}\exp\{b_{2}\left\Vert \boldsymbol{z}_{2}\right\Vert _{*}^{b_{3}}\}$,
we have, for $\left|\boldsymbol{z}_{2}\right|$ large, 
\begin{align*}
\tfrac{1}{2}\exp\bigl\{ o(1)\left\Vert \boldsymbol{z}_{2}\right\Vert _{*}^{b_{3}}\bigr\}-(1-\eta_{2})V_{2}(\boldsymbol{z}_{2}) & =\bigl[\exp\bigl\{[o(1)-b_{2}]\left\Vert \boldsymbol{z}_{2}\right\Vert _{*}^{b_{3}}\bigr\}-(1-\eta_{2})\bigr]\tfrac{1}{2}\exp\bigl\{ b_{2}\left\Vert \boldsymbol{z}_{2}\right\Vert _{*}^{b_{3}}\bigr\}\\
 & \leq-\eta_{3}V_{2}(\boldsymbol{z}_{2}),
\end{align*}
where $\eta_{3}\in(0,1)$. Hence,
\[
V_{2}(\boldsymbol{z}_{2})+\tfrac{1}{2}\exp\bigl\{ o(1)\left\Vert \boldsymbol{z}_{2}\right\Vert _{*}^{b_{3}}\bigr\}-(1-\eta_{2})V_{2}(\boldsymbol{z}_{2})\leq V_{2}(\boldsymbol{z}_{2})-\eta_{3}V_{2}(\boldsymbol{z}_{2})+\overline{M}_{2}\boldsymbol{1}_{C_{2}}(\boldsymbol{z}_{2}),
\]
where $C_{2}=\left\{ \boldsymbol{z}_{2}\in\mathbb{R}^{p-1}\,:\,\left|\boldsymbol{z}_{2}\right|\leq M_{2}\right\} $
and $M_{2}$ and $\overline{M}_{2}$ are some finite constants. Using
this inequality and the bound in (\ref{Bound_-phi2+V_2,1}) we can
bound $E\left[V(\boldsymbol{y}_{1})\,\left|\,\boldsymbol{y}_{0}=\boldsymbol{x}\right.\right]$
as follows:
\begin{align*}
E\left[V(\boldsymbol{y}_{1})\,\left|\,\boldsymbol{y}_{0}=\boldsymbol{x}\right.\right] & \leq V_{1}(z_{1})-\omega b_{3}b_{1}^{\rho/b_{3}}\left(1+\ln V_{1}(z_{1})\right)^{-\alpha}V_{1}(z_{1})\\
 & \quad+V_{2}(\boldsymbol{z}_{2})-\eta_{3}V_{2}(\boldsymbol{z}_{2})+2\overline{M}_{1}\boldsymbol{1}_{C_{1}}(z_{1})+\overline{M}_{2}\boldsymbol{1}_{C_{2}}(\boldsymbol{z}_{2}).
\end{align*}

We still need to modify the right hand side of the above inequality
to a form assumed in Condition D, and for simplicity we write this
inequality as
\[
E\left[V(\boldsymbol{y}_{1})\,\left|\,\boldsymbol{y}_{0}=\boldsymbol{x}\right.\right]\leq V_{1}(z_{1})-\omega b_{3}b_{1}^{\rho/b_{3}}\left(1+\ln V_{1}(z_{1})\right)^{-\alpha}V_{1}(z_{1})+V_{2}(\boldsymbol{z}_{2})-\eta_{3}V_{2}(\boldsymbol{z}_{2})+L,
\]
where $L\geq2\overline{M}_{1}+\overline{M}_{2}$. Next note that $V(\boldsymbol{x})\geq V_{1}(z_{1})\geq1/2$
(see (\ref{Drift_1})) so that
\[
0<\left(1+\ln V(\boldsymbol{x})\right)^{-\alpha}\leq\left(1+\ln V_{1}(z_{1})\right)^{-\alpha}\leq\left(1+\ln(1/2)\right)^{-\alpha}.
\]
Using these inequalities twice and defining $c_{\phi}=\omega b_{3}b_{1}^{\rho/b_{3}}\land\eta_{3}\left(1+\ln\frac{1}{2}\right)^{\alpha}$
($>0$) we have
\begin{align*}
 & -\omega b_{3}b_{1}^{\rho/b_{3}}\left(1+\ln V_{1}(z_{1})\right)^{-\alpha}V_{1}(z_{1})-\eta_{3}V_{2}(\boldsymbol{z}_{2})\\
 & \qquad\leq-\omega b_{3}b_{1}^{\rho/b_{3}}\left(1+\ln V(\boldsymbol{x})\right)^{-\alpha}V_{1}(z_{1})-\eta_{3}\left(1+\ln(1/2)\right)^{\alpha}\left(1+\ln V(\boldsymbol{x})\right)^{-\alpha}V_{2}(\boldsymbol{z}_{2})\\
 & \qquad\leq-c_{\phi}\left(1+\ln V(\boldsymbol{x})\right)^{-\alpha}V(\boldsymbol{x}).
\end{align*}
Denoting $h(\boldsymbol{x})=c_{\phi}\left(1+\ln V(\boldsymbol{x})\right)^{-\alpha}$
we therefore obtain 
\begin{equation}
E\left[V(\boldsymbol{y}_{1})\,\left|\,\boldsymbol{y}_{0}=\boldsymbol{x}\right.\right]\leq\left(1-h(\boldsymbol{x})\right)V(\boldsymbol{x})+L.\label{Drift_L}
\end{equation}
Because $V(\boldsymbol{x})\geq1$ and $-\alpha<0$, we have $0<h(\boldsymbol{x})\leq c_{\phi}$
and $h(\boldsymbol{x})\rightarrow0$, as $\left|\boldsymbol{x}\right|\rightarrow\infty$.
Thus, for all $\left|\boldsymbol{x}\right|$ large enough, $h(\boldsymbol{x})\leq1$,
and therefore 
\begin{align*}
\left(1-h(\boldsymbol{x})\right)V(\boldsymbol{x})+L & =\left(1-h(\boldsymbol{x})\right)^{\frac{1}{2}}V(\boldsymbol{x})\cdot\left(1-h(\boldsymbol{x})\right)^{\frac{1}{2}}\left(1+L/[\left(1-h(\boldsymbol{x})\right)V(\boldsymbol{x})]\right)\\
 & \leq\left(1-\tfrac{1}{2}h(\boldsymbol{x})\right)V(\boldsymbol{x})\cdot\left(1-h(\boldsymbol{x})\right)^{\frac{1}{2}}\left(1+L/[\left(1-h(\boldsymbol{x})\right)V(\boldsymbol{x})]\right)\\
 & \leq\left(1-\tfrac{1}{2}h(\boldsymbol{x})\right)V(\boldsymbol{x})
\end{align*}
for all $\left|\boldsymbol{x}\right|$ large enough, where the first
inequality is based on the inequality $(1-x)^{a}\leq1-ax$ (which
holds for $a,x\in[0,1]$) and the second inequality is justified by
showing that the inequality 
\[
H(\boldsymbol{x})\,\overset{def}{=}\,\left(1-h(\boldsymbol{x})\right)^{\frac{1}{2}}\left(1+L/[\left(1-h(\boldsymbol{x})\right)V(\boldsymbol{x})]\right)<1
\]
holds for all $\left|\boldsymbol{x}\right|$ large enough. To show
this, note first that
\[
H(\boldsymbol{x})=\left(1-h(\boldsymbol{x})\right)^{\frac{1}{2}}+L/[\left(1-h(\boldsymbol{x})\right)^{1/2}V(\boldsymbol{x})]\leq1-\tfrac{1}{2}h(\boldsymbol{x})+L/[\left(1-h(\boldsymbol{x})\right)^{1/2}V(\boldsymbol{x})],
\]
so that it suffices to show that, for all $\left|\boldsymbol{x}\right|$
large enough, the right hand side of the last inequality is smaller
than one or, equivalently, that $L<\tfrac{1}{2}h(\boldsymbol{x})\left(1-h(\boldsymbol{x})\right)^{\frac{1}{2}}V(\boldsymbol{x})$.
This holds for all $\left|\boldsymbol{x}\right|$ large enough due
to the definitions of $V(\boldsymbol{x})$ and $h(\boldsymbol{x})$
which imply that, as $\left|\boldsymbol{x}\right|\rightarrow\infty$,
$V(\boldsymbol{x})\rightarrow\infty$ at an exponential rate (see
(\ref{Drift_1})) whereas $h(\boldsymbol{x})\rightarrow0$ at a logarithmic
rate (see the above definition of $h(\boldsymbol{x})$).

We can therefore write inequality (\ref{Drift_L}), for all $\left|\boldsymbol{x}\right|$
large enough, as
\[
E\left[V(\boldsymbol{y}_{1})\,\left|\,\boldsymbol{y}_{0}=\boldsymbol{x}\right.\right]\leq\left(1-\tfrac{1}{2}h(\boldsymbol{x})\right)V(\boldsymbol{x}).
\]
As the right hand side is bounded when $\boldsymbol{x}$ belongs to
any compact set, this further implies that there exist positive constants
$M$ and $b$ such that for $C=\left\{ \boldsymbol{x}\in\mathbb{R}^{p}\,:\,\left|\boldsymbol{x}\right|\leq M\right\} $
and for all $\boldsymbol{x}\in\mathbb{R}^{p}$
\begin{equation}
E\left[V(\boldsymbol{y}_{1})\,\left|\,\boldsymbol{y}_{0}=\boldsymbol{x}\right.\right]\leq V(\boldsymbol{x})-\phi_{1}\left(V(\boldsymbol{x})\right)+b\boldsymbol{1}_{C}(\boldsymbol{x}),\label{eq:Th2Proof_DriftFinal}
\end{equation}
where
\begin{equation}
\phi_{1}\left(V(\boldsymbol{x})\right)=\tfrac{1}{2}h(\boldsymbol{x})V(\boldsymbol{x})=\tfrac{1}{2}c_{\phi}\left(1+\ln V(\boldsymbol{x})\right)^{-\alpha}V(\boldsymbol{x}).\label{eq:Th2Proof_DriftPhi}
\end{equation}
Now note that we can always find positive constants $v_{0}$ and $c$
such that the function $\phi(v)=c(v+v_{0})(\ln(v+v_{0}))^{-\alpha}$
is a concave increasing differentiable function for all $v\geq1$
and such that 
\[
\phi_{1}(v)=\tfrac{1}{2}c_{\phi}v(1+\ln(v))^{-\alpha}\geq c(v+v_{0})(\ln(v+v_{0}))^{-\alpha}=\phi(v)
\]
for large enough $v$. Therefore, potentially redefining $M$, $b$,
and $C$, 
\[
E\left[V(\boldsymbol{y}_{1})\,\left|\,\boldsymbol{y}_{0}=\boldsymbol{x}\right.\right]\leq V(\boldsymbol{x})-\phi\left(V(\boldsymbol{x})\right)+b\boldsymbol{1}_{C}(\boldsymbol{x}).
\]
Thus, we have verified Condition D (with $\alpha=\rho/b_{3}-1$).
The result follows from Theorem 1. 

\bigskip{}

\noindent \textbf{Part (ii)}. Now $\rho=\kappa_{0}$ and, as in the
proof of Theorem 3.3(ii) in \citet[p. 1373]{douc2004practical}, many
results in the proof of case $\rho>\kappa_{0}$ can be used. Again,
we choose $b_{3}=\kappa_{0}\land(2-\rho)$, noting that now $b_{3}=\kappa_{0}$
and that the notation $\kappa_{0}$ will be used below instead of
$b_{3}$. Also, the function $V(\boldsymbol{x})=V_{1}(z_{1})+V_{2}(\boldsymbol{z}_{2})$
is as in the case $\rho>\kappa_{0}$, and we need to bound the two
terms in (\ref{Drift_3}).

\noindent \textbf{Step 1: Bounding $E\left[V_{1}\left(g(z_{1})+\varepsilon_{1}\right)\right]$
in (\ref{Drift_3})}. Exactly as in Part (i), Step 1, it again holds
that, for $\left|z_{1}\right|>M_{0}$,

\noindent 
\begin{align*}
V_{1}\left(g(z_{1})\right)-V_{1}(z_{1}) & \leq\bigl(-b_{1}r\kappa_{0}\left|z_{1}\right|^{\kappa_{0}-\rho}+\tfrac{1}{2}b_{1}^{2}r^{2}\kappa_{0}^{2}\left|z_{1}\right|^{2(\kappa_{0}-\rho)}\bigr)V_{1}(z_{1})\\
 & =\left(-b_{1}r\kappa_{0}+\tfrac{1}{2}b_{1}^{2}r^{2}\kappa_{0}^{2}\right)V_{1}(z_{1})
\end{align*}
and, for large $\left|z_{1}\right|$, 
\begin{align*}
E\left[V_{1}\left(g(z_{1})+\varepsilon_{1}\right)\right]-V_{1}\left(g(z_{1})\right) & \leq b_{1}^{2}\kappa_{0}^{2}E\left[\varepsilon_{1}^{2}V_{1}(\varepsilon_{1})\right]\left|z_{1}\right|^{2\kappa_{0}-2}V_{1}(z_{1}).
\end{align*}
Hence, for large $\left|z_{1}\right|$,
\begin{align*}
E\left[V_{1}\left(g(z_{1})+\varepsilon_{1}\right)\right] & \leq V_{1}(z_{1})+k(z_{1})V_{1}(z_{1}),
\end{align*}
where now
\[
k(z_{1})=-b_{1}r\kappa_{0}+\tfrac{1}{2}b_{1}^{2}r^{2}\kappa_{0}^{2}+b_{1}^{2}\kappa_{0}^{2}E\left[\varepsilon_{1}^{2}V_{1}(\varepsilon_{1})\right]\left|z_{1}\right|^{2\kappa_{0}-2}.
\]
Due to Assumption \ref{assu:errors}(a) and the choice of $b_{3}$,
the condition $E\bigl[\left|\varepsilon_{1}\right|^{2}V_{1}(\varepsilon_{1})\bigr]<\infty$
can be achieved by choosing the value of $b_{1}$ small enough or,
specifically, assuming $b_{1}=\tilde{b}_{1}<\beta_{0}$. Furthermore,
as $\ensuremath{\kappa_{0}\in(0,1]}$, by choosing the value of $b_{1}$
small enough the function $k(z_{1})\in(-1,0)$ and is bounded away
from $-1$ and $0$ for any $\left|z_{1}\right|$ large enough. Therefore,
for some $\delta_{1}\in(0,1)$, 
\[
E\left[V_{1}\left(g(z_{1})+\varepsilon_{1}\right)\right]\leq V_{1}(z_{1})-\delta_{1}V_{1}(z_{1})
\]
for all sufficiently large $\left|z_{1}\right|$.\bigskip{}

\noindent \textbf{Step 2: Bounding $V_{\epsilon}(\boldsymbol{x})E\left[V_{1}\left(g(z_{1})+\varepsilon_{1}\right)\right]$
in (\ref{Drift_3})}. For the first term on the right hand side of
(\ref{Drift_3}) we obtain, for $\left|z_{1}\right|$ large,
\[
V_{\epsilon}(\boldsymbol{x})E\left[V_{1}\left(g(z_{1})+\varepsilon_{1}\right)\right]\leq\exp\left\{ b_{1}\left|\epsilon(\boldsymbol{x})\boldsymbol{x}\right|^{\kappa_{0}}\right\} (1-\delta_{1})V_{1}(z_{1})=\tfrac{1}{2}(1-\delta_{1})\exp\left\{ b_{1}\left|\epsilon(\boldsymbol{x})\boldsymbol{x}\right|^{\kappa_{0}}+b_{1}\left|z_{1}\right|^{\kappa_{0}}\right\} .
\]
Write $1-\delta_{1}=(1-\delta_{1})^{1/2}(1-\delta_{1})^{1/2}\leq(1-\tfrac{1}{2}\delta_{1})^{2}$
and note that $1-\tfrac{1}{2}\delta_{1}=\exp\left\{ \ln\left(1-\tfrac{1}{2}\delta_{1}\right)\right\} \leq\exp\left\{ -\tfrac{1}{2}\delta_{1}\right\} $
to obtain 
\[
V_{\epsilon}(\boldsymbol{x})E\left[V_{1}\left(g(z_{1})+\varepsilon_{1}\right)\right]\leq\tfrac{1}{2}\left(1-\tfrac{1}{2}\delta_{1}\right)\exp\left\{ -\tfrac{1}{2}\delta_{1}+b_{1}\left|z_{1}\right|^{\kappa_{0}}+b_{1}\left|\epsilon(\boldsymbol{x})\boldsymbol{x}\right|^{\kappa_{0}}\right\} .
\]
As Assumption \ref{assu:dynamics}(ii.a) now holds with $d=1$, we
have $\left|\epsilon(\boldsymbol{x})\right|=o(\left|\boldsymbol{x}\right|^{-1})$
and
\[
b_{1}\left|\epsilon(\boldsymbol{x})\boldsymbol{x}\right|^{\kappa_{0}}\leq o(\left|\boldsymbol{x}\right|^{-\kappa_{0}})b_{1}\left|z_{1}\right|^{\kappa_{0}}+o(\left|\boldsymbol{x}\right|^{-\kappa_{0}})b_{1}\left\Vert \boldsymbol{z}_{2}\right\Vert _{*}^{\kappa_{0}}
\]
(cf. the similar inequality in the proof of case $\rho>\kappa_{0}$,
Step 2). Therefore, for $\left|z_{1}\right|$ large,
\[
-\tfrac{1}{2}\delta_{1}+b_{1}\left|z_{1}\right|^{\kappa_{0}}+b_{1}\left|\epsilon(\boldsymbol{x})\boldsymbol{x}\right|^{\kappa_{0}}\leq\left(1-\delta_{1}(z_{1})\right)b_{1}\left|z_{1}\right|^{\kappa_{0}}+o(\left|\boldsymbol{x}\right|^{-\kappa_{0}})b_{1}\left\Vert \boldsymbol{z}_{2}\right\Vert _{*}^{\kappa_{0}},
\]
where
\[
\delta_{1}(z_{1})=\frac{\delta_{1}}{2b_{1}\left|z_{1}\right|^{\kappa_{0}}}+o(\left|\boldsymbol{x}\right|^{-\kappa_{0}})=\frac{\delta_{1}+o(1)}{2b_{1}\left|z_{1}\right|^{\kappa_{0}}}
\]
with $\delta_{1}(z_{1})\in(0,1)$ and $\delta_{1}(z_{1})^{-1}o(\left|\boldsymbol{x}\right|^{-\kappa_{0}})=o(1)$
holding (here, as well as above, the term $o(1)$ is obtained because
$\left|z_{1}\right|^{\kappa_{0}}o(\left|\boldsymbol{x}\right|^{-\kappa_{0}})=o(1)$
by arguments similar to those used in the case $\rho>\kappa_{0}$,
Step 2). 

Thus, we can conclude that, for $\left|z_{1}\right|$ large, 
\begin{align*}
 & V_{\epsilon}(\boldsymbol{x})E\left[V_{1}\left(g(z_{1})+\varepsilon_{1}\right)\right]\\
 & \quad\leq\tfrac{1}{2}\left(1-\tfrac{1}{2}\delta_{1}\right)\exp\left\{ \left(1-\delta_{1}(z_{1})\right)b_{1}\left|z_{1}\right|^{\kappa_{0}}+\delta_{1}(z_{1})\delta_{1}(z_{1})^{-1}o(\left|\boldsymbol{x}\right|^{-\kappa_{0}})b_{1}\left\Vert \boldsymbol{z}_{2}\right\Vert _{*}^{\kappa_{0}}\right\} \\
 & \quad\leq\tfrac{1}{2}\left(1-\tfrac{1}{2}\delta_{1}\right)\left(1-\delta_{1}(z_{1})\right)\exp\left\{ b_{1}\left|z_{1}\right|^{\kappa_{0}}\right\} +\tfrac{1}{2}\left(1-\tfrac{1}{2}\delta_{1}\right)\delta_{1}(z_{1})\exp\left\{ \delta_{1}(z_{1})^{-1}o(\left|\boldsymbol{x}\right|^{-\kappa_{0}})b_{1}\left\Vert \boldsymbol{z}_{2}\right\Vert _{*}^{\kappa_{0}}\right\} \\
 & \quad\leq\tfrac{1}{2}\left(1-\tfrac{1}{2}\delta_{1}\right)\exp\left\{ b_{1}\left|z_{1}\right|^{\kappa_{0}}\right\} +\tfrac{1}{2}\exp\left\{ o(1)b_{1}\left\Vert \boldsymbol{z}_{2}\right\Vert _{*}^{\kappa_{0}}\right\} ,
\end{align*}
where the second inequality is due to the convexity of the exponential
function. To state this more formally, we can find $M_{1}\geq M_{0}$
and some (finite) $\overline{M}_{1}$, such that 
\[
V_{\epsilon}(\boldsymbol{x})E\left[V_{1}\left(g(z_{1})+\varepsilon_{1}\right)\right]\leq V_{1}(z_{1})-\tfrac{1}{2}\delta_{1}V_{1}(z_{1})+\tfrac{1}{2}\exp\left\{ o(1)b_{1}\left\Vert \boldsymbol{z}_{2}\right\Vert _{*}^{\kappa_{0}}\right\} +\overline{M}_{1}\boldsymbol{1}_{C_{1}}(z_{1}),
\]
where $\delta_{1}\in(0,1)$ and $C_{1}=\left\{ z_{1}\in\mathbb{R}\,:\,\left|z_{1}\right|\leq M_{1}\right\} $
(cf. the proof of part (ii) in Douc et al. (2004, p. 1373)). Moreover,
as in case $\rho>\kappa_{0}$ (the beginning of Step 4), the term
$o(1)$ on the right hand side converges to zero as $\left|\boldsymbol{z}_{2}\right|\rightarrow\infty$.\bigskip{}

\noindent \textbf{Step 3: Bounding $V_{2}\left(\boldsymbol{\Pi}_{1}\boldsymbol{z}_{2}+z_{1}\boldsymbol{\iota}_{p-1}\right)$
in (\ref{Drift_3})}. As in the the proof of case $\rho>\kappa_{0}$,
Step 3, assume that the value of $b_{1}$ is fixed to $\tilde{b}_{1}$
specified above. Repeating the arguments in the proof of case $\rho>\kappa_{0}$,
Step 3, we first obtain
\begin{align*}
V_{2}\bigl(\mathbf{\Pi}\boldsymbol{z}_{2}+z_{1}\boldsymbol{\iota}_{p-1}\bigr) & \leq\tfrac{1}{2}\exp\left\{ b_{2,1}\left|z_{1}\right|^{\kappa_{0}}\right\} +\tfrac{1}{2}\exp\left\{ b_{2,2}\eta^{\kappa_{0}}\left\Vert \boldsymbol{z}_{2}\right\Vert _{*}^{\kappa_{0}}\right\} \overset{def}{=}V_{2,1}(z_{1})+V_{2,2}(\boldsymbol{z}_{2}),
\end{align*}
where $b_{2,1}=b_{2}\left\Vert \boldsymbol{\iota}_{p-1}\right\Vert _{*}^{\kappa_{0}}/\tau_{1}$
and $b_{2,2}=b_{2}/\tau_{2}$ with $\tau_{1}\in(0,1)$ and $\tau_{2}=1-\tau_{1}$.
Also, as in case $\rho>\kappa_{0}$, we can choose $\tau_{2}\in(\eta^{\kappa_{0}},1)$
so that $b_{2,2}\eta^{\kappa_{0}}=b_{2}\eta^{\kappa_{0}}/\tau_{2}<b_{2}$,
and the value of $b_{2}$ so small that $b_{2,1}=b_{2}\left\Vert \boldsymbol{\iota}_{p-1}\right\Vert _{*}^{\kappa_{0}}/\tau_{1}<b_{1}=\tilde{b}_{1}$
with $\tilde{b}_{1}$ as fixed above. 

We next bound $V_{2,1}(z_{1})$ and $V_{2,2}(\boldsymbol{z}_{2})$.
Arguments similar to those used in the corresponding proof of case
$\rho>\kappa_{0}$, Step 3, apply but the bound obtained for $V_{2,1}(z_{1})$
simplifies. Specifically, 
\begin{align*}
V_{2,1}(z_{1}) & \leq\epsilon V_{1}(z_{1})\quad\textrm{and}\quad V_{2,2}(\boldsymbol{z}_{2})\leq\eta_{2}V_{2}(\boldsymbol{z}_{2}),
\end{align*}
where the first inequality holds for any $\epsilon>0$ as long as
$\left|z_{1}\right|$ is large enough and the second inequality holds
for some $\eta_{2}\in(0,1)$ and $\left\Vert \boldsymbol{z}_{2}\right\Vert _{*}$
bounded away from zero. These inequalities can be written as
\[
V_{2,1}(z_{1})\leq\epsilon V_{1}(z_{1})+\overline{M}_{1}\boldsymbol{1}_{C_{1}}(z_{1})\quad\textrm{and}\quad V_{2,2}(\boldsymbol{z}_{2})\leq V_{2}(\boldsymbol{z}_{2})-(1-\eta_{2})V_{2}(\boldsymbol{z}_{2})+\overline{M}_{2}\boldsymbol{1}_{C_{2}}(\boldsymbol{z}_{2}),
\]
where, for simplicity, we have assumed that the term $\overline{M}_{1}\boldsymbol{1}_{C_{1}}(z_{1})$
can be the same as at the end of Step 2 and where $C_{2}=\left\{ \boldsymbol{z}_{2}\in\mathbb{R}^{p-1}\,:\,\left|\boldsymbol{z}_{2}\right|\leq M_{2}\right\} $
with $M_{2}$ and $\overline{M}_{2}$ some positive and finite constants.
Thus, we can conclude that
\[
V_{2}\bigl(\mathbf{\Pi}\boldsymbol{z}_{2}+z_{1}\boldsymbol{\iota}_{p-1}\bigr)\leq\epsilon V_{1}(z_{1})+V_{2}(\boldsymbol{z}_{2})-(1-\eta_{2})V_{2}(\boldsymbol{z}_{2})+\overline{M}_{1}\boldsymbol{1}_{C_{1}}(z_{1})+\overline{M}_{2}\boldsymbol{1}_{C_{2}}(\boldsymbol{z}_{2}).
\]

\noindent \textbf{Step 4: Bounding $E\left[V(\boldsymbol{y}_{1})\,\left|\,\boldsymbol{y}_{0}=\boldsymbol{x}\right.\right]$
in (\ref{Drift_3})}. The bounds obtained for $V_{\epsilon}(\boldsymbol{x})E\left[V_{1}\left(g(z_{1})+\varepsilon_{1}\right)\right]$
and $V_{2}\bigl(\mathbf{\Pi}\boldsymbol{z}_{2}+z_{1}\boldsymbol{\iota}_{p-1}\bigr)$
in Steps 2 and 3, respectively, yield
\begin{align*}
E\left[V(\boldsymbol{y}_{1})\,\left|\,\boldsymbol{y}_{0}=\boldsymbol{x}\right.\right] & =V_{\epsilon}(\boldsymbol{x})E\left[V_{1}\left(g(z_{1})+\varepsilon_{1}\right)\right]+V_{2}\bigl(\mathbf{\Pi}\boldsymbol{z}_{2}+z_{1}\boldsymbol{\iota}_{p-1}\bigr)\\
 & \leq V_{1}(z_{1})-\tfrac{1}{2}\delta_{1}V_{1}(z_{1})+\tfrac{1}{2}\exp\left\{ o(1)b_{1}\left\Vert \boldsymbol{z}_{2}\right\Vert _{*}^{\kappa_{0}}\right\} +\epsilon V_{1}(z_{1})\\
 & \quad+V_{2}(\boldsymbol{z}_{2})-(1-\eta_{2})V_{2}(\boldsymbol{z}_{2})+2\overline{M}_{1}\boldsymbol{1}_{C_{1}}(z_{1})+\overline{M}_{2}\boldsymbol{1}_{C_{2}}(\boldsymbol{z}_{2}).
\end{align*}
 As the value of $\epsilon>0$ can be made as close to zero as desired
(by only choosing $\left|z_{1}\right|$ large enough and independently
of choices made for any other parameters), we can assume that $\epsilon<\tfrac{1}{2}\delta_{1}$
so that 
\[
-\tfrac{1}{2}\delta_{1}V_{1}(z_{1})+\epsilon V_{1}(z_{1})\leq-\delta_{2}V_{1}(z_{1})
\]
holds with some $\delta_{2}\in(0,1)$. Moreover, as in the proof of
case $\rho>\kappa_{0}$, Step 4, 
\begin{align*}
\tfrac{1}{2}\exp\{o(1)\left\Vert \boldsymbol{z}_{2}\right\Vert _{*}^{b_{3}}\}-(1-\eta_{2})V_{2}(\boldsymbol{z}_{2}) & \leq-\eta_{3}V_{2}(\boldsymbol{z}_{2}),
\end{align*}
$\eta_{3}\in(0,1)$. Thus, defining $\tilde{\lambda}=\delta_{2}\land\eta_{3}\in(0,1)$
and $L\geq2\overline{M}_{1}+\overline{M}_{2}$ we find that
\begin{align*}
E\left[V(\boldsymbol{y}_{1})\,\left|\,\boldsymbol{y}_{0}=\boldsymbol{x}\right.\right] & \leq V_{1}(z_{1}(\boldsymbol{x}))-\delta_{2}V_{1}(z_{1}(\boldsymbol{x}))+V_{2}(\boldsymbol{z}_{2}(\boldsymbol{x}))-\eta_{3}V_{2}(\boldsymbol{z}_{2}(\boldsymbol{x}))+L,\\
 & \leq V(\boldsymbol{x})-\tilde{\lambda}V(\boldsymbol{x})+L.
\end{align*}
We can write the above inequality as
\begin{align*}
E\left[V(\boldsymbol{y}_{1})\,\left|\,\boldsymbol{y}_{0}=\boldsymbol{x}\right.\right] & \leq(1-\tilde{\lambda})^{\frac{1}{2}}V(\boldsymbol{x})\cdot(1-\tilde{\lambda})^{\frac{1}{2}}\bigl(1+L/[(1-\tilde{\lambda})V(\boldsymbol{x})]\bigr),
\end{align*}
from which it follows that, for all $\left|\boldsymbol{x}\right|$
large enough, $E\left[V(\boldsymbol{y}_{1})\,\left|\,\boldsymbol{y}_{0}=\boldsymbol{x}\right.\right]\leq(1-\tilde{\lambda})^{\frac{1}{2}}V(\boldsymbol{x})$,
implying that there exist positive constants $M$ and $b$ such that,
for $C=\left\{ \boldsymbol{x}\in\mathbb{R}^{p}\,:\,\left|\boldsymbol{x}\right|\leq M\right\} $,
\[
E\left[V(\boldsymbol{y}_{1})\,\left|\,\boldsymbol{y}_{0}=\boldsymbol{x}\right.\right]\leq(1-\tilde{\lambda})^{\frac{1}{2}}V(\boldsymbol{x})+b\boldsymbol{1}_{C}(\boldsymbol{x}).
\]
Defining $\lambda=1-(1-\tilde{\lambda})^{\frac{1}{2}}\in(0,1)$ we
can conclude that Condition D holds with $\phi(v)=\lambda v$ and
therefore Theorem 1(i) shows that the Markov chain $\boldsymbol{y}_{t}$
is geometrically ergodic and the convergence (\ref{f-ergodicity})
holds with $f(\boldsymbol{x})=V(\boldsymbol{x})=V_{1}(z_{1}(\boldsymbol{x}))+V_{2}(\boldsymbol{z}_{2}(\boldsymbol{x}))$.

\bigskip{}

\noindent \textbf{Case $p=1$:} When $p=1$ we have $\boldsymbol{x}=x_{1}=u$
and we simply write $x$ for any of these. In this case, model (\ref{Companion form})
reduces to $y_{t}=y_{t-1}+\tilde{g}(y_{t-1})+\varepsilon_{t}$, Assumption
1(i) becomes redundant, Assumption 1(ii.a) is automatically satisfied
with $g(x)=x+\tilde{g}(x)$, $\epsilon(x)=0$, and $d$ redundant
(as long as the condition $\left|g(x)\right|\to\infty$ as $\left|x\right|\to\infty$
still holds), and Assumptions 1(ii.b) and 2 are as in the case $p\geq2$.
In other words, the model can be written as $y_{t}=g(y_{t-1})+\varepsilon_{t}$
with $g$ satisfying Assumption 1(ii.b) as well as $\left|g(x)\right|\to\infty$
as $\left|x\right|\to\infty$. This also means that the assumptions
of Theorem 3.3 in \citet{douc2004practical} are satisfied except
for the continuity of $g$ required in their Assumption 3.4. However,
in our case this assumption is not needed because the boundedness
of $g$ on compact subsets of $\mathbb{R}$ implied by our Assumption
1(ii) actually suffices. 

First consider the case $\rho>\kappa_{0}$. Proceeding as in the proof
of Theorem 3.3(i) of \citet{douc2004practical} we can conclude that
there exist positive constants $M$ and $b$ such that, for $C=\left\{ x\in\mathbb{R}\,:\,\left|x\right|\leq M\right\} $
and for all $x\in\mathbb{R}$,
\begin{equation}
E\left[V(y_{1})\,\left|\,y_{0}=x\right.\right]\leq V(x)-\phi_{1}\left(V(x)\right)+b\boldsymbol{1}_{C}(x),\label{eq:Cor2Proof_Drift}
\end{equation}
where $\phi_{1}(V(x))=\tilde{c}_{\phi}\left(1+\ln V(x)\right)^{-\alpha}V(x)$
with $\alpha=\rho/b_{3}-1>0$ and some $\tilde{c}_{\phi}>0$ (see
the top of p.~1373 of \citet{douc2004practical} and note also our
additional assumption $\left|g(x)\right|\to\infty$ as $\left|x\right|\to\infty$).
Comparing this with (\ref{eq:Th2Proof_DriftFinal}) and (\ref{eq:Th2Proof_DriftPhi})
at the end of the proof of part (i) shows that we can continue as
therein and conclude that Condition D is satisfied with $V(x)=\exp\{b_{1}\left|x\right|^{b_{3}}\}$
and $\phi(v)=c(v+v_{0})(\ln(v+v_{0}))^{-\alpha}$ (for some $c,v_{0}>0$
and $\alpha=\rho/b_{3}-1>0$). The result of part (i) now follows
from Theorem 1(ii). 

Now consider the case $\rho=\kappa_{0}$. As in the proof of Theorem
3.3(ii) of \citet{douc2004practical} we can conclude that (\ref{eq:Cor2Proof_Drift})
holds with $\phi_{1}\left(V(x)\right)=\lambda V(x)$ and some $\lambda>0$,
and with $M$, $b$, and $C$ redefined (see the middle of p.~1373
of \citet{douc2004practical} and note again the above-mentioned additional
assumption). The result of part (ii) now follows from Theorem 1(i).\qed

\bigskip{}
\bigskip{}

\noindent \textbf{Proof of Theorem 3.} First note that our Assumption
\ref{assu:errors}(b) implies Assumptions (NSS 1) and (NSS 4) of \citet{fort2003polynomial}.
Also, in the same way as in the proof of Theorem 2 we can show that
the Markov chain $\boldsymbol{y}_{t}$ is a $\psi$-irreducible and
aperiodic $T$-chain with $\psi$ the Lebesgue measure, and that all
compact sets of $\mathcal{B}(\mathbb{R}^{p})$ are petite. This, in
turn, implies that Assumption (NSS 2) of \citet{fort2003polynomial}
holds. These facts together with Assumption 1 are used below to verify
Assumption (NSS 3) of \citet{fort2003polynomial} which enables us
to apply Lemma 3 of that paper. 

As $V(\boldsymbol{x})=1+\left|z_{1}\right|^{s_{0}}+s_{1}\left\Vert \boldsymbol{z}_{2}\right\Vert _{*}^{\alpha s_{0}}$
we have (cf.~the beginning of the proof of Theorem 2) 
\begin{align}
E\left[V(\boldsymbol{y}_{1})\,\left|\,\boldsymbol{y}_{0}=\boldsymbol{x}\right.\right] & =1+E\left[\left|\overline{g}(\boldsymbol{x})+\varepsilon_{1}\right|^{s_{0}}\right]+s_{1}\left\Vert \boldsymbol{\Pi}_{1}\boldsymbol{z}_{2}+z_{1}\boldsymbol{\iota}_{p-1}\right\Vert _{*}^{\alpha s_{0}}.\label{Drift_4}
\end{align}
In this case it appears convenient to start with bounding the latter
term on the right hand side.\bigskip{}

\noindent \textbf{Step 1:} \textbf{Bounding $s_{1}\left\Vert \boldsymbol{\Pi}_{1}\boldsymbol{z}_{2}+z_{1}\boldsymbol{\iota}_{p-1}\right\Vert _{*}^{\alpha s_{0}}$
in (\ref{Drift_4})}. First note that $\alpha=1-\rho/s_{0}\in(0,1)$
because $0<\rho<s_{0}$ is assumed. We consider separately the cases
where $\alpha s_{0}\leq1$ and $\alpha s_{0}>1$, and show that there
exist constants $\eta_{0}\in(0,1)$ and $\overline{s}_{1}>0$ such
that 
\begin{align}
s_{1}\left\Vert \boldsymbol{\Pi}_{1}\boldsymbol{z}_{2}+z_{1}\boldsymbol{\iota}_{p-1}\right\Vert _{*}^{\alpha s_{0}} & \leq s_{1}\left\Vert \boldsymbol{z}_{2}\right\Vert _{*}^{\alpha s_{0}}-\eta_{0}s_{1}\left\Vert \boldsymbol{z}_{2}\right\Vert _{*}^{\alpha s_{0}}+\overline{s}_{1}\left|z_{1}\right|^{\alpha s_{0}}\label{Inequality_z2}
\end{align}
holds for both $\alpha s_{0}\leq1$ and $\alpha s_{0}>1$. Moreover,\textcolor{black}{{}
the value of $\overline{s}_{1}$ can be chosen as close to zero as
desired.}

First consider the case $\alpha s_{0}\leq1$ and\textcolor{black}{{}
assume that $s_{1}<1$}. Denoting $\tilde{s}_{1}=s_{1}\left\Vert \boldsymbol{\iota}_{p-1}\right\Vert _{*}^{\alpha s_{0}}$
we obtain (cf. the proof of Theorem 2, the beginning of Step 3) 
\begin{align*}
s_{1}\left\Vert \boldsymbol{\Pi}_{1}\boldsymbol{z}_{2}+z_{1}\boldsymbol{\iota}_{p-1}\right\Vert _{*}^{\alpha s_{0}} & \leq s_{1}\eta^{\alpha s_{0}}\left\Vert \boldsymbol{z}_{2}\right\Vert _{*}^{\alpha s_{0}}+\tilde{s}_{1}\left|z_{1}\right|^{\alpha s_{0}}=s_{1}\left\Vert \boldsymbol{z}_{2}\right\Vert _{*}^{\alpha s_{0}}-\eta_{1}s_{1}\left\Vert \boldsymbol{z}_{2}\right\Vert _{*}^{\alpha s_{0}}+\tilde{s}_{1}\left|z_{1}\right|^{\alpha s_{0}},
\end{align*}
where $\eta\in(0,1)$ by assumption and $\eta_{1}=1-\eta^{\alpha s_{0}}\in(0,1)$
which shows that inequality (\ref{Inequality_z2}) holds with $\eta_{0}=\eta_{1}$
and $\overline{s}_{1}=\tilde{s}_{1}$. \textcolor{black}{Also, the
value of $\tilde{s}_{1}$ can be made as close to zero as desired
by choosing $s_{1}$ small enough.}

Now consider the case $\alpha s_{0}>1$. Here $s_{1}<1$ is still
assumed and $s_{0}>1$ must hold because $\alpha\in(0,1)$. Write
\begin{align*}
s_{1}\left\Vert \boldsymbol{\Pi}_{1}\boldsymbol{z}_{2}+z_{1}\boldsymbol{\iota}_{p-1}\right\Vert _{*}^{\alpha s_{0}} & =s_{1}\left(\left\Vert \boldsymbol{\Pi}_{1}\boldsymbol{z}_{2}+z_{1}\boldsymbol{\iota}_{p-1}\right\Vert _{*}^{\alpha}\right)^{s_{0}}\\
 & \leq s_{1}\left(\left\Vert \boldsymbol{\Pi}_{1}\boldsymbol{z}_{2}\right\Vert _{*}^{\alpha}+\left\Vert \boldsymbol{\iota}_{p-1}\right\Vert _{*}^{\alpha}\left|z_{1}\right|^{\alpha}\right)^{s_{0}}\\
 & \leq\bigl(s_{1}^{1/s_{0}}\eta^{\alpha}\left\Vert \boldsymbol{z}_{2}\right\Vert _{*}^{\alpha}+s_{1}^{1/s_{0}}\left\Vert \boldsymbol{\iota}_{p-1}\right\Vert _{*}^{\alpha}\left|z_{1}\right|^{\alpha}\bigr)^{s_{0}},
\end{align*}
where $\eta\in(0,1)$ again holds by assumption. Let $\tau_{1}\in(0,1)$
and $\tau_{2}=1-\tau_{1}$, and denote $s_{1,1}=s_{1}^{1/s_{0}}\left\Vert \boldsymbol{\iota}_{p-1}\right\Vert _{*}^{\alpha}/\tau_{1}$
and $s_{1,2}=s_{1}^{1/s_{0}}/\tau_{2}$. Then,
\begin{align*}
s_{1}\left\Vert \boldsymbol{\Pi}_{1}\boldsymbol{z}_{2}+z_{1}\boldsymbol{\iota}_{p-1}\right\Vert _{*}^{\alpha s_{0}} & \leq\left(\tau_{1}s_{1,1}\left|z_{1}\right|^{\alpha}+\tau_{2}s_{1,2}\eta^{\alpha}\left\Vert \boldsymbol{z}_{2}\right\Vert _{*}^{\alpha}\right)^{s_{0}}\\
 & \leq\tau_{1}s_{1,1}^{s_{0}}\left|z_{1}\right|^{\alpha s_{0}}+\tau_{2}s_{1,2}^{s_{0}}\eta^{\alpha s_{0}}\left\Vert \boldsymbol{z}_{2}\right\Vert _{*}^{\alpha s_{0}}\\
 & \leq s_{1,1}^{s_{0}}\left|z_{1}\right|^{\alpha s_{0}}+s_{1,2}^{s_{0}}\eta^{\alpha s_{0}}\left\Vert \boldsymbol{z}_{2}\right\Vert _{*}^{\alpha s_{0}},
\end{align*}
where the second inequality is justified by the convexity of the function
$\left|x\right|\mapsto\left|x\right|^{s_{0}}$ for $s_{0}>1$. 

Next, as $\eta^{\alpha}<1$, we can choose $\tau_{2}\in(\eta^{\alpha},1)$
so that $s_{1,2}^{s_{0}}\eta^{\alpha s_{0}}=s_{1}\eta^{\alpha s_{0}}/\tau_{2}^{s_{0}}<s_{1}$.
Denoting $\eta_{2}=1-\left(\eta^{\alpha}/\tau_{2}\right)^{s_{0}}$
we have $\eta_{2}\in(0,1)$ and
\[
s_{1,2}^{s_{0}}\eta^{\alpha s_{0}}\left\Vert \boldsymbol{z}_{2}\right\Vert _{*}^{\alpha s_{0}}=s_{1}\left\Vert \boldsymbol{z}_{2}\right\Vert _{*}^{\alpha s_{0}}-s_{1}\left(1-\eta^{\alpha s_{0}}/\tau_{2}^{s_{0}}\right)\left\Vert \boldsymbol{z}_{2}\right\Vert _{*}^{\alpha s_{0}}=s_{1}\left\Vert \boldsymbol{z}_{2}\right\Vert _{*}^{\alpha s_{0}}-\eta_{2}s_{1}\left\Vert \boldsymbol{z}_{2}\right\Vert _{*}^{\alpha s_{0}},
\]
and we can conclude that
\[
s_{1}\left\Vert \boldsymbol{\Pi}_{1}\boldsymbol{z}_{2}+z_{1}\boldsymbol{\iota}_{p-1}\right\Vert _{*}^{\alpha s_{0}}\leq s_{1}\left\Vert \boldsymbol{z}_{2}\right\Vert _{*}^{\alpha s_{0}}-\eta_{2}s_{1}\left\Vert \boldsymbol{z}_{2}\right\Vert _{*}^{\alpha s_{0}}+s_{1,1}^{s_{0}}\left|z_{1}\right|^{\alpha s_{0}}.
\]
Thus, inequality (\ref{Inequality_z2}) holds with $\eta_{0}=\eta_{2}$
and $\overline{s}_{1}=s_{1,1}^{s_{0}}$. Above we fixed the value
of $\tau_{2}$, and hence also the value of $\tau_{1}$, but\textcolor{black}{{}
we are still free to choose the value of $s_{1}$ and make $s_{1,1}^{s_{0}}=s_{1}\left\Vert \boldsymbol{\iota}_{p-1}\right\Vert _{*}^{\alpha s_{0}}/\tau_{1}^{s_{0}}<1$
as close to zero as desired by choosing $s_{1}$ small enough. From
now on, we assume that $\eta_{0}=\eta_{1}\land\eta_{2}$ and $\overline{s}_{1}=\tilde{s}_{1}\lor s_{1,1}^{s_{0}}$
so that inequality (\ref{Inequality_z2}) applies irrespective of
whether $\alpha s_{0}\leq1$ or $\alpha s_{0}>1$, and the value of
$\overline{s}_{1}$ can be chosen arbitrarily close to zero. }\bigskip{}

\noindent \textbf{Step 2:} \textbf{Bounding $E\left[\left|\overline{g}(\boldsymbol{x})+\varepsilon_{1}\right|^{s_{0}}\right]$
in (\ref{Drift_4})}. Consider the cases $s_{0}<1$ and $s_{0}\geq1$
separately. When $s_{0}<1$, the definition of the function $\overline{g}$,
the triangle inequality, and Assumption 1(ii.a) yield
\begin{align}
E\left[\left|\overline{g}(\boldsymbol{x})+\varepsilon_{1}\right|^{s_{0}}\right] & =E\left[\left|\overline{g}(\boldsymbol{x})-g(z_{1})+g(z_{1})+\varepsilon_{1}\right|^{s_{0}}\right]\nonumber \\
 & \leq E\left[\left|\overline{g}(\boldsymbol{x})-g(z_{1})\right|^{s_{0}}+\left|g(z_{1})+\varepsilon_{1}\right|^{s_{0}}\right]\nonumber \\
 & \leq\left|\epsilon(\boldsymbol{x})\boldsymbol{x}\right|^{s_{0}}+E\left[\left|g(z_{1})+\varepsilon_{1}\right|^{s_{0}}\right].\label{eq:Th2_PrStep2_1a}
\end{align}
When $s_{0}\geq1$, we can use Minkowski's inequality and obtain
\begin{align}
\left(E\left[\left|\overline{g}(\boldsymbol{x})+\varepsilon_{1}\right|^{s_{0}}\right]\right)^{1/s_{0}} & =\left(E\left[\left|\overline{g}(\boldsymbol{x})-g(z_{1})+g(z_{1})+\varepsilon_{1}\right|^{s_{0}}\right]\right)^{1/s_{0}}\nonumber \\
 & \leq\left|\overline{g}(\boldsymbol{x})-g(z_{1})\right|+\left(E\left[\left|g(z_{1})+\varepsilon_{1}\right|^{s_{0}}\right]\right)^{1/s_{0}}\nonumber \\
 & \leq\left|\epsilon(\boldsymbol{x})\boldsymbol{x}\right|+\left(E\left[\left|g(z_{1})+\varepsilon_{1}\right|^{s_{0}}\right]\right)^{1/s_{0}}.\label{eq:Th2_PrStep2_2a}
\end{align}

The next step is to bound the expectation $E\left[\left|g(z_{1})+\varepsilon_{1}\right|^{s_{0}}\right]$.
Assumption 1(ii.b) ensures that the function $g$ satisfies the conditions
in Assumption (NSS 3) of \citet{fort2003polynomial} which (together
with other assumptions of the theorem) implies that we can use Lemma
3 of that paper. Thus, as $\alpha s_{0}=s_{0}-\rho$, inequality (36)
in that lemma shows that
\[
E\left[\left|g(z_{1})+\varepsilon_{1}\right|^{s_{0}}\right]\leq\left|z_{1}\right|^{s_{0}}-\lambda\left|z_{1}\right|^{\alpha s_{0}}\left(1+\tilde{\epsilon}(z_{1})\right),
\]
where $\tilde{\epsilon}(z_{1})\rightarrow0$ as $\left|z_{1}\right|\rightarrow\infty$
and $\lambda>0$ (to see this, note that the cases (i)\textendash (iii)
in our Theorem 3 correspond to the cases (i)\textendash (iii) in Lemma
3 of \citet{fort2003polynomial} so that the result is obtained with
$\lambda=s_{0}r$ in cases (i) and (ii) and with $\lambda=s_{0}r-\frac{1}{2}s_{0}(s_{0}-1)E[\varepsilon_{1}^{2}]$,
which is positive by assumption, in case (iii)). Thus, the above inequality
implies that, for $\left|z_{1}\right|$ large, 
\begin{align}
E\left[\left|g(z_{1})+\varepsilon_{1}\right|^{s_{0}}\right] & \leq\left|z_{1}\right|^{s_{0}}-\tilde{\lambda}\left|z_{1}\right|^{\alpha s_{0}},\label{Inequality_E|g+eps|^s_0}
\end{align}
where $\tilde{\lambda}>0$ and, without loss of generality, we can
assume that $\tilde{\lambda}\leq1$ also holds. Note that this inequality
holds for both $s_{0}<1$ and $s_{0}\geq1$; these two cases will
be treated separately below. 

\textbf{Case $s_{0}<1$}. First recall from the proof of Theorem 2,
Step 2, that 
\[
\left|\epsilon(\boldsymbol{x})\boldsymbol{x}\right|\leq\left|\epsilon_{1}(\boldsymbol{x})\right|\left|z_{1}\right|+\left|\epsilon_{1}(\boldsymbol{x})\right|\left\Vert \boldsymbol{z}_{2}\right\Vert _{*},
\]
where $\epsilon_{1}(\boldsymbol{x})=c_{*}\epsilon(\boldsymbol{x})$,
$c_{*}>0$. Using (\ref{eq:Th2_PrStep2_1a}), (\ref{Inequality_E|g+eps|^s_0}),
and the assumption $s_{0}<1$, we find that, for $\left|z_{1}\right|$
large,
\begin{align*}
E\left[\left|\overline{g}(\boldsymbol{x})+\varepsilon_{1}\right|^{s_{0}}\right] & \leq\left|z_{1}\right|^{s_{0}}-\tilde{\lambda}\left|z_{1}\right|^{\alpha s_{0}}+\left|\epsilon(\boldsymbol{x})\boldsymbol{x}\right|^{s_{0}}\\
 & \leq\left|z_{1}\right|^{s_{0}}-\tilde{\lambda}\left|z_{1}\right|^{\alpha s_{0}}+\left|\epsilon_{1}(\boldsymbol{x})\right|^{s_{0}}\left|z_{1}\right|^{s_{0}}+\left|\epsilon_{1}(\boldsymbol{x})\right|^{s_{0}}\left\Vert \boldsymbol{z}_{2}\right\Vert _{*}^{s_{0}}\\
 & \leq\left|z_{1}\right|^{s_{0}}-\tilde{\lambda}\left|z_{1}\right|^{\alpha s_{0}}+\left|\epsilon_{1}(\boldsymbol{x})\right|^{s_{0}}\left|z_{1}\right|^{\rho}\left|z_{1}\right|^{\alpha s_{0}}+\left|\epsilon_{1}(\boldsymbol{x})\right|^{s_{0}}\left\Vert \boldsymbol{z}_{2}\right\Vert _{*}^{\rho}\left\Vert \boldsymbol{z}_{2}\right\Vert _{*}^{\alpha s_{0}},
\end{align*}
where the last inequality follows because $\alpha=1-\rho/s_{0}$ so
that $\alpha s_{0}=s_{0}-\rho$. As $\left|\epsilon_{1}(\boldsymbol{x})\right|=o(\left|\boldsymbol{x}\right|^{-\rho/s_{0}})$
by assumption, we have, for $\left|z_{1}\right|$ large, $\left|\epsilon_{1}(\boldsymbol{x})\right|^{s_{0}}\left|z_{1}\right|^{\rho}=o(\left|\boldsymbol{x}\right|^{-\rho})\left|z_{1}\right|^{\rho}<\tilde{\lambda}$,
and thus
\begin{equation}
E\left[\left|\overline{g}(\boldsymbol{x})+\varepsilon_{1}\right|^{s_{0}}\right]\leq\left|z_{1}\right|^{s_{0}}-\tilde{\lambda}_{1}\left|z_{1}\right|^{\alpha s_{0}}+o(1)\left\Vert \boldsymbol{z}_{2}\right\Vert _{*}^{\alpha s_{0}},\label{eq:Th2_PrStep2_1b}
\end{equation}
where $\tilde{\lambda}_{1}\in(0,1)$ and $o(1)\rightarrow0$ as $\left|\boldsymbol{x}\right|\rightarrow\infty$
(the upper bound of $\tilde{\lambda}_{1}$ follows because $\tilde{\lambda}\leq1$
was assumed above and the term $o(1)$ is obtained as in the proof
of Theorem 2, Step 2).

\textbf{Case} $s_{0}\geq1$. When $s_{0}\geq1$, inequalities (\ref{eq:Th2_PrStep2_2a})
and (\ref{Inequality_E|g+eps|^s_0}) imply that, for $\left|z_{1}\right|$
large, 
\begin{align*}
\left(E\left[\left|\overline{g}(\boldsymbol{x})+\varepsilon_{1}\right|^{s_{0}}\right]\right)^{1/s_{0}} & \leq\left(E\left[\left|g(z_{1})+\varepsilon_{1}\right|^{s_{0}}\right]\right)^{1/s_{0}}+\left|\epsilon(\boldsymbol{x})\boldsymbol{x}\right|\\
 & \leq(\left|z_{1}\right|^{s_{0}}-\tilde{\lambda}\left|z_{1}\right|^{\alpha s_{0}})^{1/s_{0}}+\left|\epsilon_{1}(\boldsymbol{x})\right|\left|z_{1}\right|+\left|\epsilon_{1}(\boldsymbol{x})\right|\left\Vert \boldsymbol{z}_{2}\right\Vert _{*}\\
 & =\left|z_{1}\right|(1-\tilde{\lambda}\left|z_{1}\right|^{-\rho})^{1/s_{0}}+\left|\epsilon_{1}(\boldsymbol{x})\right|\left|z_{1}\right|+\left|\epsilon_{1}(\boldsymbol{x})\right|\left\Vert \boldsymbol{z}_{2}\right\Vert _{*}\\
 & \leq\left|z_{1}\right|\Bigl(1-\frac{\tilde{\lambda}}{s_{0}}\left|z_{1}\right|^{-\rho}\Bigr)+\left|\epsilon_{1}(\boldsymbol{x})\right|\left|z_{1}\right|+\left|\epsilon_{1}(\boldsymbol{x})\right|\left\Vert \boldsymbol{z}_{2}\right\Vert _{*}.
\end{align*}
Here the equality is again due to the definition of $\alpha$ which
implies $\alpha s_{0}=s_{0}-\rho$, and the last inequality follows
because $(1-u)^{a}\leq1-au$ holds for all $0\leq u,a\leq1$. As $\left|\epsilon_{1}(\boldsymbol{x})\right|=o(\left|\boldsymbol{x}\right|^{-\rho})$
by assumption, we have, for $\left|z_{1}\right|$ large enough, $\left|\epsilon_{1}(\boldsymbol{x})\right|\left|z_{1}\right|=\left|\epsilon_{1}(\boldsymbol{x})\right|\left|z_{1}\right|^{\rho}\left|z_{1}\right|^{1-\rho}<\frac{\tilde{\lambda}}{s_{0}}\left|z_{1}\right|^{1-\rho}$,
and
\begin{equation}
\left(E\left[\left|\overline{g}(\boldsymbol{x})+\varepsilon_{1}\right|^{s_{0}}\right]\right)^{1/s_{0}}\leq\left|z_{1}\right|\bigl(1-\tilde{\lambda}_{2}\left|z_{1}\right|^{-\rho}\bigr)+\left|\epsilon_{1}(\boldsymbol{x})\right|\left\Vert \boldsymbol{z}_{2}\right\Vert _{*},\label{eq:Th2_PrStep2_2b}
\end{equation}
where $\tilde{\lambda}_{2}\in(0,1)$. As the term $1-\tilde{\lambda}_{2}\left|z_{1}\right|^{-\rho}$
in (\ref{eq:Th2_PrStep2_2b}) is positive, we can write
\[
1-\tilde{\lambda}_{2}\left|z_{1}\right|^{-\rho}=\bigl(1-\tilde{\lambda}_{2}\left|z_{1}\right|^{-\rho}\bigr)^{1/2}\bigl(1-\tilde{\lambda}_{2}\left|z_{1}\right|^{-\rho}\bigr)^{1/2}\leq\bigl(1-\tfrac{1}{2}\tilde{\lambda}_{2}\left|z_{1}\right|^{-\rho}\bigr)^{2},
\]
and arguments similar to those in the proof of Theorem 2, Step 2,
can be used. Thus, we define $\tau_{1}(z_{1})=1-\tfrac{1}{2}\tilde{\lambda}_{2}\left|z_{1}\right|^{-\rho}$
and $\tau_{2}(z_{1})=1-\tau_{1}(z_{1})$, and express inequality (\ref{eq:Th2_PrStep2_2b})
as 
\[
\left(E\left[\left|\overline{g}(\boldsymbol{x})+\varepsilon_{1}\right|^{s_{0}}\right]\right)^{1/s_{0}}\leq\tau_{1}(z_{1})\left|z_{1}\right|\bigl(1-\tfrac{1}{2}\tilde{\lambda}_{2}\left|z_{1}\right|^{-\rho}\bigr)+\tau_{2}(z_{1})\tau_{2}(z_{1})^{-1}\left|\epsilon_{1}(\boldsymbol{x})\right|\left\Vert \boldsymbol{z}_{2}\right\Vert _{*}.
\]
From this we can conclude that, for $\left|z_{1}\right|$ large,
\begin{align*}
E\left[\left|\overline{g}(\boldsymbol{x})+\varepsilon_{1}\right|^{s_{0}}\right] & \leq\bigl[\tau_{1}(z_{1})\left|z_{1}\right|\bigl(1-\tfrac{1}{2}\tilde{\lambda}_{2}\left|z_{1}\right|^{-\rho}\bigr)+\tau_{2}(z_{1})\tau_{2}(z_{1})^{-1}\left|\epsilon_{1}(\boldsymbol{x})\right|\left\Vert \boldsymbol{z}_{2}\right\Vert _{*}\bigr]^{s_{0}}\\
 & \leq\tau_{1}(z_{1})\left|z_{1}\right|^{s_{0}}\bigl(1-\tfrac{1}{2}\tilde{\lambda}_{2}\left|z_{1}\right|^{-\rho}\bigr)^{s_{0}}+\tau_{2}(z_{1})\left(\tau_{2}(z_{1})^{-1}\left|\epsilon_{1}(\boldsymbol{x})\right|\left\Vert \boldsymbol{z}_{2}\right\Vert _{*}\right)^{s_{0}}\\
 & \leq\left|z_{1}\right|^{s_{0}}\bigl(1-\tfrac{1}{2}\tilde{\lambda}_{2}\left|z_{1}\right|^{-\rho}\bigr)^{s_{0}}+\tau_{2}(z_{1})\left(\tau_{2}(z_{1})^{-1}\left|\epsilon_{1}(\boldsymbol{x})\right|\left\Vert \boldsymbol{z}_{2}\right\Vert _{*}\right)^{s_{0}}\\
 & \leq\left|z_{1}\right|^{s_{0}}\bigl(1-\tfrac{1}{2}\tilde{\lambda}_{2}\left|z_{1}\right|^{-\rho}\bigr)+\tau_{2}(z_{1})\left(\tau_{2}(z_{1})^{-1}\left|\epsilon_{1}(\boldsymbol{x})\right|\left\Vert \boldsymbol{z}_{2}\right\Vert _{*}\right)^{s_{0}}.
\end{align*}
Here the second inequality is due the convexity of the function $\left|x\right|\mapsto\left|x\right|^{s_{0}}$,
$s_{0}\geq1$, and the last one follows because $s_{0}\geq1$. By
the definition of $\tau_{2}(z_{1})$, $\tau_{2}(z_{1})^{-1}\left|\epsilon_{1}(\boldsymbol{x})\right|=(2/\tilde{\lambda}_{2})\left|z_{1}\right|^{\rho}\left|\epsilon_{1}(\boldsymbol{x})\right|$,
so that, for some positive constants $A_{1}$ and $A_{2}$,
\begin{align*}
\tau_{2}(z_{1})\left(\tau_{2}(z_{1})^{-1}\left|\epsilon_{1}(\boldsymbol{x})\right|\left\Vert \boldsymbol{z}_{2}\right\Vert _{*}\right)^{s_{0}} & \leq A_{1}\left|z_{1}\right|^{-\rho}\left|z_{1}\right|^{s_{0}\rho}\left|\epsilon_{1}(\boldsymbol{x})\right|^{s_{0}}\left\Vert \boldsymbol{z}_{2}\right\Vert _{*}^{s_{0}}\\
 & =A_{1}\left|z_{1}\right|^{s_{0}\rho-\rho}\left|\epsilon_{1}(\boldsymbol{x})\right|^{s_{0}}\left\Vert \boldsymbol{z}_{2}\right\Vert _{*}^{\rho}\left\Vert \boldsymbol{z}_{2}\right\Vert _{*}^{\alpha s_{0}}\\
 & \leq A_{2}\left|\boldsymbol{x}\right|^{s_{0}\rho-\rho}\left|\epsilon_{1}(\boldsymbol{x})\right|^{s_{0}}\left|\boldsymbol{x}\right|^{\rho}\left\Vert \boldsymbol{z}_{2}\right\Vert _{*}^{\alpha s_{0}}\\
 & =A_{2}\left|\boldsymbol{x}\right|^{s_{0}\rho}\left|\epsilon_{1}(\boldsymbol{x})\right|^{s_{0}}\left\Vert \boldsymbol{z}_{2}\right\Vert _{*}^{\alpha s_{0}}\\
 & =o(1)\left\Vert \boldsymbol{z}_{2}\right\Vert _{*}^{\alpha s_{0}}.
\end{align*}
Here the first equation is again due to the definition of $\alpha$
and the last one follows because $\left|\epsilon_{1}(\boldsymbol{x})\right|=o(\left|\boldsymbol{x}\right|^{-\rho})$
by assumption. The second inequality follows because $\left|z_{1}\right|\leq\left|\boldsymbol{z}\right|\leq c\left|\boldsymbol{x}\right|$
and similarly with $\left|z_{1}\right|$ replaced by $\left|\boldsymbol{z}_{2}\right|$
(see footnote \ref{fn:x_big_iff_z_big}). Hence, as $\alpha s_{0}=s_{0}-\rho$,
we find that, for $\left|z_{1}\right|$ large, 
\begin{equation}
E\left[\left|\overline{g}(\boldsymbol{x})+\varepsilon_{1}\right|^{s_{0}}\right]\leq\left|z_{1}\right|^{s_{0}}-\tfrac{1}{2}\tilde{\lambda}_{2}\left|z_{1}\right|^{\alpha s_{0}}+o(1)\left\Vert \boldsymbol{z}_{2}\right\Vert _{*}^{\alpha s_{0}},\label{eq:Th2_PrStep2_2c}
\end{equation}
where $o(1)\rightarrow0$ as $\left|\boldsymbol{x}\right|\rightarrow\infty$. 

To combine the cases $s_{0}<1$ and $s_{0}\geq1$, set $\tilde{\lambda}_{0}=\tilde{\lambda}_{1}\wedge\tfrac{1}{2}\tilde{\lambda}_{2}\in(0,1)$
and conclude from (\ref{eq:Th2_PrStep2_1b}) and (\ref{eq:Th2_PrStep2_2c})
that, for $\left|z_{1}\right|$ large,
\begin{equation}
E\left[\left|\overline{g}(\boldsymbol{x})+\varepsilon_{1}\right|^{s_{0}}\right]\leq\left|z_{1}\right|^{s_{0}}-\tilde{\lambda}_{0}\left|z_{1}\right|^{\alpha s_{0}}+o(1)\left\Vert \boldsymbol{z}_{2}\right\Vert _{*}^{\alpha s_{0}},\label{eq:Th2_PrStep2end}
\end{equation}
where $o(1)\rightarrow0$ as $\left|\boldsymbol{x}\right|\rightarrow\infty$.
\bigskip{}

\noindent \textbf{Step 3:} \textbf{Bounding (\ref{Drift_4})}. First
conclude from inequalities (\ref{Inequality_z2}) and (\ref{eq:Th2_PrStep2end})
that, for $\left|z_{1}\right|$ large, 
\begin{align*}
E\left[V(\boldsymbol{y}_{1})\,\left|\,\boldsymbol{y}_{0}=\boldsymbol{x}\right.\right] & \leq1+\left|z_{1}\right|^{s_{0}}-\tilde{\lambda}_{0}\left|z_{1}\right|^{\alpha s_{0}}+\overline{s}_{1}\left|z_{1}\right|^{\alpha s_{0}}\\
 & \quad+s_{1}\left\Vert \boldsymbol{z}_{2}\right\Vert _{*}^{\alpha s_{0}}-\eta_{0}s_{1}\left\Vert \boldsymbol{z}_{2}\right\Vert _{*}^{\alpha s_{0}}+o(1)\left\Vert \boldsymbol{z}_{2}\right\Vert _{*}^{\alpha s_{0}}.
\end{align*}
Furthermore, we noted earlier that the value of $s_{1}$, and hence
also the value of $\overline{s}_{1}$, can be chosen as close to zero
as desired. Therefore, for $\left|\boldsymbol{z}_{2}\right|$ large
enough and for some $\overline{\eta}\in(0,1)$,
\[
-\eta_{0}s_{1}\left\Vert \boldsymbol{z}_{2}\right\Vert _{*}^{\alpha s_{0}}+o(1)\left\Vert \boldsymbol{z}_{2}\right\Vert _{*}^{\alpha s_{0}}\leq-\overline{\eta}\left\Vert \boldsymbol{z}_{2}\right\Vert _{*}^{\alpha s_{0}}\leq-\overline{\eta}\left\Vert \boldsymbol{z}_{2}\right\Vert _{*}^{\alpha^{\text{2}}s_{0}},
\]
where the replacement of $\left\Vert \boldsymbol{z}_{2}\right\Vert _{*}^{\alpha s_{0}}$
with $\left\Vert \boldsymbol{z}_{2}\right\Vert _{*}^{\alpha^{2}s_{0}}$
is justified because $\alpha\in(0,1)$ (this replacement is needed
below). Also, as we can assume that the value of $\overline{s}_{1}$
is so small that $\tilde{\lambda}_{0}-\overline{s}_{1}>0$, we have
$-\tilde{\lambda}_{0}\left|z_{1}\right|^{\alpha s_{0}}+\overline{s}_{1}\left|z_{1}\right|^{\alpha s_{0}}=-\overline{\lambda}\left|z_{1}\right|^{\alpha s_{0}}$
where $\overline{\lambda}\in(0,1)$ (the upper bound follows because
$\tilde{\lambda}_{0}<1$, as noted above). Thus, we can conclude that,
for $\left|z_{1}\right|$ and $\left|\boldsymbol{z}_{2}\right|$ large,
\begin{align*}
E\left[V(\boldsymbol{y}_{1})\,\left|\,\boldsymbol{y}_{0}=\boldsymbol{x}\right.\right] & \leq1+\left|z_{1}\right|^{s_{0}}+s_{1}\left\Vert \boldsymbol{z}_{2}\right\Vert _{*}^{\alpha s_{0}}-\overline{\lambda}\left|z_{1}\right|^{\alpha s_{0}}-\overline{\eta}\left\Vert \boldsymbol{z}_{2}\right\Vert _{*}^{\alpha^{2}s_{0}}.
\end{align*}

Now, let $\left\Vert \boldsymbol{z}_{2}\right\Vert _{*}$ be so large
that $\overline{\eta}\left\Vert \boldsymbol{z}_{2}\right\Vert _{*}^{\alpha^{2}}\geq c>1$.
Then, 
\[
-\overline{\eta}\left\Vert \boldsymbol{z}_{2}\right\Vert _{*}^{\alpha^{2}s_{0}}=-1-\overline{\eta}\left\Vert \boldsymbol{z}_{2}\right\Vert _{*}^{\alpha^{2}s_{0}}\bigl(1-1/(\overline{\eta}\left\Vert \boldsymbol{z}_{2}\right\Vert _{*}^{\alpha^{2}s_{0}})\bigr)\leq-1-\overline{\eta}\left(1-1/c\right)\left\Vert \boldsymbol{z}_{2}\right\Vert _{*}^{\alpha^{2}s_{0}},
\]
where $\overline{\eta}\left(1-1/c\right)\in(0,1]$ and, setting $\overline{c}=\overline{\lambda}\land\left(\overline{\eta}\left(1-1/c\right)\right)$,
we have $\overline{c}\in(0,1]$ and
\[
-\overline{\lambda}\left|z_{1}\right|^{\alpha s_{0}}-\overline{\eta}\left\Vert \boldsymbol{z}_{2}\right\Vert _{*}^{\alpha^{2}s_{0}}\leq-1-\overline{c}\left|z_{1}\right|^{\alpha s_{0}}-\overline{c}\left\Vert \boldsymbol{z}_{2}\right\Vert _{*}^{\alpha^{2}s_{0}}\leq-\overline{c}\bigl(1+\left|z_{1}\right|^{\alpha s_{0}}+\left\Vert \boldsymbol{z}_{2}\right\Vert _{*}^{\alpha^{2}s_{0}}\bigr).
\]
Next note that
\[
-\overline{c}\bigl(1+\left|z_{1}\right|^{\alpha s_{0}}+\left\Vert \boldsymbol{z}_{2}\right\Vert _{*}^{\alpha^{2}s_{0}}\bigr)\leq-\overline{c}\left(1+\left|z_{1}\right|^{s_{0}}+\left\Vert \boldsymbol{z}_{2}\right\Vert _{*}^{\alpha s_{0}}\right)^{\alpha}\leq-\overline{c}\left(1+\left|z_{1}\right|^{s_{0}}+s_{1}\left\Vert \boldsymbol{z}_{2}\right\Vert _{*}^{\alpha s_{0}}\right)^{\alpha},
\]
where the first inequality follows because $\alpha\in(0,1)$ and the
second one because $s_{1}<1$ by assumption. This implies that
\[
-\overline{\lambda}\left|z_{1}\right|^{\alpha s_{0}}-\overline{\eta}\left\Vert \boldsymbol{z}_{2}\right\Vert _{*}^{\alpha^{2}s_{0}}\leq-\overline{c}\left(1+\left|z_{1}\right|^{s_{0}}+s_{1}\left\Vert \boldsymbol{z}_{2}\right\Vert _{*}^{\alpha s_{0}}\right)^{\alpha}.
\]

By the preceding discussion we can find positive (and finite) constants
$M_{i}$ and $\overline{M}_{i}$ $(i=1,2)$ such that 
\begin{align}
E\left[V(\boldsymbol{y}_{1})\,\left|\,\boldsymbol{y}_{0}=\boldsymbol{x}\right.\right] & \leq1+\left|z_{1}\right|^{s_{0}}+s_{1}\left\Vert \boldsymbol{z}_{2}\right\Vert _{*}^{\alpha s_{0}}-\overline{c}\left(1+\left|z_{1}\right|^{s_{0}}+s_{1}\left\Vert \boldsymbol{z}_{2}\right\Vert _{*}^{\alpha s_{0}}\right)^{\alpha}\nonumber \\
 & \quad\quad\quad\quad\quad\quad\quad\quad\quad+\overline{M}_{1}\boldsymbol{1}_{C_{1}}(z_{1})+\overline{M}_{2}\boldsymbol{1}_{C_{2}}(\boldsymbol{z}_{2}),\label{eq:PolynProofStep3_DriftCase1}
\end{align}
where $C_{1}=\left\{ z_{1}\in\mathbb{R}\,:\,\left|z_{1}\right|\leq M_{1}\right\} $
and $C_{2}=\left\{ \boldsymbol{z}_{2}\in\mathbb{R}^{p-1}\,:\,\left|\boldsymbol{z}_{2}\right|\leq M_{2}\right\} $.
\bigskip{}

\noindent \textbf{Step 4:} \textbf{Completing the proof}. Using the
definition $V(\boldsymbol{x})=1+\left|z_{1}(\boldsymbol{x})\right|^{s_{0}}+s_{1}\left\Vert \boldsymbol{z}_{2}(\boldsymbol{x})\right\Vert _{*}^{\alpha s_{0}}$
and letting $L\geq\overline{M}_{1}+\overline{M}_{2}$, we obtain from
(\ref{eq:PolynProofStep3_DriftCase1}) that
\begin{align*}
E\left[V(\boldsymbol{y}_{1})\,\left|\,\boldsymbol{y}_{0}=\boldsymbol{x}\right.\right] & \leq V(\boldsymbol{x})-\overline{c}V(\boldsymbol{x})^{\alpha}+L=\left(1-h(\boldsymbol{x})\right)V(\boldsymbol{x})+L,
\end{align*}
where $h(\boldsymbol{x})=\overline{c}V(\boldsymbol{x})^{\alpha-1}$.

As $\alpha\in(0,1)$ and $\overline{c}\in(0,1]$, we have $0<h(\boldsymbol{x})\leq\overline{c}$
and $h(\boldsymbol{x})\rightarrow0$, as $\left|\boldsymbol{x}\right|\rightarrow\infty$.
Comparing the above inequality with inequality (\ref{Drift_L}) (see
the proof of Theorem 2 (Part (i), Step 4)) and the properties of the
function $h(\boldsymbol{x})$ shows that we can verify Condition D
with arguments similar to those in the aforementioned proof. Specifically,
we need to show that $L<\tfrac{1}{2}h(\boldsymbol{x})\left(1-h(\boldsymbol{x})\right)^{\frac{1}{2}}V(\boldsymbol{x})$
holds for all $\left|\boldsymbol{x}\right|$ large enough. That this
holds is seen by noting that (see the definition of $h(\boldsymbol{x})$
above)
\[
\tfrac{1}{2}h(\boldsymbol{x})\left(1-h(\boldsymbol{x})\right)^{\frac{1}{2}}V(\boldsymbol{x})=\tfrac{1}{2}\overline{c}\left(1-\overline{c}V(\boldsymbol{x})^{\alpha-1}\right)^{\frac{1}{2}}V(\boldsymbol{x})^{\alpha},
\]
where $V(\boldsymbol{x})^{\alpha}\rightarrow\infty$ and $V(\boldsymbol{x})^{\alpha-1}\rightarrow0$,
as $\left|\boldsymbol{x}\right|\rightarrow\infty$.

Hence, as in the proof of Theorem 2 (Part (i), Step 4) we can conclude
that, there exist positive constants $M$ and $b$ such that, for
$C=\left\{ \boldsymbol{x}\in\mathbb{R}^{p}\,:\,\left|\boldsymbol{x}\right|\leq M\right\} $,

\[
E\left[V(\boldsymbol{y}_{1})\,\left|\,\boldsymbol{y}_{0}=\boldsymbol{x}\right.\right]\leq V(\boldsymbol{x})-\phi_{1}\left(V(\boldsymbol{x})\right)+b\boldsymbol{1}_{C}(\boldsymbol{x}),
\]
where $\phi_{1}\left(v\right)=\tfrac{1}{2}h(\boldsymbol{x})V(\boldsymbol{x})=\tfrac{1}{2}\overline{c}v{}^{\alpha}$.
This implies that Condition D holds with $\phi=\phi_{1}$. The result
follows from Theorem 1 (note that $\alpha=1-\rho/s_{0}$ so that $1-\alpha=\rho/s_{0}$).

\bigskip{}

\noindent \textbf{Case $p=1$: }As in the corresponding proof of
Theorem 2, we have $\boldsymbol{x}=x_{1}=u$, so we simply write $x$
for any of these and note the following: Model (\ref{Companion form})
reduces to $y_{t}=y_{t-1}+\tilde{g}(y_{t-1})+\varepsilon_{t}$, Assumption
1(i) becomes redundant, Assumption 1(ii.a) is automatically satisfied
with $g(x)=x+\tilde{g}(x)$, $\epsilon(x)=0$, and $d$ redundant
(as long as the condition $\left|g(x)\right|\to\infty$ as $\left|x\right|\to\infty$
still holds), and Assumptions 1(ii.b) and 2 are as when $p\geq2$.
In other words, the model can be written as $y_{t}=g(y_{t-1})+\varepsilon_{t}$
with $g$ satisfying Assumption 1(ii.b) as well as $\left|g(x)\right|\to\infty$
as $\left|x\right|\to\infty$. Note further that now $z_{1}(\boldsymbol{x})$
reduces to $x_{1}$ and we simply write $x$ in place of either of
these. Also, due to the choice $g(x)=x+\tilde{g}(x)$ we have $\overline{g}(\boldsymbol{x})=g(x)$.

We go through the changes needed in the proof of Theorem 3 in case
$p\geq2$. Note that the equality $V(\boldsymbol{x})=1+\left|z_{1}\right|^{s_{0}}+s_{1}\left\Vert \boldsymbol{z}_{2}\right\Vert _{*}^{\alpha s_{0}}$
in case $p\geq2$ reduces to $V(x)=1+\left|x\right|^{s_{0}}$ by setting
$s_{1}=0$. The beginning of the proof until (\ref{Drift_4}) remains
valid with (\ref{Drift_4}) reducing to 
\begin{align}
E\left[V(y_{1})\,\left|\,y_{0}=x\right.\right] & =1+E\left[\left|g(x)+\varepsilon_{1}\right|^{s_{0}}\right].\label{Drift_4-1}
\end{align}
Step 1 can be omitted as the term considered therein equals zero.
In Step 2, setting $\epsilon(x)=0$ inequalities (\ref{eq:Th2_PrStep2_1a})
and (\ref{eq:Th2_PrStep2_2a}) remain valid, and so does (\ref{Inequality_E|g+eps|^s_0}).
The numbered inequalities (\ref{eq:Th2_PrStep2_1b})\textendash (\ref{eq:Th2_PrStep2end})
all hold but in all of them the last term is set to zero. In Step
3, the first inequality holds with $s_{1}$, $\bar{s}_{1}$, and the
$o(1)$ term all set to zero. In the following arguments, set $\bar{\eta}=0$
and $\bar{\lambda}=\tilde{\lambda}_{0}$. Now, some slight changes
are needed. Set $\bar{c}=\bar{\lambda}/2\in(0,1)$ and assume $\left|x\right|$
is so large that $\left|x\right|^{\alpha s_{0}}\geq1/\bar{c}$. This
implies that 
\[
-\overline{\lambda}\left|x\right|^{\alpha s_{0}}\leq-1-\overline{c}\left|x\right|^{\alpha s_{0}}\leq-\overline{c}\bigl(1+\left|x\right|^{\alpha s_{0}}\bigr)\leq-\overline{c}\left(1+\left|x\right|^{s_{0}}\right)^{\alpha}
\]
similarly to the corresponding derivations in Step 3. Therefore, inequality
(\ref{eq:PolynProofStep3_DriftCase1}) holds with $s_{1}$ and $\overline{M}_{2}$
set to zero. Step 4 remains valid, so that the stated $(f,r)$-ergodicity
result is obtained from Theorem 1 with $f=V^{1-\delta(1-\alpha)}=V^{1-\delta\rho/s_{0}}=(1+\left|x\right|^{s_{0}})^{1-\delta\rho/s_{0}}$
and $\delta\in[1,1/(1-\alpha)]$. Denoting, for brevity, $\gamma=1-\delta\rho/s_{0}\in(0,1]$
note that $1+\left|x\right|^{s_{0}-\delta\rho}=1+(\left|x\right|^{s_{0}})^{\gamma}=\{[1+(\left|x\right|^{s_{0}})^{\gamma}]^{1/\gamma}\}^{\gamma}\leq\{C[1+\left|x\right|^{s_{0}}]\}^{\gamma}$
for some finite positive $C$ (due to Loève's $c_{r}$-inequality)
so that the $(f,r)$-ergodicity with $f(x)=1+\left|x\right|^{s_{0}-\delta\rho}$
follows.\qed

\bigskip{}
\bigskip{}

\noindent \textbf{Proof of Corollary to Theorem 3}. First consider
the case $p\geq2$. We find from the proof of Theorem 3 (the beginning
of Step 3) that, for $\left|z_{1}\right|$ large,
\begin{align*}
E\left[V(\boldsymbol{y}_{1})\,\left|\,\boldsymbol{y}_{0}=\boldsymbol{x}\right.\right] & \leq1+\left|z_{1}\right|^{s_{0}}-(\tilde{\lambda}_{0}-\overline{s}_{1})\left|z_{1}\right|^{\alpha s_{0}}+s_{1}\left\Vert \boldsymbol{z}_{2}\right\Vert _{*}^{\alpha s_{0}}-(\eta_{0}s_{1}-o(1))\left\Vert \boldsymbol{z}_{2}\right\Vert _{*}^{\alpha s_{0}},
\end{align*}
where $\overline{s}_{1}$ is so small that $\tilde{\lambda}_{0}-\overline{s}_{1}>0$
holds and $o(1)\rightarrow0$ as $\left|\boldsymbol{x}\right|\rightarrow\infty$.
Hence, defining $\overline{\eta}\in(0,1)$, $\overline{M}_{1}\boldsymbol{1}_{C_{1}}(z_{1})$,
and $\overline{M}_{2}\boldsymbol{1}_{C_{2}}(\boldsymbol{z}_{2})$
as in the proof of Theorem 3 (Step 3), we have
\begin{align*}
E\left[V(\boldsymbol{y}_{1})\,\left|\,\boldsymbol{y}_{0}=\boldsymbol{x}\right.\right] & \leq1+\left|z_{1}\right|^{s_{0}}+s_{1}\left\Vert \boldsymbol{z}_{2}\right\Vert _{*}^{\alpha s_{0}}-(\tilde{\lambda}_{0}-\overline{s}_{1})\left|z_{1}\right|^{\alpha s_{0}}-\overline{\eta}\left\Vert \boldsymbol{z}_{2}\right\Vert _{*}^{\alpha s_{0}}\\
 & \qquad\qquad\qquad\qquad\qquad+\overline{M}_{1}\boldsymbol{1}_{C_{1}}(z_{1})+\overline{M}_{2}\boldsymbol{1}_{C_{2}}(\boldsymbol{z}_{2}),
\end{align*}
and setting $c_{1}=(\tilde{\lambda}_{0}-\overline{s}_{1})\land\overline{\eta}$,
\begin{align*}
E\left[V(\boldsymbol{y}_{1})\,\left|\,\boldsymbol{y}_{0}=\boldsymbol{x}\right.\right] & \leq1+\left|z_{1}\right|^{s_{0}}+s_{1}\left\Vert \boldsymbol{z}_{2}\right\Vert _{*}^{\alpha s_{0}}-c_{1}\left(\left|z_{1}\right|^{\alpha s_{0}}+\left\Vert \boldsymbol{z}_{2}\right\Vert _{*}^{\alpha s_{0}}\right)+\overline{M}_{1}\boldsymbol{1}_{C_{1}}(z_{1})+\overline{M}_{2}\boldsymbol{1}_{C_{2}}(\boldsymbol{z}_{2}).
\end{align*}
As $V(\boldsymbol{x})=1+\left|z_{1}(\boldsymbol{x})\right|^{s_{0}}+s_{1}\left\Vert \boldsymbol{z}_{2}(\boldsymbol{x})\right\Vert _{*}^{\alpha s_{0}}$
and $\alpha s_{0}=s_{0}-\rho$ we can write this, for all $\boldsymbol{x}$,
as
\begin{align*}
E\left[V(\boldsymbol{y}_{1})\,\left|\,\boldsymbol{y}_{0}=\boldsymbol{x}\right.\right] & \leq V(\boldsymbol{x})-c_{1}\left(\left|z_{1}(\boldsymbol{x})\right|^{s_{0}-\rho}+\left\Vert \boldsymbol{z}_{2}(\boldsymbol{x})\right\Vert _{*}^{s_{0}-\rho}\right)+\overline{M}_{1}\boldsymbol{1}_{C_{1}}(z_{1}(\boldsymbol{x}))+\overline{M}_{2}\boldsymbol{1}_{C_{2}}(\boldsymbol{z}_{2}(\boldsymbol{x})).
\end{align*}
From Theorem 14.3.7 of \citet{meyn2009markov} we now find that $\pi\left(\left|z_{1}(\boldsymbol{x})\right|^{s_{0}-\rho}+\left\Vert \boldsymbol{z}_{2}(\boldsymbol{x})\right\Vert _{*}^{s_{0}-\rho}\right)<\infty$
and, by the equivalence of vector norms in $\mathbb{R}^{p}$, $\pi\left(\left|z_{1}(\boldsymbol{x})\right|^{s_{0}-\rho}+\left|\boldsymbol{z}_{2}(\boldsymbol{x})\right|^{s_{0}-\rho}\right)<\infty$
also holds. Furthermore, as $\left|z_{1}(\boldsymbol{x})\right|^{s_{0}-\rho}+\left|\boldsymbol{z}_{2}(\boldsymbol{x})\right|^{s_{0}-\rho}\geq c_{2}\left(\left|z_{1}(\boldsymbol{x})\right|+\left|\boldsymbol{z}_{2}(\boldsymbol{x})\right|\right)^{s_{0}-\rho}\geq c_{2}\left|\boldsymbol{z}(\boldsymbol{x})\right|^{s_{0}-\rho}$
and $\left|\boldsymbol{z}(\boldsymbol{x})\right|^{s_{0}-\rho}=\left|\boldsymbol{A}\boldsymbol{x}\right|^{s_{0}-\rho}\geq c_{3}\left|\boldsymbol{x}\right|^{s_{0}-\rho}$
for some $c_{2},\,c_{3}\in(0,\infty)$ (that depend on $s_{0}$ and
$\rho$), it follows that $\pi\left(\left|\boldsymbol{x}\right|^{s_{0}-\rho}\right)<\infty$. 

In the case $p=1$, the above arguments hold if one sets $\overline{s}_{1}=0$,
$c_{1}=\tilde{\lambda}_{0}$, and drops all the terms related to $\boldsymbol{z}_{2}$.\qed

\bigskip{}
\bigskip{}

\end{document}